\documentclass[12pt]{iopart}
\usepackage[dvips]{graphicx} 
\usepackage{setstack}
\usepackage{iopams}
\newcounter{MaxMatrixCols}
\renewenvironment{matrix}{\begin{array}{*{\value{MaxMatrixCols}}{c}}}{\end{array}}%
\renewenvironment{pmatrix}{\left(\begin{array}{*{\value{MaxMatrixCols}}{c}}}{\end{array}\right)}%
\def\beq{\begin{equation}}
\def\eeq{\end{equation}}
\def\bea{\begin{eqnarray}}
\def\eea{\end{eqnarray}}
\def\l<{\langle}
\def\r>{\rangle}
\def\ep{\epsilon}
\def\+{\!+\!}
\def\-{\!-\!}
\def\al{\alpha}
\def\be{\beta}

\def\ep{\epsilon}
		\def\Th{\Theta}

\def\sig{\sigma}	

\def\CE{{\cal E}}	

\begin{document}
\title[Disordered $O(n)$ Loop Model and Coupled Conformal Field Theories]{Disordered $O(n)$ Loop Model and Coupled Conformal Field Theories}

\author{Hirohiko Shimada}

\address{Department of Basic Sciences, University of Tokyo\\ 
Komaba 3-8-1, Meguro-ku, Tokyo 153-8902, Japan}
\ead{shimada@dice.c.u-tokyo.ac.jp}
\begin{abstract}
A family of models for fluctuating loops 
in a two dimensional random background is analyzed.
The models are formulated as $O(n)$ spin models 
with quenched inhomogeneous interactions.
Using the replica method,
the models are mapped to the $M\rightarrow 0$ limits of $M$-layered $O(n)$ models
coupled each other via $\phi_{1,3}$ primary fields.
The renormalization group flow is calculated in the vicinity of
the decoupled critical point, by an epsilon expansion around the Ising point ($n=1$),
varying $n$ as a continuous parameter. 
The one-loop beta function suggests the existence of 
a strongly coupled phase ($0<n<n_*$) near the self-avoiding walk point ($n=0$)
and a line of infrared fixed points ($n_*<n<1$) near the Ising point.
For the fixed points, the effective central charges are calculated.
The scaling dimensions of the energy operator and the spin operator
are obtained up to two-loop order. 
The relation to the random-bond $q$-state Potts model is briefly discussed.
\end{abstract}
\section{Introduction}
A variety of random, or disordered systems are known to have non-trivial universalities.
These include the universality consisting of the well-known three symmetry classes 
which appear in the weak localization effects to the conductance \cite{hikami}
or the level correlations \cite{efetov} in disordered electronic systems, 
and the universality in the wave function multifractality at the integer quantum Hall transition in which disorder plays an essential role \cite{mf};
even the universality in the distribution of the zeros of the Riemann zeta function
is expected to have its origin in some chaotic system  \cite{berry}, 
and it may possibly be closely related to some disordered system.
It is amazing that some of these systems in appropriate limits are 
phenomenologically well described by such simple ansatz as the one
 in the random matrix theory.  
It is, however, by no means certain that these phenomena are well understood theoretically.
The most fundamental problem to be understood is 
how the universalities in disordered systems emerge 
from a microscopic structure of each system.

A lot of effort has been made to investigate this direction, 
and in some cases we have a partial answer;
we can first formulate a system microscopically, then 
by using a very crude approximation (such as `dimensional reduction'
used in disordered electronic systems \cite{efetov}), 
deduce universality directly in some carefully chosen limit.
Even under such fortunate circumstances, 
it is usually the case that we do not have much knowledge 
on the way along which the system deviates from the universality away from the limit.
This is, roughly speaking, because 
a disordered system formulated on a microscopic ground typically has 
highly nonlinear interactions induced by disorder and 
becomes almost intractable without any crude approximations.
Thus, it is important to study a simple model 
which is tractable with a sensible approximation; 
analysis of such a model and ideas used there may lead to 
a practical way to obtain qualitative results on
the deviation from the universality in more complicated disordered systems
and to an insight on the emergence of the universality itself.

With this broad motivation in mind, among the diverse disordered systems, 
we take a problem in statistical physics, 
where relations between microscopic properties and macroscopic behavior  
of models are often quite intelligible.
More specifically, we study a model with quenched disorder defined on a lattice
focusing on its possible critical behavior.     
Understanding the behavior of such a system 
governed by the Hamiltonian which contains quenched inhomogeneous interactions
is especially important, both from the practical perspective that 
there are no translationally invariant perfect crystals in the real systems
and from theoretical interests inspired by the other disorder-induced phenomena. 

As is well known, the solid notion of the universality class has been established 
in the field of critical phenomena without disorder. 
Nonlinear interactions inherent in a model 
are essential to understand the existence of 
the non-trivial universality classes \cite{wilsonfisher};
one should confront with the nonlinearities in 
a more powerful framework than the mean field approximation.
The ideas of scaling and renormalization group (RG) give such a framework.
In this framework, one introduces a theory space and a vector field on it, 
which generates a flow that indicates the scale dependence of a theory considered.
Assuming critical points in the second order phase transitions are described by 
scale invariant field theories, 
they correspond to the fixed points of the flow.
The flow can be determined, step by step, by a path integration on the fluctuations 
belonging to a certain scale;
the nonlinear interactions induce the mutual coupling 
among degrees of freedom at different scales.
As a result, the global configuration of the flow can become non-trivial.  
Then we can read off the universality classes from the sets of RG eigenvalues 
which characterize the flow linearized around the fixed points.
Furthermore, it is known that the ideas of RG are not only powerful 
to resolve the universality classes in the critical phenomena of pure systems 
but also flexible enough to deal with some disordered systems.
  
For the target of this paper, a system governed by an inhomogeneous Hamiltonian, 
it is natural to consider a configuration of the interactions 
as one realization from an ensemble that respects given probability distribution.
Basic observables to be discussed can be 
related to the free energy in a large enough system, 
which is expected to be self-averaging over the disorder distribution. 
We should therefore study the quenched average: 
the average over the disorder distribution, of the free energy 
obtained by tracing over the statistical mechanical degrees of freedom.

Actually, evaluating the quenched average of the free energy 
is notoriously difficult task,
since we should evaluate the average of the logarithm of the partition function 
in a non-translational invariant realization of the disorder.
It is practical to use either one of the two methods: replica \cite{ludwig} 
or supersymmetry \cite{efetov, bernard}.
When these methods are applicable, 
we are left with 
an effective theory with the translational invariance, 
but this time, with the additional nonlinear interactions induced by the disorder.
It should be noted that the system acquires enhanced symmetry, namely,
the replica permutation symmetry or the supergroup symmetry, respectively.

For a weakly disordered system, 
one can consider the disorder-induced couplings in the effective theory as 
a perturbation to the corresponding pure theory.
A natural question is whether the induced coupling destructs
the critical phenomena of pure system, or not.
A basic and general result on this direction is known as the ``Harris criterion" 
that tells when the disorder can be neglected at large scales;
if the couplings are irrelevant in the RG sense, 
the disorder can not change the universality class of the system from that of the pure system \cite{harris}.
However when they are relevant or marginal,
one should work harder to analyze the flow, namely, 
proceed to calculation of loop diagrams formed by the disorder-induced couplings.
This procedure corresponds to the calculation of the RG beta function up to the second or higher order in the couping.  
In the case that the beta function has a zero in addition to the one corresponding
to the pure fixed point,
the flow can (depending on the sign of the beta function) transfer, at large scale,
the theory to another fixed point; 
one recognizes that the disordered system belongs to a non-trivial universality class.
The RG eigenvalues belonging to the universality class can be predicted 
perturbatively. 
In general, without a special reason to be integrable,
one can not solve given interacting statistical model.
Therefore, the importance of such a crossover from one fixed point to the other 
can not be overemphasized.

We shall study a one-parameter family of disordered models 
and discuss their universality class in two dimensions.
Thus, let us briefly note the special role of two dimensions 
in the study of the universality class of pure systems first, 
and then comment on the current status of corresponding study in disordered systems.
Working in two dimensions provides us with an ideal circumstance 
to study the crossover in depth.
First, in two dimensions the conformal symmetry is infinite dimensional,
 and assuming the conformal symmetry (instead of the scale invariance only)
enables us to classify the possible fixed points of unitary theories with minimal symmetry \cite{BPZ}.
Second, it can be shown that  
the RG flow is irreversible for unitary theories (``$c$-theorem") \cite{zamolodchikovc}. 
Third, there are known examples of the crossover 
in which, at the non-trivial fixed point
reachable via the flow from other fixed points, the realized enhanced-symmetries are known and
the non-perturbative results can be obtained \cite{KZ, witten, shankarread}.

In the disordered systems in two dimensions, on the other hand, 
fixed points created by disorder are expected to be described by CFT's, 
but by more general, non-unitary ones. 
The lack of the unitarity gives rise to the challenging problems
of determining the universality class in disordered systems. 
Major current examples concerned with the restrictive cases 
where the supersymmetry method can be used.
They typically lead to logarithmic CFT's where the disorder-induced coupling is 
marginally irrelevant \cite{bernard,gurarie,maassarani}.
In these cases, the coupling goes to zero under the RG flow, 
and the models do not cause 
any non-trivial crossover.
There is also a consideration on the running of the effective central charge
defined on the basis of the supersymmetry \cite{gurarie}, 
in analogy with the $c$-theorem in pure (unitary) systems \cite{zamolodchikovc}.   
At present, however, not much is known about the crossover cases,
and hence our understanding of the disordered critical points is quite limited.

One of a few exceptional examples showing the crossover  
is the random-bond $q$-state Potts model in two dimension\footnote{
Other exceptions includes the disordered Dirac fermion problems 
\cite{guruswamy, bernardleclair}.} 
\cite{ludwig, ludwigcardy, dotsenkopicco}.
The model at $q=2$ is the random-bond Ising model 
and does not show the crossover; 
it is equivalent to the random-mass fermion model, or using the replica method, 
to the multi-color Gross-Neveu model,
in all of which the interaction is marginally irrelevant \cite{bernard}.
Now in the $\mathbb{Z}_2$-invariant scalar field theory (without disorder),
a nontrivial fixed point emerges when the dimension $d$ is considered 
as a continuous parameter in the range $d<4$ \cite{wilsonfisher}.
In the random-bond $q$-state Potts model, it is also fundamental   
to consider the parameter $q$ as a continuous number. 
One has then a non-trivial fixed point in the region $q>2$.
The perturbative calculation of the RG eigenvalues belonging to 
this non-trivial universality class  is well under control.
In particular, the theoretical prediction for the exponent of spin-spin correlation function 
is in good agreement with the numerical simulation \cite{cardyjacobsen}.

In this paper, we introduce the disordered version of the $O(n)$ model\footnote{
The $O(n)$ model is another natural extension of the Ising model 
other than the $q$-state Potts model.} 
and study the crossover in it.
Our model has its own physical importance at certain integral values of $n$;
it corresponds to the random-bond XY $(n=2)$, the random-bond Ising $(n=1)$ 
and the polymers (or, self-avoiding walks) in random environment $(n=0)$.
But more importantly, we consider the models for continuous values of $n$; 
this leads to another example of the crossover in disordered system.

For continuous values of $n$, as explained below,
we have a family of models which describe fluctuating loops 
in a two dimensional random background.
As is well known, elementary excitations in a spin system like the  $O(n)$ model 
can be considered as non-local geometrical objects, namely, loops 
in the high-temperature expansion.
It has been repeatedly emphasized that 
these loops are analogous to the closed trajectories of some particles 
\cite{feynman, polyakov}. 
Now the parameter $n$ controls a statistics of particles 
in the random potential; 
we expect distinct, non-trivial behavior according to the value of $n$. 
In order to study these, we use the replica method as in the studies of 
the random-bond Potts model \cite{ludwig, ludwigcardy, dotsenkopicco};
the method leads us to consider several, say $M$, layers of two dimensional $O(n)$ models coupled each other via the disorder-induced coupling,
and to take the $M\rightarrow 0$ limit in the end.   
   
The loops in the pure $O(n)$ model become scale invariant at critical temperature 
for $|n|\leq 2$.
We investigate critical behavior in the disordered $O(n)$ model 
using conformal perturbation theory around the one-parameter family of CFT's 
corresponding to a line of the pure $O(n)$ critical points \cite{dotsenkofateev}.
The existence of the crossover in our model suggests that 
there is a one-parameter family of non-unitary CFT's. 
In this respect, we mention 
recent development of the stochastic Loewner evolution (SLE) \cite{bauerbernard},
and its application to non-unitary critical points \cite{ludwigwiegmann,santachiara}.
The line of the fixed points in our model may serves as 
a natural target of such study in disordered system.

The organization of the paper is the following.
In Section 2, we formulate the model on a lattice, 
and explain the types of the quenched disorder considered. 
Then we use the replica method and take the disorder average.
As a result, we reach an intriguing picture of particles 
going up and down across the two dimensional layers, 
thus forming a whole connected diagrams. 
Then we discuss the relation between the
observable on a lattice and the scaling fields in a continuum limit. 
We see that the nontrivial cases in which the disorder is relevant occur for $n<1$.
In Section 3, we perform one loop calculation and 
discuss the existence of a non-trivial fixed point.
One loop beta function suggests the existence of a threshold $n^*$;
the fixed point exists for $n^*<n<1$, while 
the strongly coupled phase exists in $0<n<n^*$. 
In Section 4, we perform the two loop calculation for $n^*<n<1$, using the full information of the four-point function provided by the $O(n)$ CFT.
The next-leading order correction for the thermal exponent and  
the lowest order correction for the spin exponent are then found.
In Section 5, we calculate the effective central charge defined in the replica formalism.
We find that this increases, along the flow, 
against the c-theorem which is responsible for unitary theories.    
We conclude in Section 6, and comment on few further directions.
Appendix A provides the formulas on the critical Liouville field theory used in the paper. 
In Appendices B and C, we describe the calculation of the integral 
in two loop calculation of the beta function and in the spin scaling dimensions, respectively. 
Finally, Appendix D is devoted to the derivation of the integral formula 
and the expansion techniques.
The main formula takes the form of the scattering amplitude
reflecting the picture that 
the particle forming the loop can propagate via intermediate states 
while going across the replica layers.
\section{Formulation of the Model}\label{formulation}
In this section, we formulate disordered $O(n)$ loop models on a lattice
and discuss their continuum limit.
In Section \ref{lattice}, 
three types of the disordered lattice models are mapped to
homogeneous coupled lattice models by the replica method.
We introduce known CFT for the homogeneous $O(n)$ model
and discuss the relation between lattice and continuum in Section \ref{homogeneous}.
In Section \ref{scalingdisordered},  the effective action for the continuum 
disordered model is derived 
making use of the operator product expansion (OPE).
\subsection{Disordered Loop and Coupled Loop Model on a Lattice}\label{lattice}
We start with the partition function of pure $O(n)$ model on a two-dimensional lattice: 
\beq
 Z(t,n) = \int\prod_i \mu(s_i) d^ns_i\ \prod_{\l< i,j \r>}\left(1+t s_i\cdot s_j\right),
 \label{ztn}
\eeq
where  $s_i$ is a $n$-component spin on a site $i$, 
and $\mu(s)$ represents a measure on the isotropic internal space; 
using the notation $\Tr_{s_i}$ for the tracing operation $\int \prod_i\mu(s_i)d^n s_i\cdot$, 
it satisfies 
$\Tr_{s_i} 1=1$, $\Tr_{s_i} s_i = 0$ and $\Tr_{s_i} (s_i\cdot s_i)=n$.
The interaction is short-ranged, and the notation $\l< i,j \r>$ refers to a link between the nearest-neighbor sites of the lattice. 

A basic idea in this paper is to continue the parameter $n$ to non-integral values
and to take advantage of known continuum properties of the corresponding $O(n)$ model
\cite{dotsenkofateev}.  
On a lattice, the partition function (\ref{ztn}) of the spin system 
with an arbitrary value of $n$ can be interpreted as a model for fluctuating loops. 
It becomes especially simple on a honeycomb lattice, and then
tracing over the spin degrees of freedom yields,
\beq
Z(t,n)=\sum_{\mathrm{config.\ of\ loops}} t^{\mathrm{\# bonds}} n^{\mathrm{\# loops}}, 
\label{tn}
\eeq
where the summation is taken over the configuration of the closed loops \cite{domany}.
Now the weight per bond of loops is $t$, while 
the weight per closed loop is $n$, which is called a fugacity
and can take non-integral values.
The model in (\ref{ztn}) shows universal critical behavior when $|n|\leqslant 2$,
and we may use the same continuum description regardless of specific lattice structures
\footnote{This is valid in the dilute phase but not in the dense phase.
For instance, the intersections which may occur in the model on a square lattice
become relevant in the dense phase 
and discriminate it from the model on the honeycomb lattice.
For an explanation of the dilute and the dense phase, see below.}. 

Another interpretation to make sense out of the non-integral values of $n$ is
 possible for $|n| \leqslant 2$.   
By orienting the loops, one can assign a complex Boltzmann weight  
$e^{i\chi}$ ($e^{-i\chi}$) to each clockwise (anti-clockwise) loop
with a phase angle $\chi$ determined from the relation $n=e^{i\chi}+e^{-i\chi}=2\cos \chi$.
This notion of the oriented loops here is standard in the coulomb gas (CG) methods 
\cite{nienhuis}, 
where the loops are mapped to the level lines of a height model \cite{nienhuis, kondev,wiegmann}.
One can also consider the loop as a closed trajectory of some particle.
The local phase factor $\exp(i\theta\chi/2\pi)$ is then associated, at a site, with 
each turn of the particle through an angle $\theta$.
This is a lattice version of the spin factor discussed in \cite{polyakov}.
In a sense, the value of $n$ controls statistics of the particles.

The qualitative behavior of the $O(n)$ model can be summarized as follows.
When $|n| \leqslant 2$, the model  has three different phases 
separated by a critical point 
$t=t_c$\footnote{On a hexagonal lattice, to be concrete, it is given by $t_c(n)=(2+\sqrt{2-n})^{-1/2}$.} .
The region $t<t_c$ corresponds to the high temperature phase of the spin model,
and the length of loop measured by the unit of the lattice spacing is finite.
On the other hand, the average length of the loop is divergent either at $t=t_c$ or in $t>t_c$.
The $O(n)$ model at the critical point $t=t_c$ is called in the ``dilute" phase, 
since the fraction of the number of sites visited by some loop is zero. 
In $t>t_c$, this fraction becomes non-zero and the corresponding phase is called ``low temperature" or ``dense" phase.
It is known that the shapes of the loops in the dilute and the dense phase are described by 
the ${\rm SLE}_{\kappa}$ with the Brownian motion amplitude 
$\kappa<4$ and $\kappa>4$, respectively \cite{bauerbernard}.  

We now give the partition function of the random model as
\beq
 Z\left[\{t\},n\right] =  \Tr_{s_i} \ \prod_{\l< i,j \r>}\left(1+t_{ij}s_i\cdot s_j\right),
\label{ztn_dis}
\eeq
where the local interactions $t_{ij}$ between spins are position dependent 
and the notation $\{t\} $ refers to some definite configuration of the interaction.
We consider the configuration $\{t\} $ as a realization taken from some ensemble
with a probability distribution functional $P[\{t\}]$.
One might assume a non-local probability distribution functional,
but here we restrict ourselves to study the case of short range correlation. 
This means that 
the interaction $t=t_{ij}$ on each link independently respects single distribution function $P(t)$.  
Later, we shall let $P(t)$ a Gaussian-like distribution.

\begin{figure}[!htbp]
\begin{center}
\includegraphics[width=5.4cm]{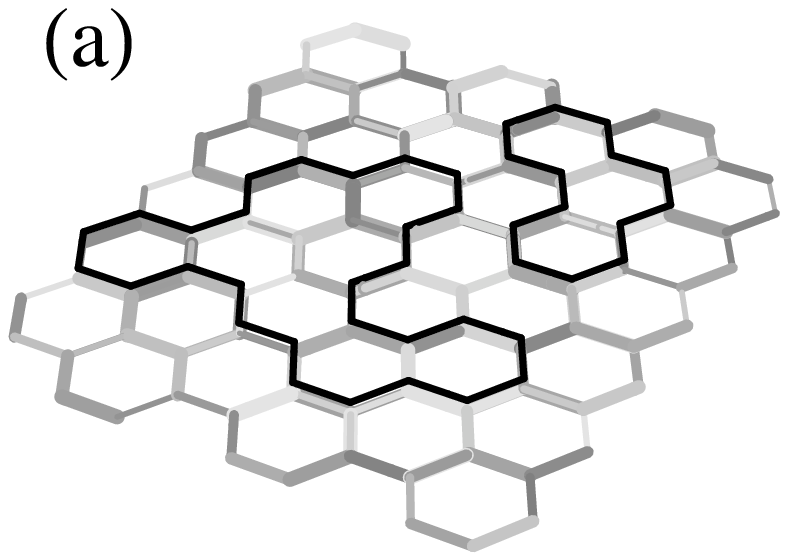}
\hspace{1cm}
\label{loop}
\includegraphics[width=7.8cm]{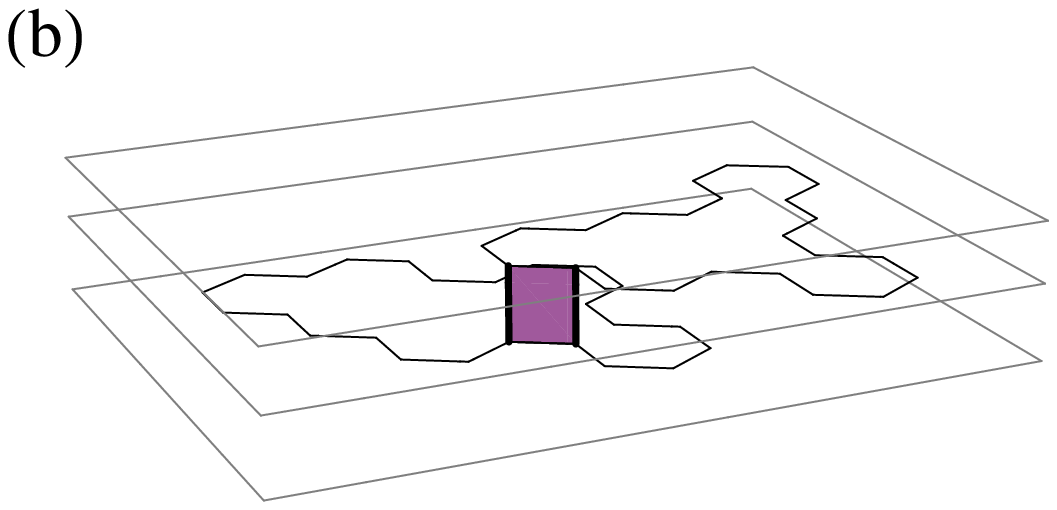}
\label{replica}
\caption{(a) Loops on a inhomogeneous lattice. 
The grayscale on each link represents the magnitude of the weight $t_{ij}$.
(b) Loops on several sheets of homogeneous lattices are coupled each other. An example of a term with the first order in the coupling $t_2$.}
\label{figurelattice}
\end{center}
\end{figure}

Given an inhomogeneous realization $\{t\}$, 
the tracing over the spin degrees of freedom is still possible: 
\beq
Z(\{t\},n)=\sum_{\mathrm{config.\ of\ loops}}n^{\mathrm{\# loops}}\prod_{l=1}^{\mathrm{\# loops}} t_{a_l(1) a_l(2)}t_{a_l(2) a_l(3)}\cdots t_{a_l(L_l) a_l(1)}, \label{ntt}
\eeq
where $a_l (i)$ ($i\in\{1,2, \cdots, L_l\}$) denotes a site 
which belongs to the $l$-th loop of a length $L_l$.  
Now the path of the particles forming loops should avoid the links with higher cost; 
we expect its behavior to change, according to the value of $n$.
The situation is, to some extent, analogous to the problem of the electrons in a random potential 
which has been studied in connection with the Anderson localization \cite{anderson}.  
As the weight $t$ is different from link to link (Figure \ref{figurelattice}-(a)), 
the exact methods applicable in the pure $O(n) $ model, 
such as the CG method, can not be applied here.

Assuming the short-range correlation between the disorder, 
we can study the self-averaging quantities such as 
the free energy and the translationally averaged correlation functions 
by calculating the quenched average \cite{cardybook}.
For example, neglecting the surface effects, the free energy of a large enough system A can be considered as 
a sum of the free energies of many macroscopically large subsystems of A each with a different realization.
Since by the short-range assumption these realizations are independent each other, 
the total free energy per site of A in the thermodynamical limit takes, with the probability one, the quenched averaged value defined by
\beq
\overline{f}=\lim_{N\rightarrow \infty}\int \prod_{\l< i,j \r>}dt_{ij}P(t_{ij})
\left(-\ln Z\left[\{t\},n\right]\right)/N.
\label{fbar}
\eeq
Here, we have used the fact that a free energy per site of a subsystem 
with $N$ sites is given by $f=(-\ln Z)/N$.
The overline is used to indicate the averaging over the distribution. 
It should be noted that, in this argument, 
we need a macroscopic number of macroscopically large subsystems.

In this paper, we use the replica method to evaluate 
the various quantities related to the quenched average of the free energy (\ref{fbar}).
This method is based on the identity:
\beq
\ln Z=\lim_{M\rightarrow 0}\frac{Z^M-1}{M}.
\eeq
Using this, the problem of calculating the quenched average $\overline{\ln Z}$ is reduced 
to the task of obtaining another quenched average $\overline{Z^M}$ correctly for $M\approx  0$.
To evaluate the latter, we first prepare the $M\in \mathbb{N}$ layers of replicated inhomogeneous $O(n)$ models with the same realization $\{t\}$,
and then take an average over the distribution.  
Since, in general, the moments of the interaction $t_{ij}$ are non-zero, 
the resulting theory is a coupled $M$ layers of $O(n)$ models.
We calculate the quantity $\overline{Z^M}$ for finite $M\in \mathbb{N}$,
take the limit $M\rightarrow 0$ in the end.
This procedure gives, at least formally, gives the desired average. 

Although we perform the detailed analysis in continuum theory, 
let us proceed, for a while, within the discrete lattice formulation. 
The purpose is two-fold: 
(i) to get some insight on underlying physics, 
as we can concretely see the elementary processes existing in the model and 
(ii) to see that there are possibly many lattice models 
described by the same continuum field theory.    
Now, we consider the $M$ copies of the model (\ref{ztn_dis}) and take the disorder average. This gives
\bea
\fl
\eqalign{
\overline{\left(Z\left[\{t\},n\right]\right)^M} &=\int \prod_{\l< i,j \r>}dt_{ij}P(t_{ij})  \prod_{a=1}^M\Tr_{s^{(a)}_i} \ \prod_{\l< i,j \r>}\left(1+t_{ij}s^{(a)}_i\cdot s^{(a)}_j\right)\\ 
&= \prod_{a=1}^M\Tr_{s^{(a)}_i} \ \prod_{\l< i,j \r>}
\left(1+t_1 \sum_{a=1}^M s^{(a)}_i\cdot s^{(a)}_j 
+t_2\sum_{a< b}^M s^{(a)}_i\cdot s^{(a)}_j \ s^{(b)}_i\cdot s^{(b)}_j \right) \\
&\hspace{1 cm} +\left(\mathrm{higher\ order\ terms}\right),}
\label{zmbart}
\eea
where in the last line, the first two moments are denoted as 
$t_1=\overline{t_{ij}}$ and $t_2=\overline{t_{ij}^2}$,
and we omit the terms with the higher moments for simplicity.
The term $t_1\,\left( s^{(a)}_i\cdot s^{(a)}_j\right)$ 
represents a walk on the same replica layer;
the generic case with $t_1\neq 0$ is discussed in the following.
In addition to this, the theory acquires nonlinear couplings.
For instance, when we look at a particular path of a loop, 
the coupling $t_2\,\left( s^{(a)}_i\cdot s^{(a)}_j \ s^{(b)}_i\cdot s^{(b)}_j\right)$ 
for $a<b$ is considered as a branching process across the replica direction
as shown in Figure \ref{figurelattice}-(b). 
On a lattice, evaluating directly the contribution of the diagrams 
with the second or higher orders in such processes 
should involve some sophisticated enumerative combinatorics.  
Notice, however, that the translational invariance of the individual $O(n)$ model is now restored.    
Thus, in the continuum limit,  we can use the knowledge of
the homogeneous $O(n)$ CFT.

The restriction on the sum to off-diagonal sector $(a<b)$ 
in the $t_2$ term 
follows from the fact that the original partition function (\ref{ztn_dis}) is linear in the bonds $(t_{ij}\,s_i\cdot s_j)$.
When we talk about the universality, however, this formal linearity and resulting off-diagonal property should be reconsidered;
we should rather consider a larger parameter space including more general couplings generated under block spin transformations. 

In order to see this, firstly, it is instructive to remind that the pure model (\ref{ztn})
corresponds to the Hamiltonian $H=-\sum \ln (1+ts_i\cdot s_j)$,
and originates from the high temperature expansion of another spin system 
which has the same $O(n)$ symmetry 
but has the different Hamiltonian $H=-\beta\sum s_i\cdot s_j$.  
From the RG perspective,
these two models are regarded as two different points in the same theory space 
determined from the $O(n)$ symmetry. 
Actually, both systems are believed to belong to the same universality class.
In fact, with the knowledge of the scaling dimensions of fields 
obtained by the CG method \cite{nienhuis},
we may argue that these two Hamiltonians are equivalent
modulo irrelevant operators.
In particular, the field $(s_i\cdot s_j)^2$ represents  
the double occupancy of the bonds in the partition function, 
and the presence of it enables the loops to overlap. 
Although such a field is not contained in the partition function (\ref{ztn}),
it is natural to think that the term $(s_i\cdot s_j)^2$ is generated, 
under block spin transformation,
from the single occupation $(s_i\cdot s_j)$.
But, as we will see in the end of Section \ref{homogeneous}, 
the term $(s_i\cdot s_j)^2$ is irrelevant
in the dilute phase, or equivalently, near the critical point.
Because of this, we can say that the linear expression (\ref{ztn})
nicely summarize the characteristics of the dilute phase in generic models 
with the $O(n)$ symmetry. 

Taking these considerations into account, 
we now briefly discuss the theory space of the coupled models 
derived from the disordered ones.
Our concern is near the decoupled critical point;
 applying the knowledge of the pure $O(n)$ model, 
we note the inessential differences that arise from different formulations 
of disordered models on a lattice.
For concreteness, let us consider another disordered model:
\beq
 Z\left[\{\beta\},n\right]=\Tr_{s_i}\prod_{\l<ij\r>}\exp\left(\beta_{ij}s_i \cdot s_j\right),
\label{zbetan}
\eeq
with a bond distribution $P(\beta)$. 
The formal operations lead to, again, a coupled model:
\bea
\fl
\eqalign{
\overline{\left(Z\left[\{\beta\},n\right]\right)^M}  
&= \prod_{a=1}^M\Tr_{s^{(a)}_i} \ \prod_{\l< i,j \r>}
\exp\left(\beta_1 \sum_{a=1}^M s^{(a)}_i\cdot s^{(a)}_j +\beta_2\sum_{a=1,\,b=1}^M s^{(a)}_i\cdot s^{(a)}_j \ s^{(b)}_i\cdot s^{(b)}_j \right) \\
&\hspace{1 cm} +\left(\mathrm{higher\ order\ terms}\right),}
\label{zmbarbeta}
\eea
where $\beta_1=\overline{\beta_{ij}}$ and $\beta_2=\frac{1}{2}\left(\overline{\beta_{ij}^2}-\overline{\beta_{ij}}^2\right)$
are the first and the second cumulants obtained from the distribution $P(\beta_{ij})$, respectively. 
Comparing this model (\ref{zmbarbeta}) with the previous one (\ref{zmbart}), 
we notice that the differences between these two are simply
the presence of the exponential function and that of the diagonal coupling: 
$\beta_2\,\left( s^{(a)}_i\cdot s^{(a)}_j \ s^{(b)}_i\cdot s^{(b)}_j\right)$ with $a=b$ 
in (\ref{zmbarbeta}). 
Both features lead to the term $\left(s_i^{(a)}\cdot s_j^{(a)}\right)^2$ {\it i.e.}
the process of the double occupancy of the bonds on the same replica plane. 
Such a term can also be generated in the model (\ref{zmbart}) 
under block spin transformation.
However, 
when the corresponding pure model is in the dilute phase, the process is irrelevant.
For this reason, we expect both of the coupled models 
(\ref{zmbart}) and (\ref{zmbarbeta}) are described by a same field theory. 

A related model, which is more suitable for taking continuum limit,
 can be formulated 
with a probability distribution $P_s(\mu)$ for disorders $\mu$ on each site:
\beq
 Z\left[\{\mu\},n\right]=\Tr_{s_i}\prod_{\l<ij\r>}
\exp\biggl[\left(\mu_i+\mu_j\right)s_i \cdot s_j\biggr].
\label{zmun}
\eeq
In this model, the bond strength $\tilde{\beta}_{ij}=\mu_i+\mu_j$
respects a distribution $\tilde{P}$ obtained by the convolution of two $P_s$ 
  {\it i.e.} $P_s * P_s=\tilde{P}$.
If this distribution $\tilde{P}(\tilde{\beta})$ is identical to the distribution $P(\beta)$
in (\ref{zbetan}), then this model becomes similar to that of (\ref{zbetan}).
Still, the model (\ref{zmun}) is different from the model (\ref{zbetan}), 
since there exists the correlation between 
the nearest-neighbor bonds (say $\tilde{\beta}_{ij}$ and $\tilde{\beta}_{ik}$) 
connected at a site.
Nevertheless, the correlation is still short range; 
we expect a similar behavior at large scales.  

We can argue that this correlation existing in the model (\ref{zmun}) 
is not essential to distinguish  it from the model (\ref{zbetan})
by examining how these two models 
are mapped into the theory space of the coupled models.
Averaging over the disorder $\mu$ in (\ref{zmun}) yields, at a site $i$,  
a term 
$\sum_{a,b,j,k}\left( s^{(a)}_i\cdot s^{(a)}_j \ s^{(b)}_i\cdot s^{(b)}_k\right)$
 as a leading nonlinear coupling.
In particular, the $j \neq k$ part of this coupling is induced by the correlation 
between the nearest-neighbor bonds in the model (\ref{zmun}). 
If this coupling were, with some special reason,
 never generated in the model (\ref{zmbarbeta}) mapped from (\ref{zbetan}), 
we would seriously distinguish (\ref{zmun}) from (\ref{zbetan}) in the first place.
But, it has already been contained in (\ref{zmbarbeta}), implicitly.
Indeed, although the expression (\ref{zmbarbeta})  
is written with respect to links,
this can be recasted in terms of sites
thanks to the discrete rotational and the discrete translational invariance of 
the coupled model on a lattice. 
Then we recognize, by expanding the exponential, 
this coupling is also present in the model (\ref{zmbarbeta}).

To summarize, in the dilute phase, there are two important processes on a lattice:
the walk on the same layer $\left(s_i^{(a)}\cdot s_j^{(a)}\right)$ 
and the branching to another layer 
$\left( s^{(a)}_i\cdot s^{(a)}_j \ s^{(b)}_i\cdot s^{(b)}_j\right)$ for $a\neq b$.
It is plausible to think that these coupled models discussed here are, at large scales,
described by a simple continuum field theory.

\subsection{Critical point of the homogeneous $O(n)$ model}\label{homogeneous}
When we approach a critical point of some lattice model defined in the dimension $d$, 
the correlation length $\xi $ of the model, 
which is usually the order of the lattice cut-off $a$,  
becomes arbitrary large.
The continuum limit (or, scaling limit) of the lattice model can be reached by 
taking $a\rightarrow 0$, while keeping $\xi$ fixed.
There is nice class of local lattice observables $\{\phi_k^{\mathrm{lat}}\}$ which have the finite limit 
\beq
\lim_{a\rightarrow 0}a^{-2(\Delta_{\phi_1}+\cdots+\Delta_{\phi_n})}\l<\phi_1^{\mathrm{lat}}(x_1)\cdots\phi_n^{\mathrm{lat}}(x_n)\r>=\l<\phi_1(x_1)\cdots\phi_n(x_n)\r>,
\eeq
for an appropriate choice of scaling dimensions $\{2 \Delta_{\phi_k}\}$.
With the idea of the block spin transformation,
the scaling fields $\phi_k(x_k)$ may be considered as coarse grained versions of $\phi_k^{\mathrm{lat}}$ 
over some region of size $L$, centered at $x_k$, with $a\ll L\ll\xi$.
These scaling dimensions are determined dynamically from the interactions in the system, 
and related to the RG eigenvalues $y_{\phi_k}$ via $y_{\phi_k}=d-2\Delta_{\phi_k}$.
When $y_{\phi_k}$ is positive (negative), the RG flow near the critical point 
is unstable (stable), and the corresponding observable and fields are relevant (irrelevant). 

Around the critical point in the theory space of the $O(n)$ model, 
there are, at least, two directions which are unstable ({\it i.e.} relevant) under 
the RG flow to the infrared.  
These directions are spanned by the coupling constants 
associated with the two most important pairs $\{\phi_k^{\mathrm{lat}},\phi_k\}$ of local lattice observables and the corresponding scaling fields. 
On the lattice side, one is the energy density $\sum_{j}s_i\cdot s_j$ and the other is the spin vector $s_i^\al$.
In the scaling limit, they renormalize to become the scaling fields called the energy operator $\CE(x)$ and the spin operator $\sig(x)$. 
The scaling dimension of $\CE(r)$ and that of $\sig(r)$ is related to the leading thermal and the leading magnetic eigenvalue of the spin system, respectively. 
These two positive eigenvalues correspond to the linearized RG flow near the critical point, 
and determine the thermodynamic exponents \cite{cardybook}.

\begin{table}
\begin{center}
\begin{tabular}{c|cccc}

 & Lattice & Continuum & primary fields & scaling dimension \\
\hline
energy & $\sum_j s_i \cdot s_j$ & $\CE(x)$ & $\phi_{1,3}$ & $-2+4\al_-^2$ \\
spin & $s_i^\al$ & $\sig(x)$ & $\phi_{p-1,p}$ & $1-\al_-^2/2-3/(8\al_-^2)$ \\
polarization & $s_i\!{}^\alpha s_i\!{}^\beta$ & $\sig_{\rm{II}}(x)$ & $\phi_{1,0}$ & $1-\al_-^2/2$ \\
encounter & $\sum_j(s_i \cdot s_j)^2$ & $\sig_{\rm{IV}}(x)$ & $\phi_{2,0}$ & $1-\al_-^2/2+ 3/(2\al_-^2)$ \\
\hline
\end{tabular}
\caption{Correspondence between lattice observables and continuum fields.
The value of $p$ is given by $p=1/(2-2\al_-^2)$.}
\label{latkac}
\end{center}
\end{table}

So far, we have described the qualitative structure of the theory space
focusing on the correspondence between the lattice observables and the continuum scaling fields.
It is tantalizing that there is no versatile scheme for a given lattice model
to establish this type of correspondence and 
to proceed to quantitative analysis using a scaling theory at hand. 
However, the $O(n)$ model was particularly successful in this respect; 
the geometric nature such as the loop representation of the partition function
made it possible to access some of the exact results on the scaling dimensions 
by the CG method \cite{nienhuis}
before the emergence of the conformal field theory \cite{BPZ}.
These exact results by the CG method, or by the other means \cite{cardyhamber}
 then led to the conjecture claiming that
a certain one-parameter family of CFT's describe the critical points of $O(n)$ model 
for the continuous value of $n\ (|n|\leq 2$) \cite{dotsenkofateev}.
This claim is further checked by the Bethe ansatz result \cite{batchelor}.

We assume this one-parameter family of CFT's \cite{dotsenkofateev} 
 in the rest of the paper. 
The correlation functions of primary fields  
are represented in terms of the vertex operators $\{e^{i\al_k \phi(x)}\}$ with 
an appropriate choice of charges $\{\al_k\}$, 
where $\phi(x)$ is a bosonic scalar field \cite{dotsenkofateev, kondev, wiegmann}. 
The necessary formulae on the vertex operator representation are given in Appendix A.
In this construction, we have two marginal operators (the screening vertex operators).
In the $O(n)$ model, the screening charge $\al_-$ is related to $n$ as 
\beq
n=-2\cos \left(\pi /\alpha_-^2\right).
\label{ncospi}
\eeq
The energy operator and the spin operator is identified with the primary fields as follows:
\beq
\CE(x)\rightarrow \phi_{1,3},\qquad \sig(x)\rightarrow \phi_{p-1,p},
\label{identification}
\eeq
where the value of $p$ is given by $p=1/(2-2\al_-^2)$.
These identifications are based on the dimensions originally obtained by 
the CG methods given in Table 1.
Using the formulas (\ref{charge}) and (\ref{dimension}), 
the dimensions of these operators are checked as
\beq
\fl2\Delta_\CE=-2\al_{1,3}\al_{\overline{1,3}}=-2(-\al_-)(\al_{+}+2\al_-)=-2+4\al_-^2,
\label{energydimension}
\eeq
\vspace{-7mm}
\bea
\fl2\Delta_{\sig}&=-2\al_{p-1,p}\al_{\overline{p-1,p}}=
-\frac{1}{2}\left((2-p)\al_{+}+(1-p)\al_-\right)\left(p\al_{+}+(1+p)\al_-\right)\nonumber\\
\fl&=1-\al_-^2/2-3/(8\al_-^2).
\label{spindimension}
\eea
where $\al_{+}\al_{-}=-1$ in (\ref{al+al-}) is used.

The relation to the two dimensional critical statistical models are summarized below.
The $O(n)$ model at $n=2$ ($\al_-^2=1$) and $n=1$ ($\al_-^2=\frac{3}{4}$) are 
the XY model and the Ising model, respectively.
The Ising model is the only minimal unitary model $\cite{BPZ}$
that belongs to the $O(n)$ family.
Note that the Ising model is also regarded as the $q$-state Potts model at $q=2$ 
\cite{ludwig}. 
The model at $n=0$ ($\al_-^2=\frac{2}{3}$) corresponds to the polymer, 
or self avoiding walk (SAW), as the qualitative result was obtained  
by $\ep=4-d$ expansion \cite{deGennes}. 
The model in two dimensions is related, under the SLE duality \cite{bauerbernard}, 
to the percolation (Potts model at $q=1$).        
Both models are non-unitary, and various scaling dimensions in the models
are numerically studied in ref. \cite{saleur}. 
Finally, the model at $n=-2$ ($\al_-^2=\frac{1}{2}$) is the loop-erased random walk (LERW).
The one-parameter family of $O(n)$ CFT's are related to the SLE by
\beq
\kappa=4\al_-^2,
\label{kappaalpha}
\eeq
where $\kappa$ is the strength of the Brownian motion which drives the evolution SLE.
These critical $O(n)$ models covers $2\leq \kappa \leq 4$.

In the rest of the sub-section, we discuss more on the correspondence between
the lattice and the continuum.
The relation 
between the loop configuration and the correlation function in the continuum theory
is given;
the latter is the object of our study henceforth.
The reader who is not interested in the lattice may skip the following.

In this paper, we restrict our consideration to the calculation of 
the RG flow in the subspace spanned by 
the coupling constant associated with the energy operator $\CE$ and 
the spin operator $\sig$.
There is, however, another important RG eigenvalue apart from 
the ones associated with these two;
the eigenvalue is responsible for the geometric property of the loops.
In other words, the thermal and the magnetic eigenvalues are not sufficient 
to characterize the local shape of the loops even at a primitive level.
To see this,  let us first remind that, in the SAW ($n=0$), the 
fractal dimension $D_F$ of the loop 
is given by the thermal eigenvalue $y_\CE$, and thus $D_F=1/\nu$ \cite{deGennes}; 
it is then given by $y_\CE=2-2\Delta_{\CE}=\frac{4}{3}$ in two dimensions. 
The generalization of this result to arbitrary $n$ $(|n|\leq 2)$ was discussed 
by relating  the $O(n)$ loops to the clusters in the percolation \cite{jankeschakel}. 
The fractal dimension for $n\neq 0$ should then be written as 
\beq
D_F=\frac{1}{\sig_F\nu},
\label{sigmaf}
\eeq with $\sig_F\neq 1$
(in the percolation theory, $\sig_F$ is called the Fisher exponent, 
and $\nu$ the correlation length exponent).
Further, the fractal dimension $D_F$ is expected to be the RG eigenvalue 
from the polarization operator (or, the two-leg operator):
$\{\phi^{\rm lat},\phi\}=\{s_i^\al s_i^\be,\sig_{\rm{II}}(x)\}$ with the vector indexes $\al\neq\beta$,
for which the scaling dimension can be obtained by the CG method (see Table 1).  
The fractal dimension of the $O(n)$ loop in the dilute phase is then given by $D_F=2-2\Delta_{\sig_{\rm{II}}}=1+\al_-^2/2$
\footnote{This fractal dimension is also derived, in a rigorous way, 
from the SLE \cite{bauerbernard}}. 

Diagrammatically, the polarization operator $\sig_{\rm{II}}(x)$,
when inserted into a correlation function, creates two curves emanating from a point $x$, 
while the spin operator $\sig(x)$ creates only one curve.
The curves created by $\sig_{\rm{II}}(x)$ repel each other; 
they have different colors $\al\neq\beta$ 
and hence can not connect by themselves.
They need to be annihilated by the other operator $\sig_{\rm{II}}(y)$.
By contrast, the insertion of the energy operator $\CE(x)$ 
require that a point $x$ be passed by some loop;
the curve passing the point will connect each end to form the loop.

However,  forming a loop becomes more and more difficult as $n$ tends to zero;
the loop segments repel each other in the SAW.
In this sense, the role of the energy operator $\CE$ approaches that of 
the polarization operator $\sig_{\rm{II}}$.
In fact, at the SAW, these two operators has the same scaling dimension $2\Delta=\frac{2}{3}$,
and this explain $\sig_F=1$ in (\ref{sigmaf}).   
In the CFT context,  by using the scaling dimensions 
obtained by the CG method,
 $\sig_{\rm{II}}(x)$ was argued to correspond to the $\phi_{1,0}$ primary field.
Then,  $\sig_F=1$ can be confirmed by the  well-known
equivalence between the primary field:
\beq
\phi_{r,s}=\phi_{m-r,m+1-s} \qquad (0\leq r \leq m,\, 0\leq s \leq m+1),
\label{rs}
\eeq
in the $m$-th minimal model with the central charge $c=1-6/m(m+1)$. 
Indeed, in the SAW, which is the $m=2$ minimal model,
$\phi_{1,3}$ (the energy operator $\CE$)
should have the same dimension as the field 
$\phi_{1,0}$ (the polarization operator $\sig_{\rm{II}}$),
since $(1,3)+(1,0)=(2,3)$.

There are, of course, many other composite fields 
constructed as products of the several local lattice observables.
A quantitative discussion on such fields is given, which is based on the OPE
and the correlation inequalities \cite{duplantierludwig}.
Their result suggests, in general situation, that the scaling dimension of a composite field exceeds the sum of those of the elementary fields
due to the repulsion between them.
This implies higher-order composite fields are more likely to be irrelevant in the RG.
In our context, we mention
the four-leg operator $\{\sum_j(s_i \cdot s_j)^2,\sig_{\rm{VI}}(x)\}$ 
as an important example \cite{nienhuis, wiegmann}.  
The insertion of the field $\sig_{\rm{VI}}(x)$ 
into a correlation function corresponds to requiring two of the loop segments to overlap; 
in the particle picture, the trajectories should encounter at the point $x$.
Again, the dimension of the field $\sig_{\rm{VI}}(x)$ is obtained by the CG method,
and identified with the primary field $\phi_{2,0}$ in the CFT.
Now, we see that it is relevant in the dense phase ($\kappa>4$) 
and irrelevant in the dilute phase ($\kappa<4$) as shown in Fig \ref{scDIM_on}.
As used in Section \ref{lattice}, the latter observation is
 a key ingredient in understanding the universality of the pure $O(n)$ model 
in the dilute phase.
Further, at the level of coupled model,
this is crucial in our expectation
that (\ref{zmbart}) and (\ref{zmbarbeta}) are, near the decoupled critical points, described by a same field theory.

\subsection{Scaling limit of the disordered $O(n)$ models}\label{scalingdisordered}
We shall go into the continuum formulation of the disordered models 
by using the CFT description of the critical $O(n)$ model.
In section \ref{lattice}, we have considered, on the lattice, three disordered models 
whose partition functions are given by (\ref{ztn_dis}), (\ref{zbetan}) and (\ref{zmun}). 
As we have seen, after the disorder averaging, 
they have much in common 
when mapped to the coupled models.
When the couplings between replicas are sufficiently small,
one could, by looking at (\ref{zmbart}) for instance, 
guess the action of the coupled models in the continuum limit.
But we shall take the other way.
Instead, we first consider the continuum limit of the disordered model (\ref{zmun}), 
and then take the disorder average in the continuum theory.
The advantage is that
we can use the OPE to deal with the composite operators.

In the following, we will use the relation:
\beq
\fl\prod_{\l< i,j \r>}(1+ts_i\cdot s_j )\sim   \exp\left[\beta \sum_{\l< i,j \r>}s_i\cdot s_j\right]  
\rightarrow \exp\left[-S_*+m\int d^2x\ \CE(x)\right].
\label{vicinity}
\eeq
The first equivalence symbol means that these two pure model
belong to the same universality class, as discussed in the paragraph above (\ref{zbetan}).
In the right hand side, we formally write the action of the $O(n)$ CFT as $S_*$.
The coupling constant $m$ is proportional to the reduced temperature, 
which is given by $(t-t_c)/t_c$, or $(\beta-\beta_c)/\beta_c$.

Using the relation (\ref{vicinity}),
it is natural to assume the continuum limit of the model (\ref{zmun}) is described by
\beq
\fl\left(Z\left[\{\mu\},n\right]\right)^M =
   \int \prod_{a=1}^M
\mathcal{D}\Phi^{(a)}(x) \ 
\exp\left[-\left(\sum_{a=1}^M S_*^{(a)}\right)+\int d^2x\ m(x)  
\left(\sum_{a=1}^M\CE^{(a)}(x) \right)\right], 
\eeq
where $S_*^{(a)}$ is the action of the $O(n)$ CFT on the replica plane $(a)$. 
Here, we formally write elementary fields (in the sense of \cite{zamolodchikov}) 
as $\Phi^{(a)}(x)$
and replace the tracing $\Tr_{s^{(a)}_i}$ in (\ref{zmun}) to $\int\mathcal{D}\Phi^a(x)$, 
which denotes the path integration over the fluctuation of $\Phi^{(a)}(x)$.
The realization $\{\mu\}$ of the local weights $\mu_i$ on lattice sites in (\ref{zmun})
are now replaced to its continuum counterpart, that is, 
a configuration of a scalar function $m(x)$.
Physically, this $m(x)$ serves as a locally fluctuating reduced temperature.
We assume this $m(x)$ respects 
a single distribution function $\hat{P}\left(m(x)\right)$ 
which is independent of the position $x$. 
Hereafter, we suppress the argument of the partition function $Z\left[\{\mu\},n\right]$.
Then the averaging operation is written as
\beq
\overline{Z^M} = \int \mathcal{D}m(x) \hat{P}\left(m(x)\right)\ Z^M,
\label{hatp}
\eeq
where $\mathcal{D}m(x)$ denotes a path integral measure $\prod_x dm(x)$.
Let $\hat{\xi}_k$ be a $k$-th cumulant of the distribution function $\hat{P}(m(x))$.
In the important case of a Gaussian distribution
$
\hat{P}(m(x))=\exp\left[-(m(x)-\hat{m}_0)^2/2\hat{g}_0\right]
$,
we have $\hat{\xi}_1=\hat{m}_0$, $\hat{\xi}_2=\hat{g}_0$ and $\hat{\xi}_k=0$ for $k\geq 3$. 
Then, by the averaging over the disorder, we obtain  
\bea
\fl\overline{Z^M}&=
   \int \prod_{a=1}^M
\mathcal{D}\Phi^a(x) \ 
\exp\left[-\sum_{a=1}^M S_*^{(a)}+\int d^2x\ 
\left(\hat{\xi}_1\sum_{a=1}^M\CE^{(a)}(x) +\hat{S}_I(x)\right)\right], \\
\fl\hat{S}_I(x)&=\hat{\xi}_2\sum_{a, b}^M \CE^{(a)}(x)\CE^{(b)}(x)
+\hat{\xi}_3\sum_{a, b ,c}^M \CE^{(a)}(x)\CE^{(b)}(x)\CE^{(c)}(x)+\cdots.
\label{beforeope}
\eea
In this formal expression, we should note that the nonlinear part $\hat{S}_I(x)$ contains the composite operators
{\it i.e.} the products of the several fields at the same point on the same replica plane.
We deal with them by using the fusion rules in the $O(n)$ model.
According to the identification $\CE\rightarrow \phi_{1,3}$ in (\ref{identification}), 
the fusion rule in the thermal sector reads
\beq
\phi_{1,3}\cdot \phi_{1,3}\sim\phi_{1,1}+\phi_{1,3}+\phi_{1,5},
\eeq
or, equivalently,
\beq
\CE\cdot \CE\sim I+\CE+\CE'.
\label{fusion}
\eeq
Here, $\CE'=\phi_{15}$ is the next-leading energy operator.
\begin{figure}[!tbp]
\begin{center}
\includegraphics{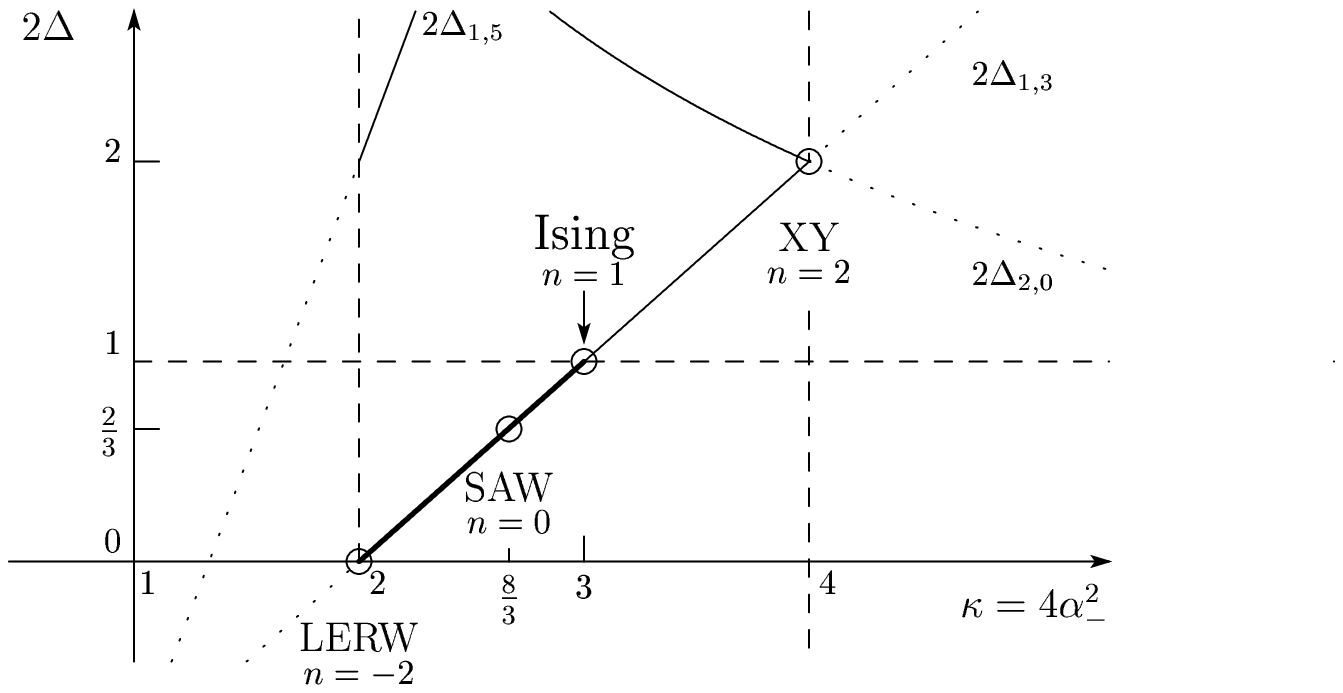}
\caption{The scaling dimensions $2\Delta$ of the primary fields $\phi_{1,3}$, $\phi_{1,5}$ and $\phi_{2,0}$
plotted as functions of the SLE parameter $\kappa=4\al_{-}$.
The horizontal line at $2 \Delta=1$ shows the marginality of 
the most relevant couping $\CE^{(a)}(x)\CE^{(b)}(x)$.}
\label{scDIM_on}
\end{center}
\end{figure}
Since this operator is irrelevant as shown in Figure \ref{scDIM_on}, we neglect 
the $\CE\cdot\CE\rightarrow\CE'$ part of the OPE (\ref{fusion}).
The identity field $I=\phi_{11}$ contributes as a trivial shift of the free energy.
The emergence of the energy operator $\CE$ in the right hand side of (\ref{fusion}) is 
a characteristic of $O(n)$ model with $n\neq1$, 
which will be discussed in the paragraph below (\ref{structure}).
Because of this contribution ($\CE\cdot\CE\rightarrow\CE$), 
qualitatively speaking, we should have a hierarchical flow of the coupling constants: 
$\cdots\rightarrow\hat{\xi}_3\rightarrow\hat{\xi}_2\rightarrow\hat{\xi}_1$.
For example, the diagonal ($a=b$) part of $\hat{\xi}_2\sum\CE^{(a)}(x)\CE^{(b)}(x)$
mixes with $\hat{\xi}_1\sum\CE^{(a)}$ 
and hence the flow $\hat{\xi}_2\rightarrow\hat{\xi}_1$ occurs 
\footnote{In the coupled model at $M\neq 0$, this flow causes shift in the critical temperature.The magnitude of this effect is proportional to 
the number of diagonal elements $M$, 
the structure constant (\ref{structure}) and $\hat{\xi_2}$.
However, since the disordered model corresponds to $M\rightarrow 0$, 
we assume it is negligible.}.
Thus, by redefining the coupling constants $\hat{\xi}_k$ , we get
\bea
\fl\overline{Z^M}& =
   \int \prod_{a=1}^M
\mathcal{D}\Phi^a(x) \ 
\exp\left[-\sum_{a=1}^M S_*^{(a)}+\int d^2x\ 
\left(m_0\sum_{a=1}^M\CE^{(a)}(x) +S_I(x)\right)\right], \\
\fl S_I(x)&=g_0\sum_{a\neq b}^M \CE^{(a)}(x)\CE^{(b)}(x)
+\xi_3\sum_{a\neq b,b\neq c,c\neq a}^M \CE^{(a)}(x)\CE^{(b)}(x)\CE^{(c)}(x)+\cdots
\label{afterope},
\eea
where we use the symbol $m_0$, $g_0$ and $\xi_k$ ($k\geq 3$) 
for the new coupling constants.
Now, the effective interaction $S_I(x)$ is defined without composite operators.
It should be noted that we restrict our considerations 
on the theory space which is replica symmetric;
we assume that the coupling constant for, say $\CE^{(a)}(x)\CE^{(b)}(x)$, 
is independent of the pair $(a,b)$.  

These coupling constants are determined by the distribution function $\hat{P}(m(x))$ 
in (\ref{hatp}), and are non-zero in general.
The massless limit ($m_0\rightarrow 0$) of the decoupled model ($g_0=0, \xi_k=0$)
remains obviously as a fixed point. 
We now change the distribution function $\hat{P}(m(x))$,
and gradually turn on the effective interaction $S_I$ in (\ref{afterope}) while keeping $m_0=0$. 
The large scale behavior is then dominated by the most relevant field: $\CE^{(a)}(x)\CE^{(b)}(x)$.
The dimension of this operator is given by $4\Delta_{\CE}$,
which is the twice the dimension of the energy operator.

As we can infer from (\ref{energydimension}), this field becomes marginal at $n=1$
($\al_-^2=\frac{3}{4}$).
Since $O(1)$ model is the Ising model, this serves as a simple check of 
the known marginality of the disorder in the random-bond Ising model \cite{ludwig, bernard}.  
Hence, we use the parametrization:
\beq
\al_-^2=\frac{3}{4}-\epsilon, 
\label{alpha34epsilon}
\eeq
to perform the epsilon expansion in the next section.

\section{Renormalization group flow in the one-loop calculation}
\setcounter{footnote}{1}
In this section, we discuss the scale dependence of the theory (\ref{afterope})
by one-loop RG calculation.
It will turn out that under certain circumstances, 
the disordered $O(n)$ model has a pair of the ultraviolet (UV) and 
the infrared (IR) fixed points,
as in the random-bond $q$-state Potts model \cite{ludwig,dotsenkopicco}.
We will investigate the properties of the IR fixed point 
by which the large scale behavior of the theory is dominated.

\subsection{The epsilon expansion and procedure of RG}\label{epsilontopology}
We shall calculate the RG flow of the coupled model (\ref{afterope}) perturbatively
in the epsilon expansion using the parameter in (\ref{alpha34epsilon}).
Generally, a flow is determined only by the most fundamental properties of a theory
such as the symmetries and the dimension of the space.
In order to grasp the topology of a theory space,
it is often useful  to think that these symmetries are dependent on some continuous parameters \cite{zamolodchikov,ludwig,dotsenkopicco,lassig}, 
or that the dimension itself is a continuous parameter \cite{wilsonfisher}.
Although the change is gradual when we look locally at a generic point as 
varying these parameters, 
the global structure of the flow, on the other hand, may change drastically at certain critical values of the parameters.
A typical topological transition is caused by a branching of one merged fixed point
into a pair of the UV fixed point and the IR fixed point. 
The idea of epsilon expansion provides us with a firm ground to perform
a perturbation calculation in the non-trivial region after the branching. 

In the parameter space, we can infer the location of the branching by 
the presence of a marginal field: the hallmark of the merged fixed points.
In the renowned example of the $\mathbb{Z}_2$-invariant scalar field theory 
\cite{wilsonfisher}, 
the most relevant interaction $\phi^4$ becomes marginal at the dimension $d=4$.
In $d=4-\ep$, the Gaussian fixed point and the Wilson-Fisher fixed point emerge; 
we can  do a solid perturbative calculation
by setting an appropriate coupling coordinate 
in which the coupling constant remains $\mathcal{O}(\ep)$. 

In the disordered $O(n)$ model, as mentioned in the previous section, 
the most relevant part of the effective interaction $S_I$ in (\ref{afterope}) 
is the term $\CE^{(a)}(x)\CE^{(b)}(x)$, which is marginal at $n=1$
and is relevant for $n<1$ (see Fig. \ref{scDIM_on}). 
The next-leading relevant term $\CE^{(a)}(x)\CE^{(b)}(x)\CE^{(c)}(x)$ is irrelevant
for $n>0$ and is marginal at $n=0$ (SAW)\footnote{ 
It may be noted that the role of this field
is analogous to that of the $\phi^6$ term in the $\mathbb{Z}_2$-invariant 
scalar field theory.
This term is irrelevant if $d>3$ and becomes marginal at $d=3$.
In $3<d<4$, we have only the two fixed points mentioned above.}. 
We restrict our analysis on $n>0$, keep the most relevant part    
\beq
S_I=g_0\sum_{a\neq b}^M \CE^{(a)}(x)\CE^{(b)}(x),
\label{perturbation}
\eeq
and discard the other higher-order terms in (\ref{afterope}).

The implementation of the RG is as follows.
We introduce a cut-off length scale $r$, which serves as an effective lattice spacing,
and determine a renormalized coupling $g(r)$ by
integrating out the short-distance degrees of freedom:
\beq
\fl
\eqalign{r^{-8\ep}g(r)\l<S_I(x)S_I(\infty)\r>_0=&g_0\l<S_I(x)S_I(\infty)\r>_0
+\frac{g_0^2}{2!}\int_{|y-x|<r} d^2y\l<S_I(x)S_I(y)S_I(\infty)\r>_0\nonumber\\
&+\frac{g_0^3}{3!}\int_{|y-x|<r,|z-x|<r} \hspace{-10mm}
d^2yd^2z\l<S_I(x)S_I(y)S_I(z)S_I(\infty)\r>_0+\cdots,}
\label{geff}
\eeq
where the insertions of the interaction fields $S_I$ are restricted onto a disk of radius $r$,
and the coupling is measured by projecting perturbative contributions on $S_I(\infty)$.
The symbol $\l<...\r>_0$ represents the unperturbed correlation function. 
A trivial scaling factor $r^{-8\ep}$ is introduced in order to make 
the renormalized coupling dimensionless. 
At this stage in RG, as is well known, a finite redefinition of the coupling is possible;
physical quantities (scaling dimensions, for instance) is invariant 
under a finite coordinate reparametrization of theory space \cite{lassig}.  
To fix this ambiguity, we choose a minimal subtraction scheme.
Then we get the renormalized coupling in the form:
\beq
g(r)=r^{8\ep}\left(g_0+G_2(\ep,r) g_0^2+G_3(\ep,r) g_0^3+\cdots\right),
\label{gr}
\eeq
where $G_2(\ep,r)$ and $G_3(\ep,r)$ have only poles in $\ep$ and no regular part
\footnote{More precisely, we keep the combination 
$r^{8\ep}/8\ep=1/8\ep+\log r+\mathcal{O}(\ep)$ and discard the other parts.}.
The beta function is then obtained by differentiating the renormalized coupling (\ref{gr}) 
with respect to $(\ln r)$.
Using the first two terms in the fusion rule (\ref{fusion}), 
we make contractions and list, in Figure \ref{betadiagrams},
the possible diagrams for the beta function. 

\subsection{One-loop beta function, a line of random fixed points and a strong coupling region}
\begin{figure}[!tbp]
\begin{center}
\includegraphics{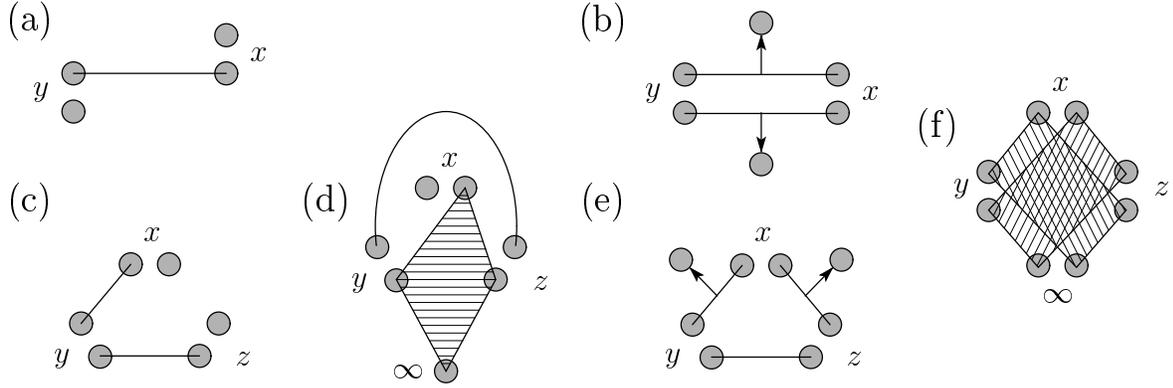}
\caption{The diagrams for the beta function.
A gray disk and a nearby pair of the disks 
represent an energy operator $\sum_a\CE^{(a)}$ and 
the interaction $S_I=\sum_{a\neq b}\CE^{(a)}\CE^{(b)}$, respectively.
An arrow corresponds to the sub-leading part of the OPE in (\ref{OPE}) and
a shaded-square represents four-point function.
The external lines to the point infinity in (\ref{geff}) are amputated.
The second order: (a) $G_{2,1}(r,\ep)$ and (b) $G_{2,2}(r,\ep)$;
the third order: (c) $G_{3,1}(r,\ep)$, (d) $G_{3,2}(r,\ep)$, 
(e) $G_{3,3}(r,\ep)$ and (f) $G_{3,4}(r,\ep)$.}
\label{betadiagrams}
\end{center}
\end{figure}

We here calculate one-loop beta function.
The process represented by the diagram in Figure \ref{betadiagrams}-(a)
involves three layers, and the number of the ways for this contraction is $4(M-2)$.
Then we have 
\bea
G_{2,1}(r,\ep)&=\frac{1}{2!}\cdot 4(M-2)\int_{|y-x|<r}d^2y \l<\CE(x)\CE(y)\r>_0\nonumber\\
&=4\pi(M-2)\left(\frac{r^{8\ep}}{8\ep}\right),
\label{g21}
\eea
where we have used $\l<\CE(x)\CE(y)\r>=|x-y|^{-4\Delta_\CE}$ 
and $4\Delta_\CE=2-8\ep$.

Besides this, up to one-loop order, there is the other contribution represented by 
Figure \ref{betadiagrams}-(b) 
which is obtained by using twice the sub-leading part of 
the OPE ($\CE\cdot \CE \rightarrow \CE$):
\bea
G_{2,2}(r,\ep)&=\frac{1}{2!}\cdot 2\int_{|y-x|<r}d^2y 
\left(\frac{C^{\CE}_{\CE\CE}(\al_-^2)}{|x-y|^{1-4\ep}}\right)^2\nonumber\\
&=2\pi\left[C^{\CE}_{\CE\CE}(\al_-^2)\right]^2 \left(\frac{r^{8\ep}}{8\ep}\right).
\label{g22}
\eea
Here, $C^{\CE}_{\CE\CE}(\al_-^2)$ is the structure constant of the CFT which appear in the OPE:
\beq
\CE(0,0)\cdot \CE(z,\bar{z})=
\frac{C_{\CE\CE}^{I}}{(z\bar{z})^{2\Delta_\CE}}I
+\frac{C_{\CE\CE}^{\CE}}{(z\bar{z})^{\Delta_\CE}}\CE
+\frac{C_{\CE\CE}^{\CE'}}{(z\bar{z})^{2\Delta_\CE-\Delta_\CE'}}\CE'+\cdots,
\label{OPE}
\eeq 
where we write just the primary operators to represent their conformal families 
including the descendants.
The structure constants are, in general, determined by  
the requirement that the operator algebra be associative. 
In practice, the crossing symmetry of the four-point function 
is strong enough to fix them \cite{BPZ}; the actual values are
obtained by using either the connection matrix of the hypergeometric function \cite{z2},
or the vertex operator representation of the four-point function \cite{dotsenkofateev2}.
This reads,
\beq
C^{\CE}_{\CE\CE}(\rho)=2(1-2\rho)^2
\frac{\gamma^{\frac{3}{2}}(\rho)}{\gamma^2(2\rho)}\frac{\gamma^{\frac{1}{2}}(2-3\rho)}{\gamma(3-4\rho)},
\label{structure}
\eeq
where we have used temporary notation $\rho=\al_-^2$ and
 $\gamma(x)=\Gamma(x)/\Gamma(1-x)$.

The structure constant $C^{\CE}_{\CE\CE}(\al_-^2)$ has information about
the  important selection rules in the pure $O(n)$ model. 
First, observe that 
the square of the structure constant (\ref{structure}) 
has a zero at $\al_-^2=\frac{3}{4}$ (the Ising model, $\ep=0$) 
and a pole at $\al_-^2=\frac{2}{3}$ (SAW, $\ep=\frac{1}{12}$).
The former zero is due to the self-duality of the Ising model \cite{cardy}.
In the vicinity of the critical point,
the duality transformation changes the sign of the reduced temperature $t$.
Since the energy operator $\CE$ couples to $t$, if the model is 
invariant under the duality, it should be odd under the duality.
Then $\CE$ is not allowed to appear in the right hand side of the fusion rule (\ref{fusion}).

The pole at $\al_-^2=\frac{2}{3}$ emerges 
because of the strong repulsion between the loop segments in the limit $n\rightarrow 0$. 
Since the normalization of operators are fixed such that $C_{\CE\CE}^{I}=1$,
what really happens is the divergence of the ratio $C_{\CE\CE}^{\CE}/C_{\CE\CE}^{I}$. 
When the two loop segments approach, they will either
(i) form a complete loop (to contribute the free energy)
or (ii) form together a joined loop segment.
Since, in the limit $n\rightarrow 0$ the process (i) is strongly suppressed by the repulsion between the loop segments,
the ratio, indeed, diverges.

\begin{figure}[!tbp]
\begin{center}
\includegraphics[width=10cm]{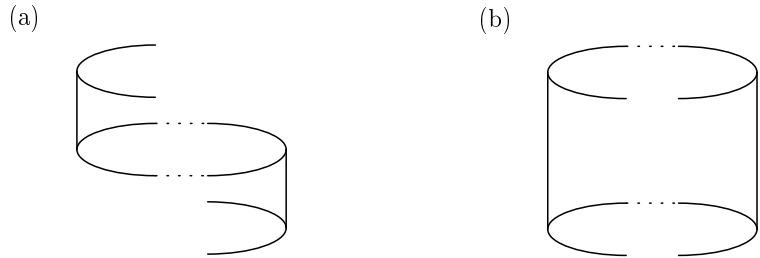}
\caption{The second order diagrams; see also Figure \ref{betadiagrams}.
 (a) $G_{2,1}(r,\ep)$: the three-layer process with a closed loop \qquad 
(b) $G_{2,2}(r,\ep)$: the two-layer process.}
\label{intuitive}
\end{center}
\end{figure}

Taking these interpretations into the account, 
more intuitive representations of the second-order contributions are possible.
The term $G_{2,1}(r,\ep)$ is represented by
the diagram in which two open segments lie on two layers and one closed loop 
on another layer (Figure \ref{intuitive}-(a)),
while the term $G_{2,2}(r,\ep)$ is represented by 
the diagram in which two open segments lie on two layers
(Figure \ref{intuitive}-(b)).

\begin{figure}[!htbp]
\begin{center}
\includegraphics[width=8cm]{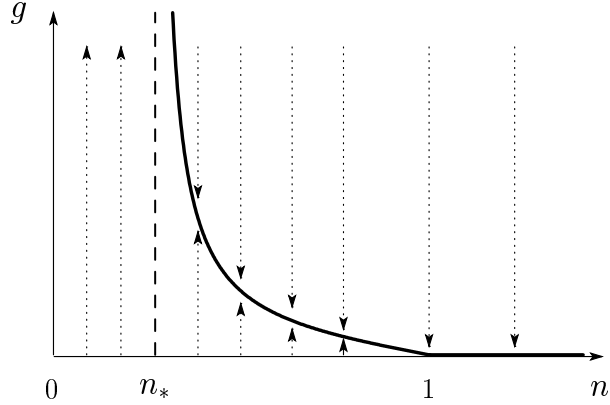}
\caption{A schematic flow diagram of the couping constant $g$ with respect to $n$ of 
the disordered $O(n)$ model suggested by the one-loop beta function (\ref{betaoneloop}). 
The bold curve represents the positions of the IR fixed points.}
\label{ng}
\end{center}
\end{figure}

Now we sum up (\ref{g21}) and (\ref{g22}) to get
\beq
g(r)=r^{8\ep}\left\{g_0+2\pi \left[2(M-2)+ \left[C^{\CE}_{\CE\CE}(\al_-^2)\right]^2 \right]
\left(\frac{r^{8\ep}}{8\ep}\right)g_0^2+\mathcal{O}(g_0^3)\right\}.
\label{gr_1}
\eeq    
By introducing $\tilde{g}=r^{8\ep}g_0$, we solve (\ref{gr_1}) with respect to $\tilde{g}$:
\beq
\tilde{g}=g(r)-\frac{2\pi}{8\ep} \left[2(M-2)+ 
\left[C^{\CE}_{\CE\CE}(\al_-^2)\right]^2 \right]
g(r)^2+\mathcal{O}\left(g(r)^3\right) .
\eeq
Using this, the beta function up to one-loop order is then obtained as\footnote{
A similar form of beta function appears in $\ep=d-2$ expansion
of the random-bond Ising model \cite{cardybook}.}
\beq
\fl\beta(g) = \frac{dg(r)}{d\ln(r)} = 8\epsilon g(r) + 2\pi \left\{2(M-2)+ 
\left[C^{\CE}_{\CE\CE}(\al_-^2)\right]^2 \right\}g(r)^2+\mathcal{O}(g(r)^3).
\label{betaoneloop}
\eeq
This beta function of the coupled model permits an IR fixed point 
when the coefficient of $g(r)^2$ is negative. 
As we take the limit $M\rightarrow 0$ for the disordered model,  
there are a competition between the terms from $G_{2,1}(r,\ep)$ and $G_{2,2}(r,\ep)$.
As a result, we get two distinct regions for $0<n<1$:
in the first case 
we have a monotonic increasing beta function;
the theory goes to the strong coupling.
In the second case we have an IR fixed point.
The one-loop beta function suggests that the IR fixed point vanishes
at $\ep=\frac{1}{12}\cdot0.770861\cdots$, which corresponds to 
$n=n_*=0.26168\cdots$.
The flow of the coupling constant is schematically shown in Fig. \ref{ng}.
It should be noted, however, that the divergence of the coupling $g$ at 
$n=n_*$ is an artifact of the one-loop calculation;
if there is a positive $g^3$ term\footnote{
Although the $\ep$ dependence of the third order coefficient 
is not calculated in this paper, we anticipate a positive $g^3$ term 
from a continuation of the result at $\ep\ll \frac{1}{12}$ in Section \ref{twoloop}. }, 
the line of the IR fixed point 
terminates at $(n_c, g_c)=(n'_*\,, g_{c*}')$ with a finite coupling $g_{c*}'$. 

The strong coupling phase of the disordered model ($M\rightarrow 0$)
near $n\approx 0$ 
is formed by the dominance of the two-layer process $G_{2,2}(r,\ep)$
over the three-layer process with a closed loop $G_{2,1}(r,\ep)$.
The qualitative behavior of the absolute ratio $\mathcal{R}(n)=|G_{2,2}/G_{2,1}|$ is  
independent of $M$ except a special case with $M=2$, 
where the three-layer process is prohibited.
Diagrammatically, a large $\mathcal{R}(n)$ implies that
the ratio of (the number of inter-layer hopping) to
(the average number of complete loops per layer)   
is also large.
Near $n\approx 0$ region, the particles, 
which are destined to form a whole connected diagram,
favor to sew layers together rather than to stay on one layer 
and to walk around avoiding their traces.

We shall henceforth proceed to the two loop calculation to investigate the IR fixed points
in the region near $n=1$ ($\ep\ll \frac{1}{12}$), where
regarding the structure constant $C_{\CE\CE}^{\CE}$ in (\ref{structure}) as $\mathcal{O}(\ep)$
is justifiable.
In this region, the calculation turns out to be parallel to that in the study of 
the random-bond Potts model by Dotsenko, Picco and Pujol \cite{dotsenkopicco}.
At one-loop level,  $C_{\CE\CE}^{\CE}=\mathcal{O}(\ep)$ implies 
$G_{2,2}=\mathcal{O}(\ep)$;
in the minimal subtraction scheme, we drop this term. 
The beta function is then 
\beq
\beta(g) = \frac{dg}{d\ln(r)} = 8\epsilon g + 4 \pi (M-2) g^2+\mathcal{O}(g^3).
\label{beta2}
\eeq
We left the two loop calculation of the intermediate region 
where $C_{\CE,\CE}^{\CE}\approx 1$ as a future problem.
\section{The scaling dimensions up to the two-loop calculation}\label{twoloop}
\subsection{The beta function at two-loop}
There are four types of third order diagrams for the beta function
as listed in Figure \ref{betadiagrams}-(c), (d), (e) and (f).
As the diagram $G_{3,3}(r,\ep)$ in Figure \ref{betadiagrams}-(e)
have extra $\ep^2$ factor from the structure constants
and the diagram $G_{3,4}(r,\ep)$ in Figure \ref{betadiagrams}-(f)
has no pole in $\ep$, we omit these terms in the following. 
The $G_{3,1}(r,\ep)$ in Figure \ref{betadiagrams}-(c) can be calculated as 
\bea
\fl
G_{3,1}(r,\epsilon) &= \frac{1}{3!}\cdot 12(M-2)(M-3) \int_{|y-x|<r,|z-x|<r} d^2yd^2z
\l<\CE(x)\CE(y)\r>_0 \l<\CE(y)\CE  (z)\r>_0  \nonumber\\
\fl&=  8\pi (M-2)(M-3)
\int_{|y|<r}dy\ y^{-1+16\epsilon} 
\int_{|z|<\frac{r}{|y|}} d^2z |z|^{-2+8\epsilon} |z-1|^{-2+8\epsilon}  \nonumber\\
\fl&=  16\pi^2 (M-2)(M-3)\frac{r^{16\epsilon}}{(8\ep)^2}+\mathcal{O}(\ep^0),\label{g31}
\eea
where we have used the asymptotic behavior of the integral for $1\ll R$:
\beq
\int_{|z|<R} d^2z |z|^{-2+8\epsilon} |z-1|^{-2+8\epsilon}=
4\pi\frac{1}{8\ep}+\mathcal{O}(R^{-2+16\ep}).
\label{zz1}
\eeq
It should be emphasized that we get no simple pole in (\ref{g31}).

Next, we have $G_{3,2}(r,\ep)$ in Figure \ref{betadiagrams}-(d) containing four-point functions: 
\bea
\fl
G_{3,2}(r,\ep)&= 
\frac{1}{3!}\cdot 24(M-2)  \int_{|y-x|<r,|z-x|<r}\hspace{-16mm}d^2y d^2z~
\l<\CE(x) \CE(y) \CE(z) \CE(\infty)\r>_0 \l<\CE(y)  \CE(z)\r>_0\label{g32_1}\\
\fl&=8\pi(M-2)\int_{y<r}\hspace{-4mm}dy\ y^{-1+16\epsilon}\,\mathcal{I}\left(r/|y|,\ep\right),
\label{g32_2}
\eea
where we define
\beq
\label{IRe}
\fl\mathcal{I}(R,\ep)=\int_{|z|<R}d^2z\l<\CE(0) \CE(1) \CE(z) \CE(\infty)\r>_0 \l<\CE(1) \CE(z)\r>_0.
\eeq
Actually, instead of $\mathcal{I}(R,\ep)$, we shall calculate  $\mathcal{I}(\infty,\ep)$ by analytic continuation in \ref{integralbeta},
and see that $\mathcal{I}(\infty,\ep)$ itself has no poles in $\ep$.
Although, in the Appendix, we prove this fact by applying the contiguity relation between 
generalized hypergeometric series, 
the fact can also be interpreted as a result of a cancellation between 
two poles with opposite sign, as explained in the following
\footnote{A similar cancellation of poles has already been observed 
in the random-bond Potts model \cite{dotsenkopicco}.}.

In the limit $|z|\rightarrow \infty$,  the integrand in (\ref{IRe}) behaves as $|z|^{-4\Delta_\CE}$
 ($4\Delta_\CE=2-8\ep$),
since the four-point function, using the identity operator $I$ in (\ref{OPE})
as the leading intermediate channel, 
decouples into two two-point functions.
Hence,  $\mathcal{I}(\infty,\ep)$ has the contributions
from both regions $|z|<R$ and $|z|>R$, each of which contains a pole 
$+2\pi R^{8\ep}/8\ep$ and $-2\pi R^{8\ep}/8\ep$, respectively;
they are canceled out each other.
Observing that $\mathcal{I}(R,\ep)$ does not have the latter pole, 
and estimating the corrections, we obtain
\beq
\mathcal{I}(R,\ep)=2\pi\frac{R^{8\ep}}{8\ep}+\mathcal{I}(\infty,\ep)+\mathcal{O}(R^{-1+8\ep}).
\label{irr}
\eeq
As the first term corresponds to the decoupling limit $y \rightarrow x$
of the four-point function in (\ref{g32_1}),
it leads to the same contribution as (\ref{g31}),
 only if we replace $(M-2)$ in (\ref{g32_2}) by $(M-2)(M-3)$.
Now, using the vertex operator representation, we write $\mathcal{I}(\infty,\ep)$ in (\ref{irr}) as 
\bea
\fl \mathcal{I}(\infty,\ep)&=\mathcal{N}
\int  d^2z d^2u d^2v\,&
\l<V_{\al_{13}}(0) V_{\al_{13}}(1) V_{\overline{\al_{13}}}(z) V_{\al_{13}}(\infty)V_{\al_{-}}(u)V_{\al_{-}}(v)\r>\l<V_{\al_{13}}(1) V_{\overline{\al_{13}}}(z)\r> \nonumber\\
\fl&=\mathcal{N}
\int d^2z d^2u d^2v\,&
|z|^{4\al_{13}\overline{\al_{13}}}|1-z|^{8\al_{13}\overline{\al_{13}}}|
(z-u)(z-v)|^{4\overline{\al_{13}}\al_{-}}\nonumber\\
\fl&&\cdot |uv(u-1)(v-1)|^{4\al_{13}\al_{-}}|u-v|^{4\al_{-}^2},
\label{ivertex}
\eea
where $\mathcal{N}$ is the normalization factor with the four-point function
given in (\ref{N_aomoto}).
Substituting (\ref{irr}) and the result (\ref{i_infty_result}) for $\mathcal{I}(\infty,\ep)$ 
into (\ref{g32_2}), we obtain 
\bea
\fl
G_{3,2}(r,\epsilon)&= 
16\pi^2(M-2)\left[\frac{1}{(8\epsilon)^2}-\frac{1}{8\epsilon}\right]r^{16\epsilon}+\mathcal{O}(\ep^0).
\label{g32_3}
\eea
We get, by collecting the third order terms (\ref{g31}) and (\ref{g32_3}), 
the renormalized coupling constant in (\ref{gr}) as
\beq\fl
g(r)=r^{8\ep}\left\{
g_0+4\pi\frac{(M-2)}{8\ep}r^{8\ep}g_0^2+16\pi^2\left[\frac{(M-2)^2}{(8\epsilon)^2}-\frac{(M-2)}{8\epsilon}\right]r^{16\epsilon}g_0^3\right\}+\mathcal{O}(g_0^4).
\label{gr_3}
\eeq
Then (\ref{gr_3}) is solved with respect to $\tilde{g}=r^{8\ep}g_0$ as
\beq
\fl\tilde{g}=g(r)-4\pi\frac{(M-2)}{8\ep}g(r)^2+16\pi^2\left[\frac{(M-2)^2}{(8\epsilon)^2}
+\frac{(M-2)}{8\epsilon}\right]g(r)^3.
\label{solvedg}
\eeq
Consequently, we obtain the RG beta function 
\beq\fl
\beta(g) = \frac{dg(r)}{d\ln r} = 8\epsilon g + 4\pi (M-2)g^2-16\pi^2(M-2)g^3+\mathcal{O}(g^4).
\label{beta3}
\eeq
As is expected from the known equivalence of the random-bond Ising model 
to the random-mass fermion model and the Gross-Neveu model \cite{bernard, ludwig}, 
at the Ising point $\ep=0$ ($n=1$),
the expression (\ref{beta3}) reduces to the two-loop beta function 
for the Gross-Neveu model \cite{wetzel}. 
Solving $\beta(g_c)=0$ for the disordered model ($M\rightarrow 0$), 
we obtain the coupling constant at the random fixed point as
\beq
g_c=\frac{\ep}{\pi}+\frac{4\ep^2}{\pi}+\mathcal{O}(\ep^3).
\label{gc}
\eeq  

\subsection{The correction to the scaling dimensions}
In our perturbation theory, the two-point correlation function between fields 
$\mathcal{O}(0)$ and $\mathcal{O}(\infty)$ is calculated as
\beq
\fl
\eqalign{
\l<\mathcal{O}(0)\mathcal{O}(\infty)\r>=\l<\mathcal{O}(0)\mathcal{O}(\infty)\r>_0
&+g_0\int_{|y|<r} \hspace{-2mm}d^2x\,\l<S_I(y)\mathcal{O}(0)\mathcal{O}(\infty)\r>_0\\
&+\frac{g_0^2}{2!}\int_{|y|<r,~|z|<r} \hspace{-10mm}d^2y\,d^2z\,\l<S_I(y)S_I(z)\mathcal{O}(0)\mathcal{O}(\infty)\r>_0+\cdots}.
\label{soo}
\eeq
Again, as in (\ref{geff}), we restrict the insertion of the interaction $S_I$ 
around the field $\mathcal{O}(0)$
within the disk of radius $r$.
From this, we can determine a renormalization constant $Z_\mathcal{O}$ 
for the field $\mathcal{O}$ 
perturbatively as   
\beq
\l<\mathcal{O}(0)\mathcal{O}(\infty)\r>
=Z_{\mathcal{O}}\l<\mathcal{O}(0)\mathcal{O}(\infty)\r>_{0}.
\label{zoo}
\eeq
The two-point functions in the theories with different 
cut-offs $r$ and $sr$ are related as
\bea
\fl\l<\mathcal{O}(0)\mathcal{O}(sR)\r>_{sr, g(sr)}&=
\frac{Z_{\mathcal{O}}^2(g(sr))}{Z_{\mathcal{O}}^2(g(r))}s^{-4\Delta_\mathcal{O}}
\l<\mathcal{O}(0)\mathcal{O}(R)\r>_{r,g(r)}\nonumber\\
\fl&=\exp\left[2\int_{g(r)}^{g(sr)} dg~
\frac{\gamma_\mathcal{O}(g)}{\beta(g)} \right]
s^{-4\Delta_\mathcal{O}} \l<\mathcal{O}(0)\mathcal{O}(R)\r>_{r,g(r)},
\label{osror}
\eea
where, in the second line, we have introduced the anomalous dimension 
\beq
\gamma_\mathcal{O}(g)=\frac{\mathrm{d} Z_{\mathcal{O}}(g)}{\mathrm{d}\ln r}.
\eeq
Now consider the large $s$ behavior of (\ref{osror}) 
and let the upper limit $g(rs)$ of the integral tend to $g_c$ of the random fixed point; 
the beta function $\beta(g)$ tends to zero,
while the anomalous dimensions $\gamma_{\mathcal{O}}(g)$, as we will see, 
remain finite both for the energy operator $\CE$ and the spin operator $\sig$.
This means the integral is dominated by the contribution from the region $g\approx g_c$,
and we obtain
\beq
2\left(\Delta_\mathcal{O}^{\mathrm{IR}}-
\Delta_\mathcal{O}^{\mathrm{UV}}\right)=-\gamma_\mathcal{O}(g_c),
\label{deldelgamma}
\eeq
where we mean, by $2\Delta_\mathcal{O}^{\mathrm{IR}}$ and $2\Delta_\mathcal{O}^{\mathrm{UV}}$,
 the scaling dimension of the field $\mathcal{O}$ at 
the IR (the random) fixed point and the UV fixed point, respectively.
\begin{figure}[!htbp]
\begin{center}
\includegraphics[width=12 cm]{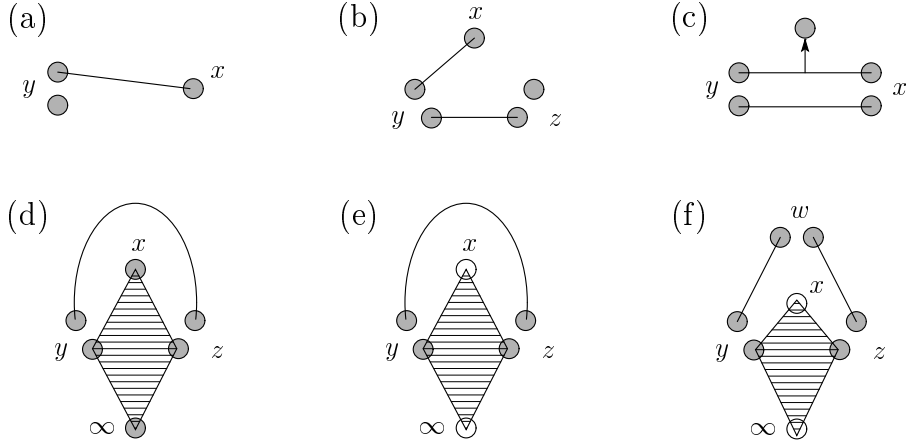}
\caption{The contributions for the scaling dimensions.
The diagrams (a), (b) (c) and (d) are for $Z_\CE$, 
while (e) $\mathcal{S}_{2}(r,\ep)$ and (f) $\mathcal{S}_{3}(r,\ep)$ are for $Z_\sig$.
The white circle represents the spin operator $\sig^{(a)}$;
for symbols, see also the caption in Figure \ref{betadiagrams}. }
\label{gammadiagrams}
\end{center}
\end{figure}
\subsection{Scaling dimension of the energy operator}
The first and second order diagrams 
for the renormalization constant of the energy operator $Z_\CE$
are listed in Figure \ref{gammadiagrams}-(a), (b), (c) and (d).
As the integral in the diagram in Figure \ref{gammadiagrams}-(c)
does not have pole in $\ep$,
we omit this term from the calculation. 
Since the integrals are same as those for the two loop beta function,
by adapting combinatorial factors, we have  
\bea
\fl
Z_\CE&=1+g_0 \cdot\frac{2(M-1)}{2(M-2)}  G_{2,1}
+\frac{g_0^2}{2!}\left[\frac{8(M-1)(M-2)}{2(M-2)(M-3)}G_{3,1}
+\frac{4(M-1)}{4(M-2)}G_{3,2}\right]\nonumber\\
\fl&=1+4\pi\tilde{g}\frac{(M-1)}{8\ep}
+8\pi^2 \tilde{g}^2\left[\frac{(M-1)(2M-3)}{(8\ep)^2}-\frac{(M-1)}{16\ep}\right],
\eea
where the results (\ref{g21}), (\ref{g31}) and (\ref{g32_3}) are used in the second line.
Again, by using the expression for $\tilde{g}$ in (\ref{solvedg}), we get
\beq
\fl
\gamma_\CE(g)=\frac{\mathrm{d}\ln Z_\CE}{\mathrm{d}\ln r}
=4\pi(M-1)g(r)-8\pi^2(M-1)g(r)^2+\mathcal{O}\left(g(r)^3\right).
\label{gamma_energy}
\eeq
Consequently, by using (\ref{gc}) in (\ref{deldelgamma}), we obtain
\beq
2\left(\Delta_\CE^{\mathrm{IR}}-
\Delta_\CE^{\mathrm{UV}}\right)=-\gamma_\CE(g_c)=4\ep+8\ep^2+\mathcal{O}(\ep^3).
\label{energy_correction}
\eeq

\subsection{Scaling dimension of the spin operator}
We shall here calculate the renormalization constant for spin operator $Z_\sig$
up to the third order in $g_0$, 
which gives the lowest order correction to the scaling dimension $2\Delta_\sig$
at the IR fixed point.
According to the operator algebra (see the diagrams in Figure \ref{gammadiagrams}), there are no contribution at the first order,    
and one contribution at the second order $\mathcal{S}_{2}(r,\ep)$
and the other one at the third order $\mathcal{S}_{3}(r,\ep)$:  
\beq
\fl
Z_\sig=1+\mathcal{S}_{2}(r,\ep)+\mathcal{S}_{3}(r,\ep)+\mathcal{O}(g_0^4).
\eeq
The second order diagram in Figure \ref{gammadiagrams}-(e) is given by
\beq
\fl
\mathcal{S}_{2}(r,\epsilon)= \frac{g_0^2}{2!}\cdot 4(M-1) 
\int_{|y-x|,~|z-x|<r} \hspace{-15mm}d^2y\, d^2z
 \,\l<\sig(x) \CE(y) \CE(z) \sig(\infty)\r>_0 \l<\CE(y)  \CE(z)\r>_0,
\label{s2}
\eeq
while the third order diagram in Figure \ref{gammadiagrams}-(f) is given by
\bea
\fl
\mathcal{S}_{3}(r,\epsilon)= \frac{g_0^3}{3!}\cdot 24 &(M-1)(M-2) \nonumber\\  
\fl &\cdot\int_{|y-x|<r,~|z-x|<r,~|w-x|<r}\hspace{-33mm} d^2y\, d^2z\, d^2w
\,\l<\sig(x) \CE(y) \CE(z) \sig(\infty)\r>_0 \l<\CE(y)  \CE(w)\r>_0 \l<\CE(w)  \CE(z)\r>_0.
\label{s3}
\eea
In evaluating the integrals (\ref{s2}) and (\ref{s3}), we keep only the terms with
the lowest powers in $\ep$.
For this reason, it turns out that we can add certain extra regions to these integrals. 
First, by adding the region $|z-x|>r$ to (\ref{s2}), we get 
\bea
\fl\mathcal{S}_{2}(r,\epsilon)&= 4 \pi g_0^2(M-1) \,  \int_{y<r} dy\ y^{-1+16\epsilon}~ 
\mathcal{K}_2\left(\infty,\ep\right),
\label{s2k2}
\eea
with $\mathcal{K}_2(\infty,\ep)$ defined as
\beq
\label{k2}
\fl\mathcal{K}_2(\infty,\ep)=\int_{\mathbb{C}}d^2z\l<\sig(0) \CE(1) \CE(z) \sig(\infty)\r>_0
\l<\CE(1) \CE(z)\r>_0.
\eeq
Second, if we add the region $|w-x|>r$ and $|z-x|>r$ to (\ref{s3}),
we see that the third order contribution $\mathcal{S}_{3}(r,\epsilon)$ has 
the similar structure as the second order one $\mathcal{S}_{2}(r,\epsilon)$ in (\ref{s2}).
In fact, under a trivial change of variables, we obtain the factorized form: 
\beq
\fl\mathcal{S}_{3}(r,\epsilon)= 8 \pi g_0^3(M-1)(M-2) 
\int_{y<r} dy\ y^{-1+24\epsilon}\!
\int_{\mathbb{C}} d^2w |w (w-1)|^{-4\Delta_{\CE}}
\mathcal{K}_3\left(\infty,\ep\right),
\label{s3k3}
\eeq
with the definition
\beq
\label{k3}
\fl\mathcal{K}_3(\infty,\ep)=
\int_{\mathbb{C}}d^2z\, |1-z|^{2-8\Delta_{\CE}}\l<\sig(0) \CE(1) \CE(z) \sig(\infty)\r>_0.
\eeq
Then, we write the integrals (\ref{k2}) and (\ref{k3}) using the vertex operator representation:
\bea
\fl \mathcal{K}_2(\infty,\ep)&=\mathcal{N}\!
\int  d^2zd^2ud^2v\,& |1-z|^{-4\Delta_{\CE}}
\l<V_{\overline{\al_{p-1,p}}}(0) V_{\al_{13}}(1) V_{\al_{13}}(z) V_{\al_{p-1,p}}(\infty)
V_{\al_{-}}(u)V_{\al_{-}}(v)\r>\hspace{-7mm} \nonumber\\
\fl&=\mathcal{N}
\int d^2zd^2ud^2v\,&~
|z|^{4\overline{\al_{p-1,p}}\al_{13}}|1-z|^{4\al_{13}\al_{13}-4\Delta_{\CE}}
|uv|^{4\overline{\al_{p-1,p}}\al_{-}}|u-v|^{4\al_{-}^2}\nonumber\\
\fl&&~\cdot |(1-u)(1-v)(z-u)(z-v)|^{4\al_{13}\al_{-}},
\label{k2vertex}
\eea
\bea
\fl \mathcal{K}_3(\infty,\ep)&=
\mathcal{N}
\int d^2zd^2ud^2v\,&~
|z|^{4\overline{\al_{p-1,p}}\al_{13}}|1-z|^{4\al_{13}\al_{13}+2-8\Delta_{\CE}}
|uv|^{4\overline{\al_{p-1,p}}\al_{-}}|u-v|^{4\al_{-}^2}\nonumber\\
\fl&&~\cdot |(1-u)(1-v)(z-u)(z-v)|^{4\al_{13}\al_{-}}.
\label{k3vertex}
\eea
and give the results in Appendix C. 

Now we obtain, by collecting the contributions (\ref{s2k2}) and (\ref{s3k3}),
the renormalization constant for the spin operator as
\bea
\fl \ln Z_{\sig}
=4\pi g_0^2(M-1)\frac{r^{16\ep}}{16\ep}\mathcal{K}_2(\infty,\ep)
+8\pi g_0^3(M-1)(M-2)\left[\frac{r^{24\ep}}{24\ep}\frac{\pi}{2\ep}\right]\mathcal{K}_3(\infty,\ep)
+\mathcal{O}(g_0^4),\nonumber
\eea
where we have used (\ref{zz1}) to obtain the third order term.
Differentiating this with respect to $(\ln r)$ and 
using the result (\ref{kstar}) for the integrals $\mathcal{K}_2(\infty,\ep)$ and $\mathcal{K}_3(\infty,\ep)$, 
we obtain the anomalous dimension for the spin operator as  
\bea
\fl
\gamma_\sig(g)&=
4\pi(M-1)\left\{g(r)-(M-2)\left(\frac{\pi}{2\ep}\right)g(r)^2\right\}^2\cdot 2\mathcal{N}
\pi\ep\left[U^2+W^2-8YZ\pi\ep\right]\nonumber\\
\fl&\qquad +8\pi(M-1)(M-2)\left(\frac{\pi}{2\ep}\right)g(r)^3~\cdot 2\mathcal{N}
\pi\ep\left[U^2+W^2-12YZ\pi\ep\right]+\mathcal{O}(g(r)^4)\nonumber\\
\fl&=8\mathcal{N}\pi^2(M-1)
\left\{\ep g(r)^2\left[U^2+W^2-8YZ\right]-4g(r)^3(M-2)YZ\right\}+\mathcal{O}(g(r)^4),
\label{gamma_sigma_m0}
\eea
where the expression for the bare coupling constant (\ref{solvedg}) is used in the second equality. 
For the definition of the constants $\mathcal{N}$, $U$, $W$, $Y$ and $Z$, see  (\ref{N_aomoto})
and (\ref{xyzuvw}). 
The limit $M\rightarrow 0$ for the disordered $O(n)$ model yields,
\beq
\fl\gamma_\sig(g)=-8\mathcal{N}\pi^2\ep g(r)^2\left[U^2+W^2\right]+\mathcal{O}(g(r)^4).
\eeq
Now, from (\ref{deldelgamma}), we obtain
\bea
2\left(\Delta_\sigma^{\rm IR}-\Delta_\sigma^{\rm UV}\right)&=-\gamma_\sig(g_c)\nonumber\\
&=8\frac{\Gamma^4(\frac{1}{4})}{\pi^4}\ep^3+\mathcal{O}(\ep^4)\nonumber\\
&=128\frac{K^2(\sin\frac{\pi}{4})}{\pi^3}\ep^3
+\mathcal{O}\left(\ep^4\right),
\label{correction_on}
\eea
where, in the last line, we have used 
the elliptic integral of the first kind defined by
\beq
K(k)=\int_0^{\pi/2} \frac{d\phi}{\sqrt{1-k^2\sin^2 \phi}},
\label{ellipticintegral}
\eeq
for the purpose of the comparison 
with the result in the random-bond Potts model \cite{dotsenkopicco};
this will be discussed in Section \ref{discussionconclusion}.

\section{Effective central charge}\label{centralcharge}
The central charge, in general, characterize a system by counting 
the number of the critical fluctuations in it.
If the replica method is used, however, the central charge becomes rather trivial;
we are always left with the vanishing central charge $c=0$
because of the formal limit $M\rightarrow 0$ in the end.
Nevertheless, we can define a so-called effective central charge 
characterizing a random fixed point as follows \cite{ludwigcardy}.
Consider the $M$ layers of the coupled model and the relation
\beq
c_{\mathit{IR}}(M)=M c_{\mathit{UV}}+\Delta c(M),
\eeq
where $c_{\mathit{UV}}$ and $c_{\mathit{IR}}(M)$ 
are the central charge of the pure system
and that of the non-trivial fixed point, respectively.  
The effective central charge is then defined  as
\beq
c_{\mathit{eff}}= \frac{\mathrm{d}\,c_{\mathit{IR}}(M)}{\mathrm{d} M}\Bigr|_{M=0}.
\label{ceff_def}
\eeq

In order to obtain the $c_{\mathit{IR}}(M)$, it is useful to recall the definition of the $C$-function
and the differential equation satisfied by it \cite{zamolodchikovc, cardybook}.
To this end, we consider the vicinity of some critical theory $S_*$ in the theory space 
spanned by the coupling constants $\{g^i\}$:  
\beq
S=S_* - \int d^2x~ g^i \Phi_i(x),
\label{ssgphi}
\eeq
and then specialize to our disordered $O(n)$ model.
The response of the action under the transformation $z^\mu\rightarrow z^\mu+\alpha^\mu(z)$
can be expressed by the stress tensor $T_{\mu\nu}$ as
\beq
\delta S=-\frac{1}{2\pi}\int d^2 r\  T_{\mu\nu}\partial^{\mu}\alpha^{\nu}. 
\label{sresponce}
\eeq
As usual, a particular component of the stress tensor
 $T_{zz}=\frac{1}{4}(T_{11}-T_{22}-2iT_{12})$ and
the trace of it $4T_{z\bar z}=T_{11}+T_{22}$ are denoted as $T$ and $\Th$, respectively.
As the RG flow $dg^i=\beta^i dt$ occurs under the dilatation $z^\mu\rightarrow (1+dt)z^\mu$, 
the definition (\ref{ssgphi}) and (\ref{sresponce}) leads to the relation: 
\beq
\Th(x)=2\pi \beta^i \Phi_i(x).
\label{thbp}
\eeq
From $T$ and $\Th$, one can consider three-types of the two-point functions:
\bea
\l<T(z,\bar z)T(0,0)\r>=&F(\tau)/z^4, \label{ftau}\\
\l<T(z,\bar z)\Th(0,0)\r>=&H(\tau)/z^3\bar{z},\\
\l<\Th(z,\bar z)\Th(0,0)\r>=&G(\tau)/z^2\bar{z}^2,
\label{thth}
\eea
which are measured at the scale $\tau=\ln(z\bar z)$. The $C$-function is then defined as
\beq
C=2F-H-\frac{3}{8}G.
\label{cfunction_def}
\eeq
The conservation of the stress tensor $\partial^\mu T_{\mu\nu}=0$ yields
$
\partial_{\bar{z}}T(z,\bar z)+\frac{1}{4}\partial_z \Th(z,\bar z)=0,
$
and one can easily show 
\beq
\frac{\mathrm{d}}{\mathrm{d}\tau}C=-\frac{3}{4}G.
\label{c34g}
\eeq
After fixing the reference scale $\tau$ to $\tau=0$, 
the $C$-function depends only on the coupling constants $g^i$.  
We introduce the Zamolodchikov metric on the theory space as
\beq
\mathcal{G}_{ij}=(z\bar z)^2\l<\Phi_i(z,\bar z)\Phi_j(0,0)\r>\Bigr|_{z\bar z=1}.
\label{metric}
\eeq
From (\ref{thbp}) and  (\ref{c34g}), we have
\beq
\frac{1}{2}\beta^i\frac{\partial}{\partial g^i}C=
-\frac{3}{4}(2\pi)^2\ \mathcal{G}_{ij}\beta^i\beta^j.
\label{bcgbb}
\eeq
For unitary theories, the positivity condition implies that the metric should be positive definite.
The $C$-function is stable at the point with $\beta^i=0$ {\it i.e.} a fixed point
and take the same value as the central charge 
as we can see from the definition (\ref{ftau})-(\ref{cfunction_def}).

In our case, the perturbing field is quadratic in energy operator
as in  (\ref{perturbation}): 
\beq
\Phi=S_I=\sum_{a\neq b} \CE^{(a)}(x)\CE^{(b)}(x),
\label{perturbing}
\eeq
and hence the metric (\ref{metric}) is
\beq
\mathcal{G}=2M(M-1).
\eeq
This metric becomes negative in the limit $M\rightarrow 0$, thereby violating the $c$-theorem.
Note also that our normalization for the energy operator $\CE$ is 
such that $C_{\CE\CE}^I=1$. 
Substituting the beta function (\ref{beta2}) into (\ref{bcgbb}), we obtain
\bea
\Delta C(M)&=-6\pi^2\cdot2M(M-1)\cdot\frac{(8\ep)^3}{6\left[4\pi(M-2)\right]^2}
+\mathcal{O}(\ep^4)\nonumber\\
&=-64\ep^3\frac{M(M-1)}{(M-2)^2}+\mathcal{O}(\ep^4).
\eea
From the definition of the effective central charge (\ref{ceff_def}), we have
\beq
c_{\mathit{eff}}^{\rm IR}-c_{\mathit{eff}}^{\rm UV}= 16\ep^3+\mathcal{O}(\ep^4).
\eeq
Since we have the IR fixed points in $\ep>0$, this always increase under the RG flow
as expected.
\section{Discussion and Conclusions}\label{discussionconclusion}

Before concluding this paper, we discuss the relation to the known results
and then comment on possible future directions.

The RG flow in our disordered $O(n)$ model near the disordered Ising point, 
which is inferred from the results 
(\ref{beta3}), (\ref{gamma_energy}) and (\ref{gamma_sigma_m0}),
is
qualitatively similar to that in the random-bond $q$-state Potts model 
\cite{ludwig, dotsenkopicco}.
The crossover to an IR fixed point
occurs only when 
the most relevant interaction,
which is quadratic in energy operators as in (\ref{perturbing}),
becomes relevant;
the region correspond to $n<1$ for the disordered $O(n)$ model and 
$q>2$ for the random-bond $q$-state Potts model. 
This can be understood by recalling how 
the two families of the pure models, namely, 
the $O(n)$ models and the $q$-state Potts models
coincide at the Ising point ($n=1$, $q=2$) \cite{dotsenkofateev, cardy}.

\begin{figure}[!tbp]
\begin{center}
\includegraphics{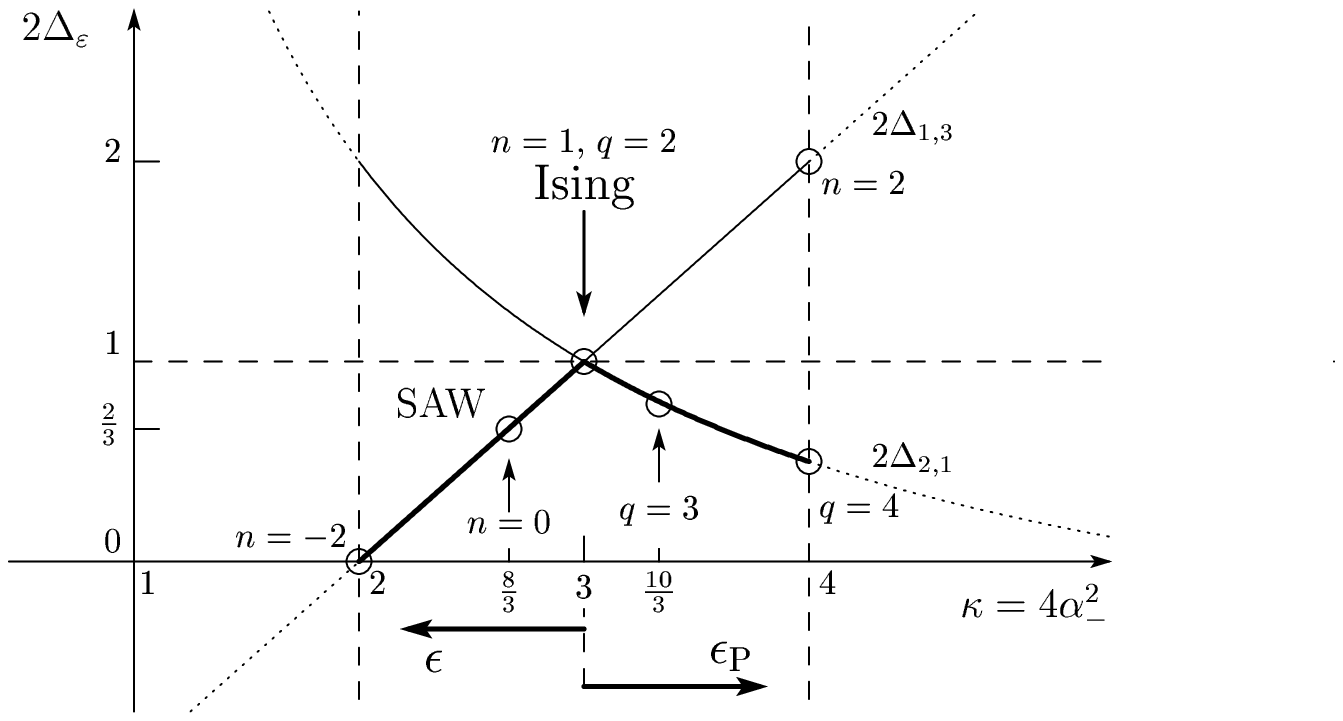}
\caption{The scaling dimension of the energy operators 
for the $O(n)$ model ($2\Delta_{1,3}$) and for the $q$-state Potts model (2$\Delta_{2,1}$)
plotted as functions of the SLE parameter $\kappa=4\al_-^2$.
The horizontal line at $2\Delta_\CE=1$ shows the marginality of the disorder-induced coupling.
The arrows indicate the directions of the expansion parameters $\ep$ and $\ep_P$.}
\label{scDIM_on_potts}
\end{center}
\end{figure}

The energy operator in the $O(n)$ model is identified with the primary field $\phi_{1,3}$ 
as we have stated in (\ref{identification}),
while the energy operator in the Potts model is identified with the field $\phi_{2,1}$.
These primary fields are, in a general minimal CFT, different objects 
as one expects from the fact that  
a four-point function of primary fields $\phi_{r,s}$ is determined from 
an ordinary differential equation of order $rs$.
Accordingly,  in the vertex operator representation of the four-point functions,
we should include two of the screening operators $V_-$ in the $O(n)$ model
and one of the other screening operator $V_+$ in the Potts model, respectively. 
However, the field $\phi_{1,3}$ reduces to the other field $\phi_{2,1}$ at the Ising point 
(the $m=3$ minimal model) because of the equivalence (\ref{rs}).
At the level of the scaling dimensions, the reduction can be summarized
 in Figure \ref{scDIM_on_potts},
in which the line of $2\Delta_{1,3}$ and the curve of $2\Delta_{2,1}$ 
intersect at the Ising point ($\kappa=3$).

Up to one-loop order, when the deviation parameter $\ep$ 
from the Ising point defined in (\ref{alpha34epsilon}) is small
and neglecting the structure constant (\ref{structure}) in the $O(n)$ model can be justified,
the beta function is determined only from the scaling dimension of the energy operator
as in (\ref{beta2});
thus similarity with the random-bond $q$-state Potts model is natural from Figure 
\ref{scDIM_on_potts}.
By contrast, the reduction of the two-loop beta function (\ref{beta3}), which involves
four-point functions, to that of 
the $q=2$ random-bond Potts model
in the limit $n\rightarrow 1$ serves as a non-trivial check
of the vertex operator representations.

Remarkably, two lines of the IR fixed points lie the opposite sides of the Ising point;
if we introduce a parameter 
\beq
\ep_\kappa=3-\kappa,
\eeq
the random fixed points of the disordered $O(n)$ models and  
those of the random-bond $q$-state Potts models exist 
in the region $\ep_\kappa>0$ and $\ep_\kappa<0$, respectively.
Another natural parameter $\ep_P$ for the Potts models 
can be taken as $\al_+^2=\frac{4}{3}-\ep_P$, 
where the IR fixed points emerges in the region $\ep_P>0$
\footnote{It should be noted that yet another parameter 
$\ep$ defined as $\al_+^2=\frac{4}{3}+\ep$ is used in \cite{dotsenkopicco}. }. 
The definition (\ref{alpha34epsilon}), the relation (\ref{kappaalpha}) 
and $\al_+^2\al_-^2=1$ from (\ref{al+al-}) lead to relations:
\beq
\epsilon=
\frac{1}{4}\ep_\kappa=-\frac{9}{16}\, \ep_{\mathrm{P}} -\frac{27}{64}\, 
\ep_{\mathrm{P}}^2+O(\ep_{\mathrm{P}}^3).
\label{trivial}
\eeq
Then the results for the scaling dimension of the energy operator 
in the random fixed point of the disordered $O(n)$ model (\ref{energy_correction}) and 
that of the random-bond $q$-state Potts models \cite{ludwig, dotsenkopicco}
can be summarized in one expression as 
\beq
\fl
2\Delta_\CE^{\rm IR}=1+\frac{1}{2}(1-2\Delta_\CE)^2+\mathcal{O}\left((1-2\Delta_\CE)^3\right)
=\cases{1+\frac{1}{2}\ep_\kappa^2+\mathcal{O}(\ep_\kappa^3)&
$(\ep_\kappa>0;\ O(n))$\\
1+\frac{2}{9}\ep_\kappa^2+\mathcal{O}(\ep_\kappa^3)&
$(\ep_\kappa<0;\ \mathrm{Potts})$,\\}
\eeq
where both results are measured from the Ising value $2\Delta_\CE=1$.
Naturally, the scaling dimensions at these IR fixed points 
imply faster decays in the correlation functions: $\Delta_\CE<\Delta_\CE^{\rm IR}$ 
as one expects with a general RG flow \cite{zinn}.
The ratio between the coefficients in $\ep_\kappa^2$ 
is $\frac{9}{4}$, which is the square of the ratio 
of the derivative of the line $2\Delta_{1,3}$ at the Ising point to that of
the curve $2\Delta_{2,1}$ in Figure \ref{scDIM_on_potts}. 

Contrastingly, the coefficients in the spin scaling dimensions are transcendental numbers.
Using the relation (\ref{trivial}), 
our result (\ref{correction_on}) for the disordered $O(n)$ model is written as
\bea
\fl 2\left(\Delta_\sigma^{\rm IR}-\Delta_\sigma^{\rm UV}\right)_{O(n)}
&=2\frac{K^2(\sin\frac{\pi}{4})}{\pi^3}\ep_\kappa^3
+\mathcal{O}\left(\ep_\kappa^4\right) \qquad (\ep_\kappa>0).
\label{period1}
\eea
Then we quote, from ref. \cite{dotsenkopicco}, 
the result for the random-bond $q$-state Potts model:
\beq
\fl
\eqalign{2\left(\Delta_\sigma^{\rm IR}-\Delta_\sigma^{\rm UV}\right)_{\mathrm{Potts}}
&=\frac{27}{32}
\frac{\Gamma\left(-\frac{2}{3}\right)^2\Gamma\left(\frac{1}{6}\right)^2}
{\Gamma\left(-\frac{1}{3}\right)^2\Gamma\left(\frac{1}{6}\right)^2}\epsilon_P^3+\mathcal{O}(\epsilon_P^4) \qquad (\ep_P>0)\nonumber\\
&=\frac{2}{81\pi^2}
\left[K\left(\sin \frac{\pi}{12}\right)K\left(\cos \frac{\pi}{12}\right)\right]^2
\left(-\ep_\kappa\right)^3
+\mathcal{O}\left(\ep_\kappa^4\right)\quad (\ep_\kappa<0),}
\label{period2}
\eeq
where, in the second line, we have used the elliptic integral defined in (\ref{ellipticintegral}).
As a first observation, notice that both coefficients are consisting of the numbers 
$K(\cos \theta)$ known as ``singular values" of the elliptic integrals 
for which the modular ratios $K(\cos \theta)/K(\sin \theta)$ belong to
quadratic irrational numbers.
The ratios here are $\sqrt{1}$ and $\sqrt{3}$
for the disordered $O(n)$ model and 
the random-bond $q$-state Potts model, respectively\footnote{The 
two singular values have number theoretic origins 
and belong to the sequence of the values derived 
by Chowla-Selberg formula \cite{chowla};
the relevant part of the formula is obtained from the functional equations of 
the Epstein's zeta functions on the quadratic number fields 
$Q(\sqrt{-1})$ and $Q(\sqrt{-3})$. }.
These modular ratios are exceptional in that 
the modular angles $\theta$ are commensurate to $\pi$, namely, 
$\theta=\pi/4$ and $\theta=\pi/12$.

The coefficient in (\ref{period1}) and that in (\ref{period2})    
characterize the universality classes of the disordered models 
as well as of the corresponding coupled two-dimensional CFT's in the zero-layer limits.
Now, it is noteworthy that they 
are simply related to the quantities originating from 
the specific structures of three dimensional lattices;
the coefficient in (\ref{period1}) for the disordered $O(n)$ model
and that in (\ref{period2}) for the random-bond $q$-state Potts model
correspond to
the Watson's triple integrals \cite{watson,glasser}
which represent the special values of Green's functions  
on the body centered cubic (bcc) lattice 
and on the face centered cubic (fcc) lattice, respectively. 
For our disordered $O(n)$ model, to be concrete, 
the coefficient in (\ref{period1}) can be written as 
\beq
2\frac{K^2(\sin\frac{\pi}{4})}{\pi^3}=\frac{1}{2\pi}\left(\frac{1}{\pi^3}\int_0^\pi\int_0^\pi\int_0^\pi  
\frac{dudvdw}{1-\cos u \cos v \cos w}\right),
\label{watson_on}
\eeq
where the quantity in the parenthesis is the Watson integral for the bcc lattice.
It would be interesting to see whether this coefficient 
can be derived, asymptotically at large scales,  
by a combinatoric analysis of coupled loop model on a lattice such as 
the one defined from the partition function (\ref{zmbart}).

These remarks are also useful to give representations  
for the two coefficients in terms of  {\it rational} integrals 
in order to see they are ``periods" proposed in ref. \cite{period}.
Looking back the arguments, both coefficients are obtained 
not as products but as ratios between the complex Selberg integrals \cite{aomoto} 
and the non-symmetric extensions of them such as 
the one obtained in \ref{integrals}, 
and thus it is not a priory clear that 
they are periods
even if each of these Selberg-type integrals independently turns out to be a period. 
But if we start from the Watson integrals, 
they can be transformed into rational integrals \cite{watson},
and hence the coefficients in (\ref{period1}) and that in (\ref{period2}),
characterizing the two universality classes interconnected
at the disordered Ising CFT, belong to, in fact, periods\footnote{
More precisely, as suggested in the expression (\ref{watson_on}), 
the coefficient in (\ref{period1}) for the disordered $O(n)$ model
 belongs to the $1/\pi$-extended period ring 
$\mathcal{P}[1/\pi]$ defined also in ref. \cite{period}.}.

Now, we list future directions in the following:

(i) {\it The fractal dimensions and SLE with the replica symmetry.}
We have seen that there is a line of IR fixed points for $n_*<n<1$.
An immediate issue to be discussed is 
the fractal dimension $D_F$ of the loops in the IR fixed points. 
One may figure out the corrections for the Fisher exponent in (\ref{sigmaf})
by perturbative calculation like the one performed in this paper.
Related result has recently appeared for the random-bond 
$q$-state Potts model \cite{jacobsenpicco}.
Further, the effective scale invariance on the line of the IR fixed points suggests
the possibility of constructing one-parameter family of new SLE's.
In order to describe the shape of the disordered loops,
stochastic motion over the effective space enhanced by 
the replica permutation symmetry may be considered 
in analogy with the SLE driven by 
a composition of the stochastic motion in the real space and that 
in the internal space such as an SU(2) \cite{ludwigwiegmann} 
or a $Z_N$ \cite{santachiara}.

(ii) {\it The strong coupling phase.}
We have seen that the coupled $O(n)$ model has moved to strongly-coupled theory
by the two-layer process in Figure \ref{intuitive}-(b) in the region $0<n<n_*$.
The physical picture of the strong coupling region 
is important for the polymers in disordered environment 
\cite{ledoussal}.  
Especially for this region, consideration on the theory space with 
the replica symmetry breaking (RSB) is desirable. 
Although the RSB-RG flow has been proposed in the random-bond $q$-state Potts model 
\cite{dotsenkodotsenko},  numerical studies there supports 
the result of the replica symmetric fixed point rather than that of the RSB fixed point.
We anticipate our model has more reason to favor the RSB situation 
due to formations of replica bound states 
boosted by the presence of the two-layer process. 
We should add that a large class of disordered models
which includes strongly-coupled layered $O(n)$ models  
has recently been discussed from the view point of the AdS/CFT duality \cite{fujitaryu}.
Actually, the focus of their analysis is on large-$n$, 
while our strong coupling region lies at $n\approx 0$;
although a direct application seems to be difficult,
this direction may be valuable in order to grasp the nature of  
strong coupling phases induced by disorder.  

(iii) {\it Non-local observables.}
We have studied the scaling dimensions of the fields 
corresponding to the local segments of loops.
In the pure $O(n)$ model, some results are known on non-local observables such as 
the area of loops \cite{cardyarea}
or the linking numbers around multiple points \cite{cardylinking}.
A challenging problem is 
the study of the off-diagonal pair-correlation between the global shapes of loops
in the disordered model.  
The whole shape of a loop itself is, of course, difficult to handle.
But extracting an essential part of such a correlation between non-local objects 
in a simple disordered system may possibly provide us with the basis of other important problems.
In chaotic systems, for instance, the off-diagonal correlation between periodic orbits, 
which are analogous to the disordered loops studied in this paper,
is already a central issue in understanding the quantum level statistics
\cite{bogomolnykeating,richter}.
It would be interesting if the possible relation between the disordered loops and periodic orbits in a chaotic system could be clarified.  

In conclusion, we have studied the disordered $O(n)$ models in two dimensions.
We formulated the inhomogeneous loop model on a lattice.
After adopting the replica method in order to take the disorder average,
the models were mapped to the coupled layers of the homogeneous loop models. 
Some of the possible differences in the lattice formulation of the disordered models
were argued to be irrelevant using the properties of the pure model in the dilute phase.
Then we discussed the continuum limit of the disordered model  
assuming the identification between the energy operator $\CE$
and the primary field $\phi_{1,3}$ in the vicinity of the pure critical $O(n)$ model.
We considered the RG flow in the replica symmetric theory space.
Since the most relevant interaction $\CE^{(a)}\CE^{(b)}$ becomes marginal at 
the disordered Ising model $n=1$,
we used the epsilon expansion methods near $n=1$ to perform a perturbative calculation.
At one loop level, the RG flow suggests that there exists a critical value $n=n_*$,
the strong disorder region for $0<n<n_*$ and the random fixed point for $n_*<n<1$.
These differences are largely controlled by the behavior of 
the structure constant $C_{\CE\CE}^{\CE}$
which encodes the selection rules in the pure $O(n)$ model.
We interpreted the OPE intuitively 
and explained the picture of the strongly-coupled phase.
Then we perform the two loop calculation near $n=1$,
where $C_{\CE\CE}^{\CE}$ is $\mathcal{O}(\ep)$. 
The beta function is equivalent to that of the Gross-Neveu model and thus checked the previously known result in the random-bond Ising model. 
The correction to scaling dimensions of the energy operator and 
the spin operator was calculated up to two loop.
Finally, We calculated the effective central charge and see this increase under the RG flow.

\ack
The author would like to thank S. Hikami for his continuous encouragement,
stimulating discussion on replica methods,
and reading of the manuscript.
The author has greatly benefited from enlightening discussions
with S. Ryu, T. Takebe and D. Bernard
about disordered systems and conformal invariance. 
It is pleasure to thank A. Kuniba, K. Fukushima, T. \& H. Shimada and H. Seki 
for various comments and encouragement.
The author would also like to thank T. Yoshimoto, T. Kimura, H. Katsura and K. Ino
for useful discussions and their interest in the work.

\newpage
\appendix
\setcounter{footnote}{1}
\section{The vertex operators in the critical Liouville field theory}\label{coulombgasrep}

The correlation functions in the $O(n)$ model can be described by 
the vertex operators in the critical Liouville theory
\footnote{Although it is often referred as the ``Coulomb gas representation" 
\cite{cardy, dotsenkopicco},
we call it  the ``vertex operator representation" in the 
critical Liouville theory \cite{kondev} in distinction with the 
``Coulomb gas method" \cite{nienhuis}.}.
The basic idea is to deform the gaussian free field theory
by putting the background charge $(-2\alpha_0)$ at the infinity.
This is accomplished by coupling the scalar field $\varphi$ to the scalar curvature $R$. 
The action of the theory is formally given by
\beq 
S = \frac{1}{16\pi} \int d^2 x \,\left[(\nabla\varphi)^2 
                    + 4i \al_0 R\varphi +\lambda_+ e^{i\alpha_+ \varphi} +\lambda_-  e^{i\alpha_- \varphi}\right].
\label{liouvilleaction}
\eeq
The corresponding central charge is determined from 
the OPE of the stress-energy tensor and turns out to be
\beq
c=1-24\alpha_0^2.
\eeq
The correlation function of the vertex operators $V_{\al_k}=e^{i\al_k \varphi}$ 
in this theory is non-zero 
only if the overall charge neutrality $\sum_k \al_k=2\al_0$ is satisfied. 
The screening vertex operators $V_\pm=e^{i\al_\pm \varphi}$ are incorporated
into the correlation functions in order to ensure the neutrality.  
The charges of the screening operators are chosen such that they are marginal, 
because otherwise the conformal invariance of the gaussian theory would be violated.
This condition determine the charges:
\beq
\al_\pm=\al_0\pm \sqrt{\al_0^2+1} .
\label{al+al-}
\eeq
The charge $\al_-$ in the $O(n)$ model  
and in the $m$-th minimal model are given by
\beq
n=-2\cos \left(\pi /\alpha_-^2\right) 
\qquad\mathrm{and}\qquad \al_-^2=\frac{m}{m+1} \label{mmplus1},
\label{minimalmm1}
\eeq
respectively.
In this paper, we fix our convention concerning the sign of the charges such that $\al_0>0$ and $\al_0<0$ 
describe the dilute (critical) and the dense (low-temperature) phase of  $O(n)$ model, respectively.
The charge $\al_{r,s}$ and its conjugate $\al_{\overline{r,s}}$ is defined as 
a linear combination of $\alpha_+$ and $\alpha_-$:
\bea
\alpha_{r,s}&\equiv &\frac{1}{2} (1-r) \alpha_+ + \frac{1}{2} (1-s) \alpha_- ,\nonumber\\
\alpha_{\overline{r,s}}  &\equiv & \frac{1}{2} (1+r) \alpha_+ + \frac{1}{2} (1+s) \alpha_-= 2\alpha_0 -\alpha_{r,s}\ .
\label{charge}
\eea
The scaling dimensions of the primary operators $\phi_{r,s}$ is given by
\beq
2\Delta_{r,s} = -2\alpha_{r,s}\al_{\overline{r,s}} = \frac{1}{2}\left[(r\alpha_+ + s\alpha_-)^2-(\alpha_+ +\alpha_-)^2\right].
\label{dimension}
\eeq
In the two-loop calculation in Section \ref{twoloop}, 
we use the vertex operator representations of the four-point functions
$\l<\CE(0) \CE(1) \CE(z) \CE(\infty)\r>_0$ for (\ref{ivertex})
and $\l<\sig(0) \CE(1) \CE(z) \sig(\infty)\r>_0$ for (\ref{k2vertex}) and (\ref{k3vertex}).
In order to satisfy the neutrality, we should include two of the screening operator $V_{-}$
in the vertex correlation functions.
The normalization factors in the four-point functions are determined in a decoupling limit,
such that the structure constant $C_{\CE\CE}^{I}$ is fixed to unity.
The value can be calculated by the complex Selberg integral \cite{aomoto}
as well as by the method described in \ref{integrals}; 
the results are the same for both types of the four-point functions:
\beq
\label{N_aomoto}
\mathcal{N}^{-1}=\frac{\Gamma\left(\frac{1}{4}\right)^8}{8\pi^2}=8\varpi^4+\mathcal{O}(\ep),
\eeq
where 
$\varpi=\Gamma\left(\frac{1}{4}\right)^2/\left(2^{3/2}\pi^{1/2}\right)=2.62205755\cdots$ 
is the lemniscate constant. 

\section{Some details of the integral for the two-loop beta function 
and the scaling dimension of the energy operator}\label{integralbeta}
In this Appendix, we thoroughly use the result of \ref{integrals}.
We read off the exponents in (\ref{I}) 
from the vertex operator representation 
(\ref{ivertex}).
$2a=4\al_{13}\al_{\overline{1,3}},2b=8\al_{13}\al_{\overline{1,3}}$,
$2a'=2b'=4\al_{13}\al_{-}$,
$2f=4\al_{\overline{1,3}}\al_{-}, 2g=4\al_{-}^2$
For convenience, we interchange the values of $a$ and $b$, and those of $a'$ and $b'$
(the latter is trivial, since here $a'=b'$).
These interchanges are possible by an obvious symmetry 
between $0$ and $1$ in the integral (\ref{I}).

As a result, we have $a=2b=-2f=4\al_{13}\al_{\overline{1,3}}=-2+8\ep$,
$a'=b'=-g=2\al_{13}\al_{-}=-\frac{3}{2}+2\ep$.
From these parameters, we determine the integral 
by the ``scattering amplitude" formula (\ref{i_amplitude}).
The scattering matrix $\mathcal{M}$ are given by (\ref{matrixelements}) and read,
\bea 
\mathcal{M}=&
\begin{pmatrix}
-4\pi\ep &0&4\pi\ep\\
4\pi\ep-16\pi^2\ep^2&-4\pi\ep&0\\
-4\pi\ep+16\pi^2\ep^2&4\pi\ep-16\pi^2\ep^2&-4\pi\ep
\end{pmatrix}+\mathcal{O}(\ep^3).
\label{energym}
\eea
From (\ref{j_parameters}), 
the initial and the final state basis in (\ref{i_amplitude}) are given by 
\bea
\fl 
J_1^+&=\left\langle\begin{matrix}
\frac{3}{2}+2\epsilon     & -\frac{1}{2}+2\epsilon   & -1+4\epsilon\\
-\frac{1}{2}+6\epsilon   & \frac{5}{2}-2\epsilon    & 1-4\epsilon\\
-1+8\epsilon  &          2-4\epsilon                       & \frac{3}{2}-2\epsilon\end{matrix}\right\rangle,\, &
J_1^-=\left\langle\begin{matrix}
\frac{5}{2}-10\epsilon & 4\epsilon   & -1+4\epsilon\\
\frac{1}{2}-6\epsilon   & 2-4\epsilon    & \frac{3}{2}-2\epsilon\\
-\frac{1}{2}+2\epsilon  & \frac{5}{2}-2\epsilon  & \frac{3}{2}-2\epsilon
\end{matrix}\right\rangle ,\nonumber\\
\fl
J_2^+&=\left\langle\begin{matrix}
\frac{3}{2}+2\epsilon     & -\frac{1}{2}+2\epsilon   & -\frac{3}{2}+2\epsilon\\
-\frac{1}{2}+6\epsilon   &   2-4\epsilon     & \frac{3}{2}-2\epsilon\\
-\frac{1}{2}+2\epsilon  &          2-4\epsilon                       & 1-4\epsilon
\end{matrix}\right\rangle,\,&
J_2^-=\left\langle\begin{matrix}
\frac{5}{2}-10\epsilon & -\frac{1}{2}+2\epsilon & -\frac{3}{2}+2\epsilon\\
\frac{1}{2}-6\epsilon   & 2-4\epsilon    & \frac{3}{2}-2\epsilon\\
-\frac{1}{2}+2\epsilon  & 2-4\epsilon  & 1-4\epsilon
\end{matrix}\right\rangle ,\nonumber\\
\fl
J_3^+&=\left\langle\begin{matrix}
\frac{3}{2}+2\epsilon     & 4\epsilon   & -1+4\epsilon\\
\frac{1}{2}+2\epsilon   &   2-4\epsilon     & \frac{3}{2}-2\epsilon\\
-\frac{1}{2}+2\epsilon  & \frac{5}{2}-2\epsilon    & \frac{3}{2}-2\epsilon
\end{matrix}\right\rangle,\,&
J_3^-=\left\langle\begin{matrix}
\frac{5}{2}-10\epsilon & -\frac{1}{2}+2\epsilon   & -1+4\epsilon\\
\frac{1}{2}-6\epsilon   & \frac{5}{2}-2\epsilon    & 1-4\epsilon\\
-4\epsilon  & 2-4\epsilon                                & \frac{3}{2}-2\epsilon
\end{matrix}\right\rangle .
\eea
where, $J_l^\pm\ (l=1, 2, 3)$ are real triple integrals defined in (\ref{J1+})-(\ref{J3-}),
and the symbol $\l< ... \r>$ is explained in (\ref{symbol}).
We should discuss the order of these integrals $J_l^\pm$ in $\ep$;
for this purpose, we can use the series representation derived in \ref{hypergeometric}.
From (\ref{gamma_l}), the prefactors $\gamma_l^\pm$ are calculated as, 
\beq
\fl
\eqalign{(\gamma_1^+,\, \gamma_2^+,\, \gamma_3^+)=
\begin{pmatrix}
-\frac{3\pi^2}{16\epsilon}+\mathcal{O}(\ep^0),\,-16\pi+\mathcal{O}(\ep),\,-\frac{\pi}{2\epsilon}+\mathcal{O}(\ep^0)
\end{pmatrix},\\
(\gamma_1^-,\, \gamma_2^-,\, \gamma_3^-)=
\begin{pmatrix}
-\frac{\pi}{2\epsilon}+\frac{16\pi}{3}+\mathcal{O}(\ep),\,
8\pi+\mathcal{O}(\ep),\,\frac{9\pi^2}{64\epsilon}+\frac{9\pi^2}{16}\left(-1+8\log 2\right)+\mathcal{O}(\ep)
\end{pmatrix}\hspace{-1.2mm}.}
\label{energygamma}
\eeq
Naively, we expect the leading terms in $J_l=\gamma_l S_l$ is the same order in $\ep$ as the $\gamma_l$,
since all the triple series $S_l$ contains the $(i, j, k)=(0,0,0)$, which is unity and hence $\mathcal{O}(1)$. 
Actually, it is not the case for $J_1^+$; the series $S_1^+$ is $\mathcal{O}(\ep)$ 
because of the non-trivial cancellation between the constant terms.

Since this observation makes the crucial point in the calculation,
we describe the calculation of $J_1^+$ in detail.
First, from (\ref{s_l}), we have
\beq\label{s1+}
\fl
S_1^+ =
\sum_{i=0}^{\infty}\sum_{j=0}^{\infty}\sum_{k=0}^{\infty}
\frac{(\-1\+4\ep)_{i} (1\-4\ep)_{j} (\frac{3}{2}\-2\ep)_{k}}{i!\ \ j!\ \ k!} 
\frac{(\frac{3}{2}\+2\ep)_{j+k}(\frac{\-1}{2}\+6\ep)_{i+j+k}(\-1\+8\ep)_{i+j}}
{(1\+4\ep)_{j+k}\ (2\+4\ep)_{i+j+k} (1\+4\ep)_{i+j}}.
\eeq
Denoting the summand in (\ref{s1+}) as $S_{i,j,k}$,
we observe, from the definition of the Pochhammer symbol $(x)_{k}=x(x+1)\cdots(x+k-1)$,
that $S_{i,j,k}$ gets a factor of $\mathcal{O}(\ep)$
whenever each of two conditions $\left\{i\geq 2,\, i+j\geq 2\right\}$ is satisfied.
Thus, we let  
\bea
 a_1=\sum_{k\geq 0} S_{0,0,k}, \qquad&a_2=\sum_{k\geq 0} S_{1,0,k},\qquad&a_3=\sum_{k\geq 0} S_{0,1,k},\nonumber\\
 a_4=\sum_{j\geq 2,\,k\geq 0}S_{0,j,k},&a_5= \sum_{j \geq 1,k\geq 0}S_{1,j,k},&
\eea
where the leading order of $a_1, a_2, a_3$ and $a_4,a_5$ are $\mathcal{O}(\epsilon^0)$ and $\mathcal{O}(\epsilon^1)$, respectively. 
To be specific, at the accuracy of $\mathcal{O}(\ep)$, we are left with the following: 
\bea
\fl
a_1&=\qquad \ _3 F_2
\Bigl(\begin{matrix}\frac{3}{2}-2\ep&\frac{3}{2}+2\ep&\frac{-1}{2}+6\ep\\ &1+4\ep&2+4\ep\end{matrix};1\Bigr) &= 
1\cdot\frac{-\Gamma(\frac{1}{4})^4+48\Gamma(\frac{3}{4})^4}{12\pi^3}+\mathcal{O}(\epsilon),
\label{a1}\\
\fl
a_2&=\left(\frac{-1}{4}+\frac{15}{2}\ep\right)\cdot _3\! F_2
\Bigl(\!\!\begin{matrix}\frac{3}{2}-2\ep &\frac{3}{2}+2\ep&\frac{1}{2}+6\ep\\ &1+4\ep&3+4\ep\end{matrix}\!\!\Bigr)  &=
\frac{-1}{4}\cdot\frac{\Gamma(\frac{1}{4})^4+48\Gamma(\frac{3}{4})^4}{3\pi^3}+\mathcal{O}(\epsilon),
\label{a2}\\
\fl
a_3&=\left(\frac{3}{8}-\frac{49}{4}\ep\right)\cdot _3\! F_2
\Bigl(\!\!\begin{matrix}\frac{3}{2}-2\ep&\frac{5}{2}+2\ep&\frac{1}{2}+6\ep\\ &2+4\ep&3+4\ep\end{matrix}\!\!\Bigr)  &= 
\frac{3}{8}\cdot\frac{4\Gamma(\frac{1}{4})^4}{9\pi^3}+\mathcal{O}(\epsilon),
\label{a3}
\eea
where the values of generalized hypergeometric functions $_3 F_2$ at unity are used.
The arguments, which is always taken at unity, are suppressed in the first equalities 
in (\ref{a2}), (\ref{a3}) and henceforth.
Now, we see the aforementioned cancellation at $\mathcal{O}(\epsilon^0)$, and get 
\beq
a_1+a_2+a_3=\mathcal{O}(\ep).
\label{contiguity}
\eeq
This is an example of identities known as
the ``contiguity relation" between hyper-geometric functions.
Note that this type of the leading order cancellation occurs only for $S_1^+$.
We note that the cancellations at this order guarantees a consistency in the RG scheme.
Resulting $\mathcal{O}(\ep)$ term can be evaluated numerically, using 
\beq
(x+\epsilon)_k=(x)_k\cdot\left\{1+\ep\left[\psi(x+k)-\psi(x)\right]+\mathcal{O}(\ep^2)\right\},
\eeq
where $\psi(x)$ is the di-gamma function
\footnote{Actually, there have recently been extensive studies on 
the closed form evaluation of the expansion of the hypergeometric series  
in algorithmic approach, for example \cite{yost}. 
However, to our knowledge, the desired expansions here have not been published. }.
The other terms contributing $\mathcal{O}(\ep)$ in $S_1^+$ are 
\bea
a_4+a_5=&\frac{\epsilon}{16}\sum_{i,j}\frac{\left(\frac{3}{2}\right)_i(1)_j\left(\frac{3}{2}\right)_{i+j}\left(\frac{5}{2}\right)_{i+j}}
{(1)_i(3)_j(4)_{i+j}(3)_{i+j}}+\mathcal{O}(\ep^2).
\label{4plus5}
\eea
To evaluate this double series, it is useful to note the following identity:
\beq
\sum_{i,j}\frac{(a)_i(b)_j(c)_{i+j}(d)_{i+j}}
{(1)_i(1)_j(e)_{i+j}(f)_{i+j}}
=\ _3 F_2\Bigl(\begin{matrix}a+b&c&d\\ &e&f\end{matrix}\Bigr).
\eeq
This holds because $(a+b)_k$ has the same binomial expansion as $(a+b)^k$.
By comparing the order $\delta$ terms of the following expression,
\beq
\sum_{i,j}\frac{\left(\frac{3}{2}\right)_i\left(-1+\delta\right)_j\left(\frac{-1}{2}\right)_{i+j}\left(\frac{1}{2}\right)_{i+j}}
{(1)_i(1)_j(2)_{i+j}(1)_{i+j}}
=\ _3 F_2\Bigl(\begin{matrix}\frac{1}{2}+\delta&\frac{-1}{2}&\frac{1}{2}\\ &2&1\end{matrix}\Bigr),
\eeq
one can express the right hand side of (\ref{4plus5}) in terms of a single series:
\beq
\eqalign{a_4+a_5=\mathcal{Q}_1\cdot\epsilon
+\mathcal{O}(\epsilon^2),\\
\mathcal{Q}_1=\ _3 F_2\Bigl(\begin{matrix}\frac{1}{2}&\frac{3}{2}&\frac{3}{2}\\ &3&2\end{matrix}\Bigr)
+8F_\delta\Bigl(\begin{matrix}\frac{1}{2}+\delta&\frac{-1}{2}&\frac{1}{2}\\ &2&1\end{matrix}\Bigr),}
\label{a45}
\eeq
where $F_{\delta}(...)$ denotes the coefficients of the order $\delta$ terms in the $_3F_2$ function at unity. 
In this case,
\beq\fl
\ _3 F_2\Bigl(\begin{matrix}\frac{1}{2}+\delta&\frac{-1}{2}&\frac{1}{2}\\ &2&1\end{matrix}\Bigr)
-\ _3 F_2\Bigl(\begin{matrix}\frac{1}{2}&\frac{-1}{2}&\frac{1}{2}\\ &2&1\end{matrix}\Bigr)
=\delta \cdot F_\delta\Bigl(\begin{matrix}\frac{1}{2}+\delta&\frac{-1}{2}&\frac{1}{2}\\ &2&1\end{matrix}\Bigr)
+\mathcal{O}(\delta^2).
\eeq
We use this notation hereafter. 
Now, collecting the $\mathcal{O}(\ep)$ parts of (\ref{a1})-(\ref{a3}) and (\ref{a45}),  we obtain
\beq
\fl
J_1^+ = \gamma_1^+  \cdot  S_1^+=\left(\frac{-3\pi^2}{16\ep}+\mathcal{O}(\ep^0)\right)\left(0\cdot 
\ep^0+\tilde{A}\ep+\mathcal{O}(\ep^2)\right)=A+\mathcal{O}(\ep) ,
\eeq
where we have used (\ref{energygamma}) for $\gamma_1^+$. The constant $A$ is given by
\bea
\fl
A=&\frac{-3\pi^2}{16}\left[\frac{15}{2}  \, _3 F_2\Bigl(\begin{matrix}\frac{3}{2}&\frac{3}{2}&\frac{1}{2}\\ &1&3\end{matrix}\Bigr)-\frac{49}{4}\, _3 F_2\Bigl(\begin{matrix}\frac{3}{2}&\frac{5}{2}&\frac{1}{2}\\ &2&3\end{matrix}\Bigr)+F_{\delta}\Bigl(\begin{matrix}\frac{3}{2}&\frac{3}{2}&\frac{-1}{2}+6\delta\\ &1+4\delta&2+4\delta\end{matrix}\Bigr)\right.
\nonumber\\
\fl &\hspace{8mm} \left. -\frac{1}{4}F_{\delta}\Bigl(\begin{matrix}\frac{3}{2}&\frac{3}{2}&\frac{1}{2}+6\delta\\ &1+4\delta&3+4\delta\end{matrix}\Bigr)
+\frac{3}{8}F_{\delta}\Bigl(\begin{matrix}\frac{3}{2}-2\delta&\frac{5}{2}+2\delta&\frac{1}{2}+6\delta\\
&2+4\delta&3+4\delta\end{matrix}\Bigr)
+\mathcal{Q}_1\right].
\label{aconstant}
\eea
This completes the calculation of $J_1^+$.
Since the linear relations (\ref{relations}) are solved as
\bea
J_1^+=\frac{1}{2}\left(J_1^{-}-2J_2^{-}+J_3^-\right)-
2\pi\ep\left(J_1^{-}+J_3^-\right)+\mathcal{O}(\ep),\label{energyrel1}\\
J_2^+=-J_2^{-}-4\pi\ep\left(J_1^- + J_2^-\right)+\mathcal{O}(\ep),\label{energyrel2}\\
J_3^+=J_1^{-}-J_2^{-}+4\pi\ep\left(-J_2^- + J_3^-\right)+\mathcal{O}(\ep),\label{energyrel3},
\eea
the knowledge of the other two bases are sufficient in the formula (\ref{i_amplitude}).
Actually, we can calculate $J_1^-$ and $J_3^-$ in similar manners.
Let us define constants $D$ and $F$ as follows:
\bea
\label{hyp1-}
\fl
S_1^- &=
\sum_{i,j,k=0}^{\infty}
\frac{(-1+4\ep)_{i}(\frac{3}{2}-2\ep)_{j} (\frac{3}{2}-2\ep)_{k}}{i!\ \ j!\ \ k!} 
\frac{(\frac{5}{2}-10\ep)_{j+k}(\frac{1}{2}-6\ep)_{i+j+k}(-\frac{1}{2}+2\ep)_{i+j}}
{(\frac{5}{2}-6\ep)_{j+k}\ (\frac{5}{2}-2\ep)_{i+j+k} \ (2)_{i+j}}\nonumber \\
\fl &=\frac{\Gamma(\frac{1}{4})^4}{8\pi^2}+\tilde{D}\ep+\mathcal{O}(\ep^2),\\
\fl J_1^-&=\gamma_1^-\cdot S_1^-=-\frac{\,\Gamma(\frac{1}{4})^4}{16\pi\ep}+D+\mathcal{O}(\ep)
\label{energyD},
\eea
\bea
\label{hyp3-}
\fl
S_3^- &=
\sum_{i,j,k=0}^{\infty}
\frac{(-1+4\ep)_{i}(1-4\ep)_{j} (\frac{3}{2}-2\ep)_{k}}{i!\ \ j!\ \ k!} 
\frac{(\frac{5}{2}-10\ep)_{j+k}(\frac{1}{2}-6\ep)_{i+j+k}(-4\ep)_{i+j}}
{(2-8\ep)_{j+k}\ (3-8\ep)_{i+j+k} \ (2-8\ep)_{i+j}}\nonumber \\
\fl &=\frac{4\Gamma(\frac{1}{4})^4}{9\pi^3}+\tilde{F}\ep+\mathcal{O}(\ep^2),\\
\fl J_3^-&=\gamma_3^-\cdot S_3^-=\frac{\,\Gamma(\frac{1}{4})^4}{16\pi\ep}+F+\mathcal{O}(\ep),
\label{energyF}
\eea
where, again, (\ref{energygamma}) has been used for $\gamma_1^-$ and $\gamma_3^-$. 
Then, $D$ and $F$ are given by
\bea\fl
D=&\frac{2\Gamma\left(\frac{1}{4}\right)^4}{3\pi}+\frac{\pi^2}{8}\left(1+3\log 2\right)
\, _3 F_2\Bigl(\begin{matrix}\frac{1}{2}&\frac{3}{2}&\frac{3}{2}\\ &3&2\end{matrix}\Bigr)
-\frac{3\pi^2}{16}\mathcal{Q}_1-\frac{\pi}{10}\mathcal{Q}_2 \nonumber\\
\fl&
-\frac{3\pi^2}{64}F_{\delta}
\Bigl(\!\begin{matrix}\frac{3}{2}-2\delta&\frac{3}{2}+2\delta&\frac{1}{2}+2\delta\\ &2+4\delta&3\end{matrix}\!\Bigr)
-\frac{3\pi^2}{8}F_{\delta}
\Bigl(\!\begin{matrix}\frac{3}{2}-2\delta&\frac{1}{2}+2\delta&\frac{-1}{2}+2\delta\\ &1+4\delta&2\end{matrix}\!\Bigr), 
\label{dconstant}
\eea
\beq
\fl F=\frac{\Gamma\left(\frac{1}{4}\right)^4}{4\pi}\left( -1+8\log 2\right)+
\frac{9\pi^2}{64}F_{\delta}
\Bigl(\begin{matrix}\frac{1}{2}-6\delta&\frac{3}{2}-2\delta&\frac{5}{2}-10\delta\\ &2-8\delta&3-8\delta\end{matrix}\Bigr) 
-\frac{3\pi^2}{16}\mathcal{Q}_1,
\label{fconstant}
\eeq
where $\mathcal{Q}_2$ denotes well-converging double series
\beq\fl
\mathcal{Q}_2=\sum_{j,k}\frac{1}{\frac{5}{2}+j+k}
\frac{\left(\frac{1}{2}\right)_j\left(\frac{3}{2}\right)_j\left(\frac{3}{2}\right)_k\left(\frac{3}{2}\right)_{j+k}}{(3)_j(1)_j(1)_k\left(\frac{7}{2}\right)_{j+k}}.
\eeq
Now, by substituting (\ref{energyD}), (\ref{energyF}) and the matrix elements (\ref{energym}) 
into the formula (\ref{i_amplitude}),
we obtain
\bea
\fl(-2\mathcal{N})^{-1}\,\mathcal{I}(\infty,\ep)&=\frac{\Gamma\left(\frac{1}{4}\right)^4}{4}(2A+D+F-J_2^+-J_2^-)
+16\pi^2\ep^2\left(\frac{\Gamma\left(\frac{1}{4}\right)^4}{16\pi\ep}\right)^2
+\mathcal{O}(\ep)\nonumber\\
\fl&=\frac{\Gamma\left(\frac{1}{4}\right)^4}{4}(2A+D+F)+\mathcal{O}(\ep),
\label{gamma2adf}
\eea
where, in the first and the second line, we have used (\ref{energyrel2}) and (\ref{energyrel3}), respectively.
Our concern is to evaluate this, from (\ref{aconstant}), (\ref{dconstant}) and (\ref{fconstant}), within high numerical accuracy. 
To this end, we have used asymptotic behavior of the di-gamma function 
for the evaluation of $F_\delta$.
We obtained the value
\bea
2A+D+F=27.50074327(21),
\eea
which is in good agreement with another value 
$\Gamma(1/4)^4/2\pi=4\varpi^2=27.50074327208\cdots$.
Assuming the latter value in (\ref{gamma2adf}) and using (\ref{N_aomoto}), we obtain
\bea
\fl\mathcal{I}(\infty,\ep)&=-2\mathcal{N}\,\frac{\Gamma\left(\frac{1}{4}\right)^8}{8\pi}+\mathcal{O}(\ep)
\nonumber\\
\fl&=-2\pi+\mathcal{O}(\ep).
\label{i_infty_result}
\eea

\section{Integrals for the correction coefficient of the spin field dimension}\label{integralspin}
We here calculate the integrals $\mathcal{K}_2(r,\ep)$ in (\ref{k2})
and $\mathcal{K}_3(r,\ep)$ in (\ref{k3}).
From (\ref{k2vertex}) and (\ref{k3vertex}),  we read off the exponents defined in (\ref{I}).
After interchanging the values of the parameters 
$a\leftrightarrow b$ and $a'\leftrightarrow b'$ (as in \ref{integralbeta}), 
we have
$a=2\al_{13}^2-2\Delta_\CE=\frac{1}{2}+2\ep$, 
$b=2\al_{\overline{p-1,p}}\al_{13}=-\frac{1}{4}+\ep$,
$a'=f=2\al_{13}\al_{-}=-\frac{3}{2}+2\ep$,
$b'=2\al_{\overline{p-1,p}}\al_{-}=\frac{1}{4}-\ep$,
and $g=2\al_{-}^2=\frac{1}{2}-2\ep$ for $\mathcal{K}_2$;
we have almost the same set of the parameters as $\mathcal{K}_2$
but $a=2\al_{13}^2+1-4\Delta_\CE=\frac{1}{2}+6\ep$  for $\mathcal{K}_3$.
We substitute these into (\ref{matrixelements})
and get, both for $\mathcal{K}_2$ and $\mathcal{K}_3$, the same matrix:
\beq 
\mathcal{M}=\frac{1}{\sqrt{2}}
\begin{pmatrix}
2\pi\ep &-2\pi\ep&2\pi\ep\\
-1+3\pi\ep&2\pi\ep&-2\pi\ep\\
-1+3\pi\ep&-1+3\pi\ep&2\pi\ep
\end{pmatrix}+\mathcal{O}(\ep^2).
\label{spinmatrix}
\eeq
It is useful to adopt a temporary notation $[x]$ which takes the value
$x$ and $x+4$ for $\mathcal{K}_2$ and $\mathcal{K}_3$, respectively.
From (\ref{j_parameters}), the basis in (\ref{i_amplitude}) is then determined as
\bea
\fl
J_1^+&=\left\langle\begin{matrix}
-1+[8]\epsilon     & \frac{5}{4}-\epsilon   & \frac{3}{2}-2\epsilon\\
-\frac{1}{2}+[6]\epsilon   & \frac{5}{2}-2\epsilon    & \frac{1}{4}-\epsilon\\
\frac{3}{2}+[2]\epsilon  & -\frac{1}{2}+2\epsilon     & -\frac{1}{4}+\epsilon\end{matrix}\right\rangle,&
J_1^-=\left\langle\begin{matrix}
\frac{3}{4}-[7]\epsilon & \frac{3}{4}+\epsilon   & \frac{3}{2}-2\epsilon\\
2-4\epsilon   & -\frac{1}{2}+2\epsilon    & -\frac{1}{4}+\epsilon\\
\frac{1}{4}-\epsilon  & \frac{5}{2}-2\epsilon  & -\frac{1}{4}+\epsilon
\end{matrix}\right\rangle ,\label{spinj1}\\
\fl
J_2^+&=\left\langle\begin{matrix}
-1+[8]\epsilon     & \frac{5}{4}-\epsilon   & -\frac{3}{2}+2\epsilon\\
-\frac{1}{2}+[6]\epsilon   &   -\frac{1}{2}+2\epsilon     & -\frac{1}{4}+\epsilon\\
-\frac{1}{2}+2\epsilon  &    -\frac{1}{2}+2\epsilon & \frac{1}{4}-\epsilon
\end{matrix}\right\rangle\!,&
J_2^-=\left\langle\begin{matrix}
\frac{3}{4}-[7]\epsilon & \frac{5}{4}-\epsilon & -\frac{3}{2}+2\epsilon\\
\frac{1}{2}-[6]\epsilon   & -\frac{1}{2}+2\epsilon   & -\frac{1}{4}+\epsilon\\
\frac{1}{4}-\epsilon  &  -\frac{1}{2}+2\epsilon  & \frac{1}{4}-\epsilon
\end{matrix}\right\rangle ,\label{spinj2}\\
\fl
J_3^+&=\left\langle\begin{matrix}
-1+[8]\epsilon     & \frac{3}{4}+\epsilon   & \frac{3}{2}-2\epsilon\\
\frac{1}{2}+2\epsilon   &   -\frac{1}{2}+2\epsilon     & -\frac{1}{4}+\epsilon\\
-\frac{1}{2}+2\epsilon  &  \frac{5}{2}-2\epsilon  & -\frac{1}{4}+\epsilon
\end{matrix}\right\rangle,&
J_3^-=\left\langle\begin{matrix}
\frac{3}{4}-[7]\epsilon & \frac{5}{4}-\epsilon   & \frac{3}{2}-2\epsilon\\
\frac{1}{2}-[6]\epsilon   & \frac{5}{2}-2\epsilon    & \frac{1}{4}-\epsilon\\
\frac{7}{4}-[7]\epsilon  & -\frac{1}{2}+2\epsilon     & -\frac{1}{4}+\epsilon
\end{matrix}\right\rangle ,
\label{spinj3}
\eea
For convenience of the presentation, we use 
the following notation both for $\mathcal{K}_2$ and $\mathcal{K}_3$:
\beq
(X,Y,Z)\equiv (J_1^+,J_2^+,J_3^+), \qquad (U,V,W)\equiv (J_1^-,J_2^-,J_3^-).
\label{xyzuvw}
\eeq
From (\ref{spinj1}), (\ref{gamma_l}) and (\ref{s_l}), we first observe that $X$ is $\mathcal{O}(\ep^0)$;
this is guaranteed by the same contiguity relation as in (\ref{contiguity}).
As a result, we realize that all of $X$, $Y$ and $Z$ are regular in $\ep$.
By the combination of this observation and the linear relations (\ref{relations}),
we infer $V=\mathcal{O}(\ep)$, which suggests a non-trivial cancellation at $\mathcal{O}(\ep^{-1})$ occurs
in the triple series $S_2^-$ determined from (\ref{spinj2}) and (\ref{s_l}). 
Now, by noting $V=\mathcal{O}(\ep)$, the linear relations (\ref{relations}) can be casted as
\bea
X-Z&=W/\sqrt{2}+\mathcal{O}(\ep),\label{spinrel1}\\
Y+Z&=\sqrt{2}\pi\ep(U+W)+\mathcal{O}(\ep^2),\\
\Omega \pi \ep~Y&=\sqrt{2}\pi\ep(U+W)-V/\sqrt{2}+\mathcal{O}(\ep^2)\label{spinrel3},
\eea
where we have introduced the variable $\Omega$, which takes the values:
\beq
\label{omega}
\Omega=\cases{8&for $\mathcal{K}_2$\\
12&for $\mathcal{K}_3$.\\}
\eeq
It should be noted that the leading order part of both $U$ and $W$ are $\mathcal{O}(1)$, 
and are common for $\mathcal{K}_2$ and $\mathcal{K}_3$.
Using the formula (\ref{i_amplitude}) for (\ref{spinmatrix}) and (\ref{xyzuvw}), 
with the help of (\ref{spinrel1})-(\ref{spinrel3}),
we obtain
\beq
\mathcal{K}_{*}=2\pi\ep\left(U^2+W^2-\Omega YZ \right)+\mathcal{O}(\ep^2),
\label{kstar}
\eeq
where the notation $\Omega$ in (\ref{omega}) is used to express 
both $\mathcal{K}_2$ and $\mathcal{K}_3$
in parallel.
Actually, only the combination $U^2+W^2$ is necessary for the disordered model
($M\rightarrow 0$).
Substituting the parameters in (\ref{spinj1}) and (\ref{spinj3}) 
into (\ref{gamma_l}) and (\ref{s_l}), we obtain
\bea
\fl U&=-\frac{32\sqrt{2}\pi}{7}\sum_{i=0}^{\infty}\sum_{j=0}^{\infty}\sum_{k=0}^{\infty}
\frac{(\frac{3}{2})_{i} (-\frac{1}{4})_{j} (-\frac{1}{4})_{k}}{i!\ \ j!\ \ k!} 
\frac{(\frac{3}{4})_{j+k}\ \ (2)_{i+j+k}\ \ (\frac{1}{4})_{i+j}}
{(\frac{3}{2})_{j+k}\ \ (\frac{3}{2})_{i+j+k}\ \ (\frac{11}{4})_{i+j}},\label{u_num}\\
\fl W&=-\frac{9\pi^{7/2}}{8\Gamma(\frac{1}{4})^2} \sum_{i=0}^{\infty}\sum_{j=0}^{\infty}\sum_{k=0}^{\infty}
\frac{(\frac{3}{2})_{i} (\frac{1}{4})_{j} (-\frac{1}{4})_{k}}{i!\ \ j!\ \ k!} 
\frac{(\frac{3}{4})_{j+k}\ \ (\frac{1}{2})_{i+j+k}\ \ (\frac{7}{4})_{i+j}}
{(2)_{j+k}\ \ (3)_{i+j+k}\ \ (\frac{5}{4})_{i+j}}\label{w_num}.
\eea
We obtain numerically 
\beq
U^2+W^2=671.0\pm 0.3,
\label{uw_numerical}
\eeq
which is compatible, within error bar ($\pm 0.05\%$), with the value
\beq
\frac{\Gamma(\frac{1}{4})^{12}}{8\pi^6}=64\frac{\varpi^6}{\pi^3}= 670.78\cdots.
\label{w6pi3}
\eeq 
Unfortunately, the convergence of the triple series $U$ is slow, and thus 
our numerical accuracy is not good here.
Nevertheless, we assume, for $U^2+W^2$, the value in (\ref{w6pi3}).

\section{Integrals}\label{integrals}
In our calculation of  the RG functions,
we should deal with a multiple integral over $\mathbb{C}^3$:

\bea
I=\int \!\!\!\int\!\!\!\int\! d^2zd^2ud^2v\ \  &|z|^{2a}|1-z|^{2b}|v-z|^{2f} |u-z|^{2f}\nonumber\\
&\cdot |u|^{2a'}|1-u|^{2b'}|u-v|^{2g}\ |v|^{2a'}|1-v|^{2b'} .
\label{I}
\eea
Since this form of the integral comes from the correlation functions of the vertex operators
(or, 
more plainly, from the interaction between charged particles in a two-dimensional plane),
it seems to be ubiquitous in physics and mathematics.
For example, the integral can be interpreted as a 
six-particle closed string amplitude \cite{shapiro}.
The integral in a special, {\it symmetric case of parameters} ($a=a'$, $b=b'$ and $f=g$) 
is well studied in the   
context of twisted cohomology, and known as the ``complex Selberg integral" \cite{aomoto}.
What we need, however, is the formula in {\it a non-symmetric case}, 
when doing perturbation theory around a conformal fixed point. 
In this respect, a formula for two variables was used in the study of 
random-bond Potts model by Dotsenko et al. \cite{dotsenkopicco}.
We extend their results and 
derive a formula for (\ref{I}) in a systematic way. 
The formula, obtained in (\ref{i_amplitude}), takes form of a scattering amplitude.

\subsection{Regularization of the one-dimensional intervals} \label{regularization}
We encounter with strong algebraic singularities that make the multiple integrals divergent.
For this reason, we consider an analytic continuation in the parameter of the integrals. 
In order to keep the discussion clear, it is helpful to make the way of the analytic continuation explicit.
The analytic continuation is achieved by the use of a ``regularization" of the intervals \cite{aomoto,mimachi},
which we now describe using the following one-dimensional simple example.

\begin{figure}[!htbp]
\begin{center}
\includegraphics[width=5cm]{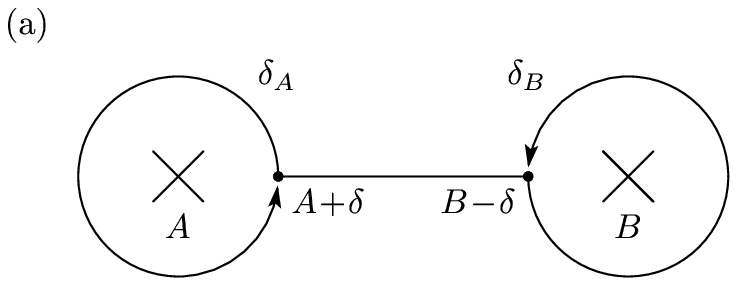}
\includegraphics[width=5cm]{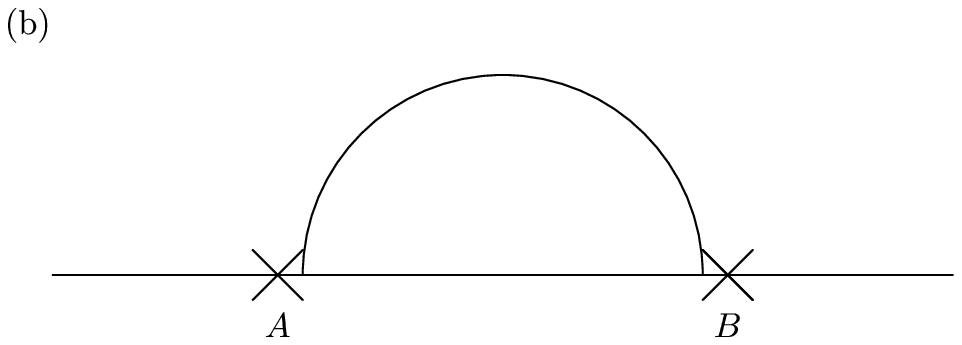}
\caption{(a) Two circles $\delta_A$ and $\delta_B$ in the definition (\ref{reg}) of the regularization of a interval $[A, B]$.\qquad
(b) The same regularization but in a simpler representation used in \ref{App_relations}.}
\label{reg_circles}
\end{center}
\end{figure}

Consider an integral on a real interval 
\beq
\int_A^B dx\ (x-A)^p(B-x)^q f(x),
\label{AB_integral}
\eeq
where $f(x)$ is an analytic function on a neighborhood of the interval $[A, B]$.
If $f(A)\neq 0$ and $f(B)\neq 0$, the condition $\mathrm{Re}\ p<-1$ or $\mathrm{Re}\ q<-1$ makes  the integral divergent.
The regularization of the interval  $[A, B]$ in the integral (\ref{AB_integral}) is given by a replacement 
\beq
\mathrm{reg:\:} \left[ A, B\right]\longmapsto  \frac{-\delta_A}{1-\exp(2\pi i p)} +\left[ A+\delta, B-\delta\right]+
\frac{+\delta_B}{1-\exp(2\pi i q)}. 
\label{reg}
\eeq
Here, as in figure \ref{reg_circles}, $\delta_A$ and $\delta_B$ are positively oriented circles of radius $\delta$ which have centers $A$ and $B$, and start at $A+\delta$ and $B-\delta$, respectively.
By replacing the interval $[A, B]$ by the regularized one ``$\mathrm{reg}\, [A, B]$" and taking the limit $\delta\rightarrow 0$,
the value of the integral (\ref{AB_integral}) remains same for $\mathrm{Re}\, p>-1$ and $\mathrm{Re}\, q>-1$ 
since the contributions from two additional circles vanish, 
and is now finite also for $\mathrm{Re}\, p<-1$ or $\mathrm{Re}\, q<-1$ unless $p\in \mathbb{Z}$ or $q\in \mathbb{Z}$.
In the latter case, adding two circles corresponds to the subtraction of infinite quantities, 
and the resulting finite value is what is known as the ``Hadamard finite part" of the integral
\footnote{Actually, the definition (\ref{reg}) of the regularization of a interval $[A, B]$ is proportional to well-known ``Pochhammer contour" which is used to define the analytic continuation of the hypergeometric functions.}.

The definition here is natural in the following sense.
Starting with the multiple integral (\ref{I}) and introducing an infinitesimal imaginary part,
we shall reach an iterated integral 
in which each integral is on {\it the regularization of the interval} rather than on the usual real interval. 
This regularization comes from the pairing of the paths on the upper and the lower half planes
both of which detour the branch points.
All the one-dimensional integrals in this Appendix 
should be regarded as the integral over the regularization of the interval.

\subsection{Decomposition of the multiple integral}\label{decomposition}
We shall consider the integrand in (\ref{I}) as a function defined on $\mathbb{C}^3\times \mathbb{C}^3$ rather than on $\mathbb{C}^3$;
writing $z=z_1+iz_2$ in the variable $z$, to be specific, 
the first factor in the integrand in (\ref{I}): $|z|^{2a}=(z_1^2+z_2^2)^a$ is now defined on   
$(z_1, z_2)\in \mathbb{C}\times\mathbb{C}$. Then
the path of $z_2$ is rotated by the angle $\pi/2-2\eta$, without hitting any singularity,
for a positive infinitesimal number $\eta$. 
By introducing a new variable $z_0$, we rewrite $|z|^{2a}$ as follows:  
\bea
|z|^{2a}=(z_1^2+z_2^2)^a\longrightarrow& \left(z_1^2+(ie^{-2i\eta}z_0)^2\right)^a\nonumber\\
&=\left(z_+-i\eta[z_+-z_-]\right)^a\left(z_-+i\eta[z_+-z_-]\right)^a,
\eea
where the notation $z_\pm=z_1\pm z_0$ is introduced.
In the following, we call $\{z_+,u_+,v_+\}$ and $\{z_-,u_-,v_-\}$, respectively, holomorphic and antiholomorphic variables.
Using a notation $X_\eta=\eta[X_+-X_-]$ for a quantity $X$, 
we decompose the integral (\ref{I}) as
\bea
\fl
I&=\left(\frac{i}{2}\right)^3\int\!\!\!\!\int\!\!\!\!\int\!\!\!\!\!\!\!\!\!\!\int\limits_{[-\infty,\infty]^6}\!\!\!\!\!\!\!\!\!\!\int\!\!\!\!\int 
dz_+ du_+ dv_+dz_- du_- dv_-\ \mathcal{J}_+ \cdot \mathcal{J}_-,\\
\fl\mathcal{J}_+&=(z_+\-i z_\eta)^a(z_+\-1\-i z_\eta)^{b}(z_+\-u_+\-i[z\-u]_\eta)^f(z_+\-v_+\-i [z\-v]_\eta)^f\nonumber\\
\fl&\cdot (u_+\-i u_\eta)^{a'}(u_+\-1\-iu_\eta)^{b'}(u_+\-v_+\-i[u\-v]_\eta)^g(v_+\-iv_\eta)^{a'}(v_+\-1\-iv_\eta)^{b'}\nonumber\\
\fl\mathcal{J}_-&= (z_-\+iz_\eta)^a(z_-\-1\+iz_\eta)^{b}(z_-\-u_-\+i[z\-u]_\eta)^f(z_-\-v_-\+i[z\-v]_\eta)^f\nonumber\\
\fl&\cdot(u_-\+iu_\eta)^{a'}(u_-\-1\+iu_\eta)^{b'}(u_-\-v_-\+i[u\-v]_\eta)^g(v_-\+iv_\eta)^{a'}(v_-\-1\+iv_\eta)^{b'}.
\label{J+J-}
\eea
Here, $\mathcal{J}_+$ and $\mathcal{J}_-$ are weakly dependent, through an infinitesimal number $\eta$, 
on the antiholomorphic ($-$) variables and the holomorphic ($+$) variables, respectively.
If we fix the holomorphic variables first, the dependence of $\mathcal{J}_-$ on the holomorphic variables 
determines relative positions of the integration paths and the two branch points $0$ and $1$ 
on the complex planes of the antiholomorphic variables.
For this relative positions, we can observe there are two very distinct cases. 
\begin{figure}[htbp]	
	\begin{center}
		\begin{minipage}{.45\linewidth} 
			\includegraphics[width=1.0 \linewidth]{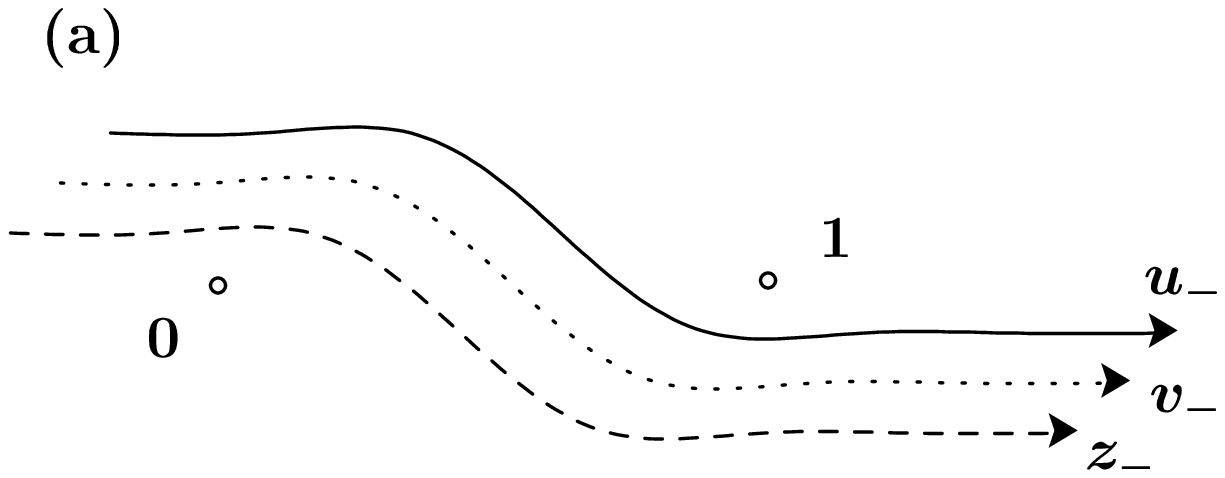} 
		\end{minipage}
		\hspace{1.0pc}   
		\begin{minipage}{.45\linewidth} 
			\includegraphics[width=1.0 \linewidth]{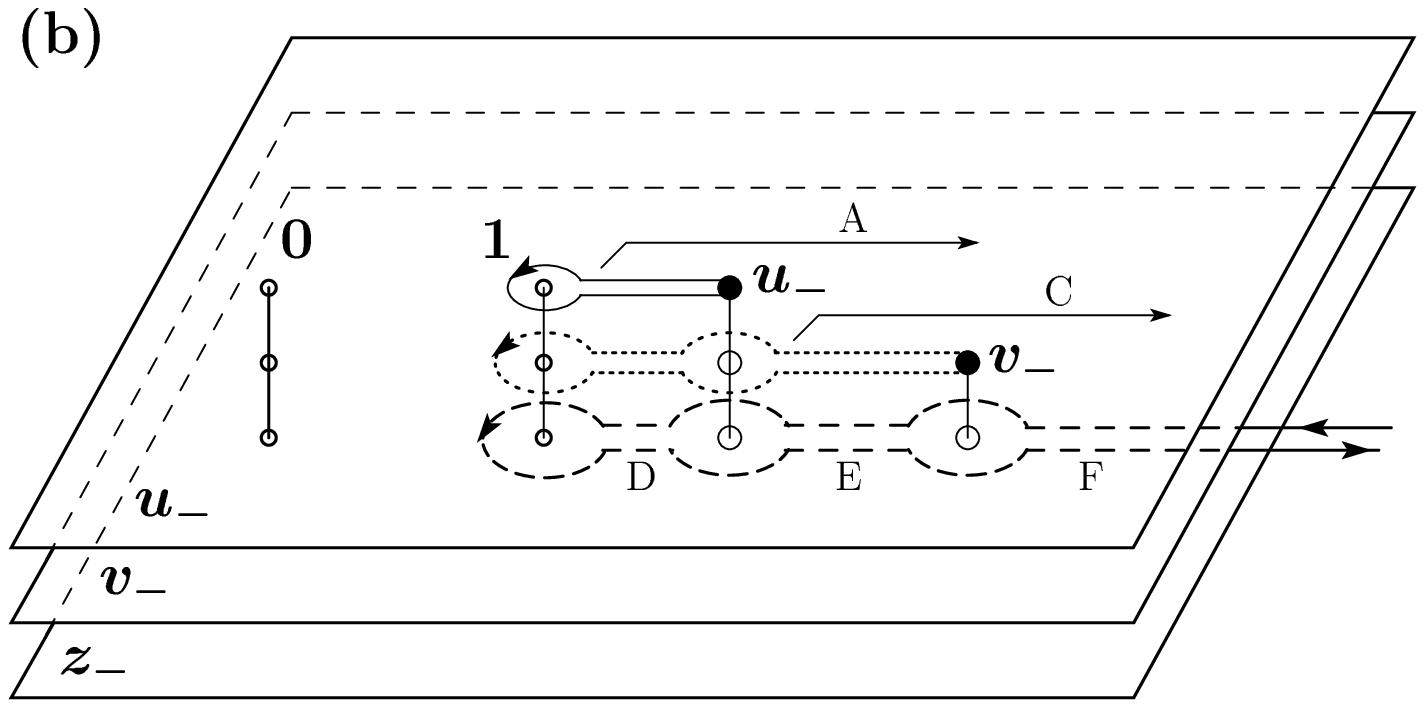}
		\end{minipage}
	\end{center}
	\caption{Deformation of the paths on the antiholomorphic planes.
Both figures correspond to the sequence of the holomorphic variables $z_+<v_+<u_+$.
(a) Before the deformations. \qquad 
(b) After the deformations. The five symbols A, C, D, E and F are assigned to each 
regularization of the interval. The case with $v_- < u_-$ is omitted here,
 and thus the regularization B in Figure \ref{tree} does not appear.}
	\label{paths}
\end{figure}

The first case is trivial, and corresponds to a choice of the holomorphic variables such that 
at least one of the $z_+,u_+$ and $v_+$ lies outside the interval $[0, 1]$.
Then, according to (\ref{J+J-}), one can find at least one integration path on the antiholomorphic plane such that 
all of the points $0$, $1$ and the other paths projected are located 
on the same side of the plane.
Therefore, the path can be contracted to a point by adding a semicircle of infinite radius and the corresponding integral should vanish. 

There is, however, the second case in which all of the $z_+,u_+$ and $v_+$ 
belong to the interval $(0, 1)$.
On the antiholomorphic planes,  
this means all the paths of $z_-, u_-$ and $v_-$ intersect the segment $[0, 1]$ 
in the same sequence as $z_+,u_+$ and $v_+$ lie on the interval $(0, 1)$.
Take a sequence: $z_+<v_+<u_+$ as in Figure \ref{paths}-(a), for instance.
Now, we deform the paths so that 
the resulting paths encircle the point $1$, the real axis from the upper and the lower side
(Figure \ref{paths}-(b)). 
Since the path nearest to the point $1$ is that of $u_-$, 
we first deform it, then that of $v_-$, and finally that of $z_-$.
Note, in deforming the path of, say $v_-$,  
the fixed variable $u_-$ is projected on the $v_-$-plane as a branch point.  

As a result of the deformation, we have a pairing of the paths 
in the upper and the lower half planes for each variable $z_-, u_-$ and $v_-$.
Each paring of the path can be decomposed into 
a sum of the integral over the regularizations of intervals 
(see \ref{regularization} for the definition of the regularizations
and ref. \cite{aomoto} for the summation of them).

Taking the presence of the branch cuts into account,
 we now attach the appropriate factors on these regularizations.
Since each factor comes as a difference of two phase factors on the upper and the lower half plane,
it takes the form of a $\sin$-function.
In the following, we assign the numbers $\{1,2,3\}$ for the sequences of variables $\{(z_\pm<v_\pm<u_\pm), (v_\pm<z_\pm<u_\pm), (v_\pm<u_\pm<z_\pm)\}$, respectively.

\begin{figure}[!tbp]
\begin{center}
\includegraphics[width=14cm]{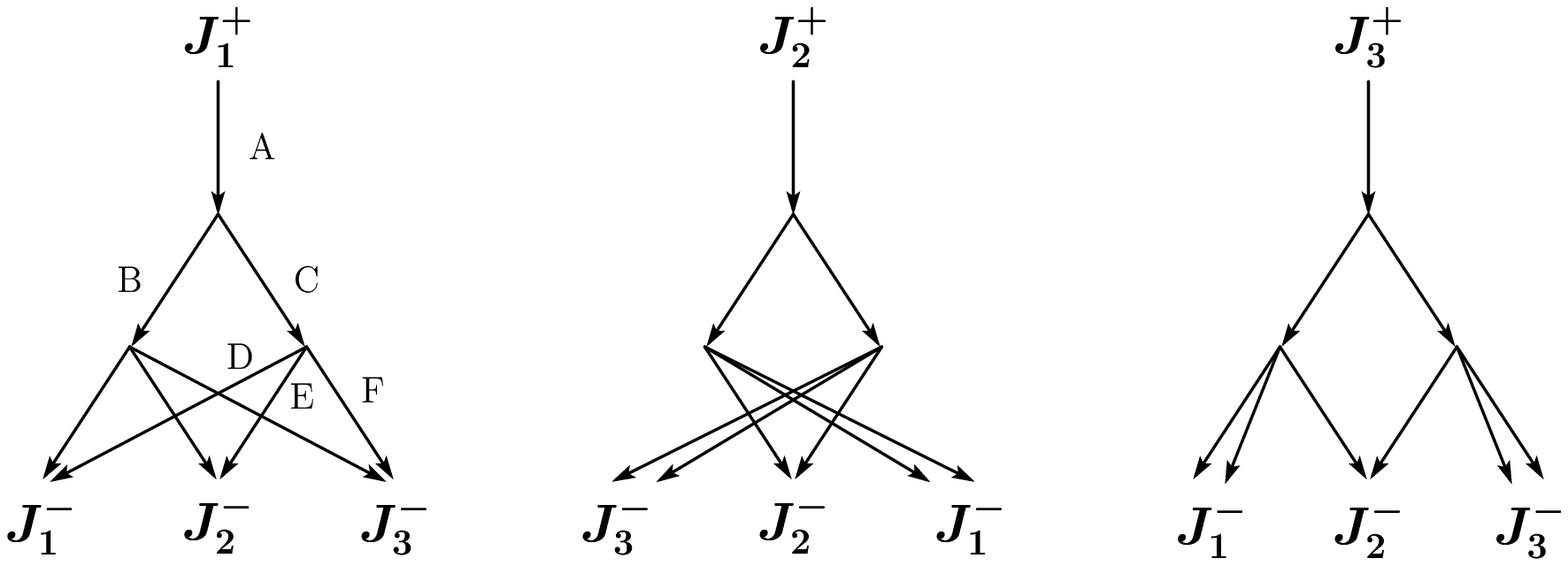}
\end{center}
\caption{Pairings of the paths (the regularizations of the intervals)
that contribute the matrix elements of $\mathcal{M}$. 
Each branch correspond to the regularization 
(see also Figure \ref{paths} for the branches A, C, D, E and F). 
The branches $B$ and $C$, for instance, 
correspond to the regularization in the $v_-$-plane 
with the sequence $v_-<u_-$ and $u_-<v_-$, respectively.
Each regularization has a certain factor on it due to the presence of the branch cuts.}
\label{tree}
\end{figure}

Consequently, from (\ref{J+J-}), we obtain a scattering-type formula
\beq
I=(-2)\cdot \begin{pmatrix}J_1^+& J_2^+& J_3^+\end{pmatrix}
\  \mathcal{M} \ 
\begin{pmatrix}J_1^-\\ J_2^-\\ J_3^-\end{pmatrix},
\label{i_amplitude}
\eeq
with
\bea
\fl\mathcal{M}=\!\!\left(\!\!
\begin{array}{lll}
\substack{ s(b) s\left(b'\right) \left[s\left(b'\right)+s\left(g+b'\right)\right]} &
\substack{ s(b+f) s\left(b'\right) \left[s\left(b'\right)+s\left(g+b'\right)\right]}   &
\substack{ s(b+2 f) s\left(b'\right) \left[s\left(b'\right)+s\left(g+b'\right)\right]} \\
\substack{ s(b) s\left(b'\right) \left[s\left(f+b'\right)+s\left(f+g+b'\right)\right]} &
\substack{ s(b) s\left(b'\right)^2+s(b+f) s\left(f+g+b'\right)}& 
\substack{ s(b+f)s\left(b'\right) \left[s\left(b'\right)+s\left(g+b'\right)\right]} \\
\substack{ s(b) s\left(f+b'\right) \left[s\left(f+b'\right)+s\left(f+g+b'\right)\right]}&
\substack{ s\left(b'\right) s\left(f+b'\right)+s(b) s\left(b'\right)s\left(f+g+b'\right)}&
\substack{ s(b) s\left(b'\right) \left[s\left(b'\right)+s\left(g+b'\right)\right]}
\end{array}
\!\!\right)\!\!,
\label{matrixelements}
\eea
where a notation $s(x)=\sin(\pi x)$ is used 
\footnote{It would be interesting if a possible relation 
between the scattering matrix $\mathcal{M}$ in (\ref{i_amplitude}) and 
$\mathcal{M}_\pm$ in (\ref{relations}) in this paper, and 
the intersection matrix and the monodromy invariant hermitian form
 in \cite{mimachi} was elaborated.}.
The dimension of the basis is $(3!)/2=3$,
 where $2$ comes from the symmetry between $u$ and $v$.
The factor 2 in (\ref{i_amplitude}) is necessary because of  
the same symmetry in $J^+$ basis. 
Each matrix element of $\mathcal{M}$ is a sum of two term;
each term is a product of three $\sin$-functions 
attached onto the regularization of the intervals
on $z_-, u_-$ and $v_-$-planes. 
Further, we have defined the initial and the final state basis as
\bea
J_1^+=\int_0^1\!\!\! du\int_0^u\!\!\! dv\int_0^v\!\!\! dz &z^a(1-z)^b (v-z)^f (u-z)^f H_+(u,v), \label{J1+}\\ 
J_2^+=\int_0^1\!\!\! du\int_0^u\!\!\! dz\int_0^z\!\!\! dv &z^a(1-z)^b (z-v)^f (u-z)^f H_+(u,v),\\ 
J_3^+=\int_0^1\!\!\! dz\int_0^z\!\!\! du\int_0^u\!\!\! dv &z^a(1-z)^b (z-v)^f (z-u)^f H_+(u,v),
\eea
and
\bea
J_1^-=\int_1^\infty \!\!\!\!\! dz\int_z^\infty\!\!\!\!\! dv\int_v^\infty\!\!\!\!\! du &z^a(z-1)^b (v-z)^f (u-z)^f H_-(u,v), \\ 
J_2^-=\int_1^\infty \!\!\!\!\! dv\int_v^\infty\!\!\!\!\! dz\int_z^\infty\!\!\!\!\! du &z^a(z-1)^b (z-v)^f (u-z)^f H_-(u,v),\\ 
J_3^-=\int_1^\infty \!\!\!\!\! dv\int_v^\infty\!\!\!\!\! du\int_u^\infty\!\!\!\!\! dz &z^a(z-1)^b (z-v)^f (z-u)^f H_-(u,v),
\label{J3-}
\eea
where we have used the notation
\bea
H_+(u,v)&=u^{a'}(1-u)^{b'}(u-v)^g\ v^{a'}(1-v)^{b'},\\
H_-(u,v)&=u^{a'}(u-1)^{b'}(u-v)^g\ v^{a'}(v-1)^{b'}.
\eea
The matrix elements of $\mathcal{M}$ are conveniently understood 
if we draw the tree-diagrams which show the sequences of the variables
under the process of the deformation 
(Figure \ref{tree}). 

\subsection{Representation of the basis through triple hypergeometric series}\label{hypergeometric}
By a suitable change of variables, the integral in (\ref{J1+}) is transformed into 
an integral over the unit cube:
\bea\fl
J_1^+=\int_0^1\!\!\! d\tilde{u}\int_0^1\!\!\! d\tilde{v}\int_0^1\!\!\! d\tilde{z}
\ \ \tilde{u}^{2+a+2a'+2f+g}(1-\tilde{u})^{b'}&
\tilde{v}^{1+a+a'+f}(1-\tilde{v})^{g}
\tilde{z}^{a}(1-\tilde{z})^{f}
\nonumber \\
&\cdot
(1-\tilde{v}\tilde{z})^{f}
(1-\tilde{u}\tilde{v}\tilde{z})^{b}
(1-\tilde{u}\tilde{v})^{b'}.
\eea
In general, all of $J_l^\pm\ (l=1, 2, 3)$ in (\ref{J1+})-(\ref{J3-}) can be cast into the same form,
namely,
\bea\fl
J_l^\pm=\int_0^1\!\!\! d\tilde{u}\int_0^1\!\!\! d\tilde{v}\int_0^1\!\!\! d\tilde{z}
\ \ \tilde{u}^{\lambda-1}(1-\tilde{u})^{\lambda'-1}&
\tilde{v}^{\mu-1}(1-\tilde{v})^{\mu'-1}
\tilde{z}^{\nu-1}(1-\tilde{z})^{\nu'-1}
\nonumber \\
&\cdot
(1-\tilde{v}\tilde{z})^{-p}
(1-\tilde{u}\tilde{v}\tilde{z})^{-q}
(1-\tilde{u}\tilde{v})^{-r},
\label{J*}
\eea
with each different set of exponents $\{\lambda, \mu, \nu,\lambda', \mu', \nu',p, q, r\}$.
In this paper, the value of the triple integral in the right hand side of (\ref{J*}) is 
denoted as a compact symbol:
\beq
\left\langle\begin{matrix}
\lambda     & \lambda'   & p\\
\mu      & \mu'    & q\\
\nu  & \nu' & r
\end{matrix}\right\rangle .
\label{symbol}
\eeq
Then each of the base $J_l^\pm\ (l=1, 2, 3)$ in (\ref{J1+})-(\ref{J3-}) looks like, 

\bea
\fl
J_1^+&=\left\langle\begin{matrix}
\xi     & 1\+b'   & -f\\
2\+a\+a'\!\+f                      & 1\+g    & -b\\
1\+a  & 1\+f                                & -b'
\end{matrix}\right\rangle,\qquad&
J_1^-=\left\langle\begin{matrix}
-\xi\-b\-2b'     & 1\+b   & -f\\
1-\xi+a+b                        & 1\+f    & -b'\\
-1\-a'\!\-b'\!\-f\-g              & 1\+g   &-b'
\end{matrix}\right\rangle ,\nonumber\\
\fl
J_2^+&=\left\langle\begin{matrix}
\xi     & 1\+b'   & -g\\
2\+a\+a'\!\+f                      & 1\+f    & -b'\\
1\+a'  & 1\+f                                & -b
\end{matrix}\right\rangle,&
J_2^-=\left\langle\begin{matrix}
-\xi\-b\-2b'     & 1\+b'   & -g\\
1-\xi+a'\!+b'         &1\+f              & -b'\\
-1\-a'\!\-b'\!\-f\-g                              & 1\+f &-b
\end{matrix}\right\rangle ,\nonumber\\
\fl
J_3^+&=\left\langle\begin{matrix}
\xi     & 1\+b   & -f\\
2+2a'\!+g\                       & 1\+f    & -b'\\
1\+a'  & 1\+g                                & -b'
\end{matrix}\right\rangle,&
J_3^-=\left\langle\begin{matrix}
-\xi\-b\-2b'     & 1\+b'   & -f\\
1-\xi+a'\!+b'                        & 1\+g    & -b\\
-1-a-b-2f                              & 1\+f &-b'
\end{matrix}\right\rangle,\nonumber\\
\fl  & 
\label{j_parameters}
\eea
where, for brevity, a notation $\xi=3+a+2a'+f+2g$ is used.

Performing the binomial expansion for the last three factors in (\ref{J*}), we get the following triple hypergeometric series: 
\bea
\fl
J_l^\pm&=\gamma_l^\pm \cdot S_l^\pm,\\
\fl
\gamma_l^\pm&=\frac{\Gamma(\lambda)\Gamma(\lambda')}{\Gamma(\lambda\+\lambda')}
\frac{\Gamma(\mu)\Gamma(\mu')}{\Gamma(\mu\+\mu')}\frac{\Gamma(\nu)\Gamma(\nu')}{\Gamma(\nu\+\nu')},
\label{gamma_l}\\
\fl
S_l^\pm
&=\sum_{i=0}^{\infty}\sum_{j=0}^{\infty}\sum_{k=0}^{\infty}
\frac{(p)_{i}(q)_{j}(r)_{k}}{i!\ \ j!\ \ k!}
\frac{(\lambda)_{j+k}\ \ \ (\mu)_{i+j+k}\ \ \ (\nu)_{i+j}}
{(\lambda+\lambda')_{j+k}(\mu+\mu')_{i+j+k}(\nu+\nu')_{i+j}},
\label{s_l}
\eea
where the Pochhammer symbol $(x)_k=x(x+1)\cdots(x+k-1)$ is used.
The parameters $\{\lambda, \mu, \nu,\lambda', \mu', \nu',p, q, r\}$ is related
to the exponents $\{a,b,a',b',f,g\}$ in (\ref{I}) as indicated in (\ref{j_parameters}).
This series representation is particularly useful when some of $\{p, q, r, \lambda, \mu, \nu\}$
are non-positive integer plus $\mathcal{O}(\ep)$.
In that case, a separation of the order in $\ep$ occurs.

\subsection{Linear relations between the basis}\label{App_relations}
For generic values of the parameters $\{a, b, a', b', g, f\}$, 
there exist three independent linear relations between the triple integral basis 
$\{J^+,J^-\}$.
Although in principle we can evaluate the coefficients in the epsilon expansion 
of these integrals using the series expressions, 
but in practice some of the base are happened to be difficult to expand in $\epsilon$, 
while the others to be more straightforward. 
Hence, these relations are necessary in our epsilon expansion calculation of the RG functions.

The explicit form of the relations are,
\beq
\mathcal{M}_+\begin{pmatrix}J_1^+\\ J_2^+ \\ J_3^+\end{pmatrix}
=-\mathcal{M}_-\begin{pmatrix}J_1^-\\ J_2^- \\ J_3^-\end{pmatrix},
\label{relations}
\eeq
with
\bea 
\fl\mathcal{M}_+=\begin{pmatrix}
\substack{s(a)s(a')\cr \cdot2s(a'\+g/2)c(g/2)} & 
\substack{s(a\+f)s(a')\cr \cdot2s(a'\+g/2)c(g/2)} &
\substack{s(a\+2f)s(a')\cr \cdot2s(a'\+g/2)c(g/2)} \vspace{2mm}\\ 
\substack{s(a)s(a')\cr \cdot2s(a'\+f\+g/2)c(g/2)} & 
\substack{s(a')\cr \cdot\left[s(a)s(a')+s(a\+f)s(a'\+g\+f)\right]} & 
\substack{s(a\+f)s(a')\cr \cdot2s(a'\+g/2)c(g/2)}\vspace{2mm}\\
\substack{s(a)s(a'\+f)\cr \cdot2s(a'\+f\+g/2)c(g/2)} & 
\substack{s(a)s(a')\cr \cdot2s(a'\+f\+g/2)c(g/2)} & 
\substack{s(a)s(a')\cr \cdot2s(a'\+g/2)c(g/2)}
\end{pmatrix},& \\ \nonumber \\
\fl\mathcal{M}_-=\begin{pmatrix} 
\substack{s(a\+b\+2f)s(a'\+b'\+g\+f)\cr \cdot2s(a'\+b'\+f\+g/2)c(g/2)} & 
\substack{s(a\+b\+2f)s(a'\+b'\+g\+f)\cr \cdot 2s(a'\+b'\+g/2)c(g/2)} &
\substack{s(a\+b\+2f)s(a'\+b'\+g)\cr \cdot2s(a'\+b'\+g/2)c(g/2)} \vspace{2mm}\\ 
\substack{s(a\+b\+f)s(a'\+b'\+g\+f)\cr \cdot2s(a'\+b'\+f\+g/2)c(g/2)} & 
\substack{s(a'\+b'\+g\+f)\cr \cdot\left[ s(a\+b\+2f)s(a'\+b'\+g\+f)\right.\cr \left.\ \ \ \ +\ s(a\+b\+f)s(a'\+b')\right]} & 
\substack{s(a\+b\+2f)s(a'\+b'\+g\+f)\cr \cdot2s(a'\+b'\+g/2)c(g/2)} \vspace{2mm}\\
\substack{s(a\+b)s(a'\+b'\+g\+f)\cr \cdot2s(a'\+b'\+f\+g/2)c(g/2)} & 
\substack{s(a\+b\+f)s(a'\+b'\+g\+f)\cr \cdot2s(a'\+b'\+g/2)c(g/2)} & 
\substack{s(a\+b\+2f)s(a'\+b'\+g\+f)\cr \cdot2s(a'\+b'\+f\+g/2)c(g/2)} 
\end{pmatrix},&
\eea
where we have used the notation $c(x)=\cos(\pi x)$.
We now derive the first one of these three relations for an illustration.
Since it is the analyticity of the integrand on the region except the branch cuts 
that makes the relation valid, 
the basic strategy is to deform successively each integration path 
defined on each complex plane. 
To keep track of the successive deformations of the integration paths on 
the three complex planes, 
we use a simple semicircular diagram to represent the regularization of the interval 
(see Figure \ref{reg_circles}).

\begin{figure}[!htbp]
\begin{center}
\includegraphics[width=8cm]{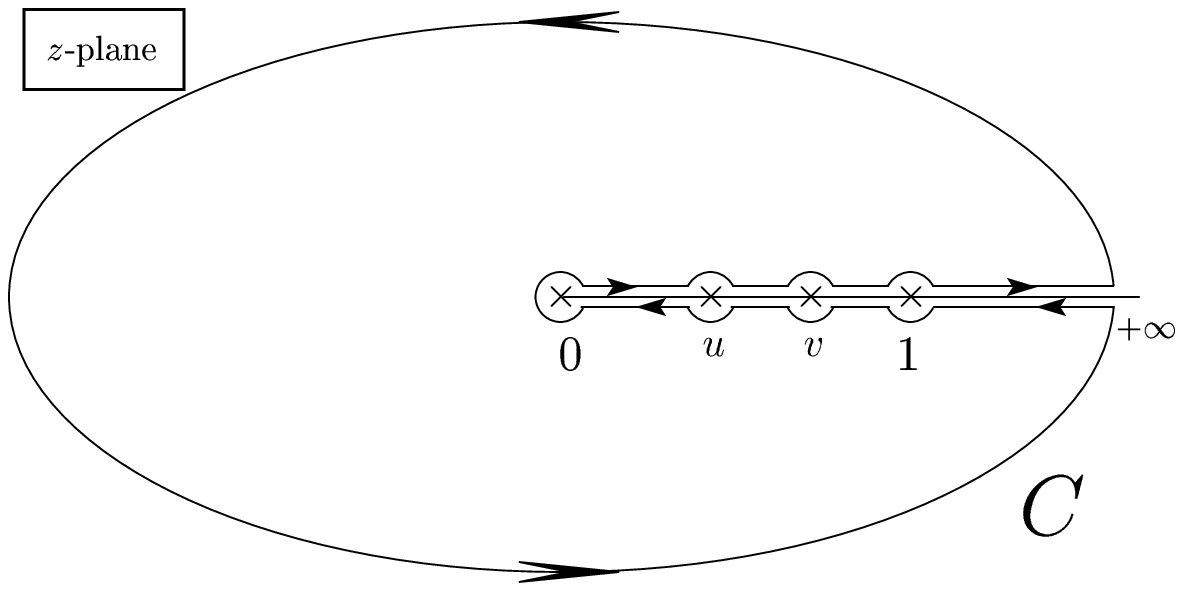}
\caption{The contour C defined on the $z$ complex plane.}
\label{contour_cc}
\end{center}
\end{figure}

Consider the $z$-plane in which 
all the branch cuts are taken along the real axis from each branch point to positive infinity. 
Since the integrand has no branch points except the non-negative real axis,
the integral along the contour $C$ in Figure \ref{contour_cc} is zero. 

\begin{figure}[!h]
\begin{center}
\includegraphics[width=12cm]{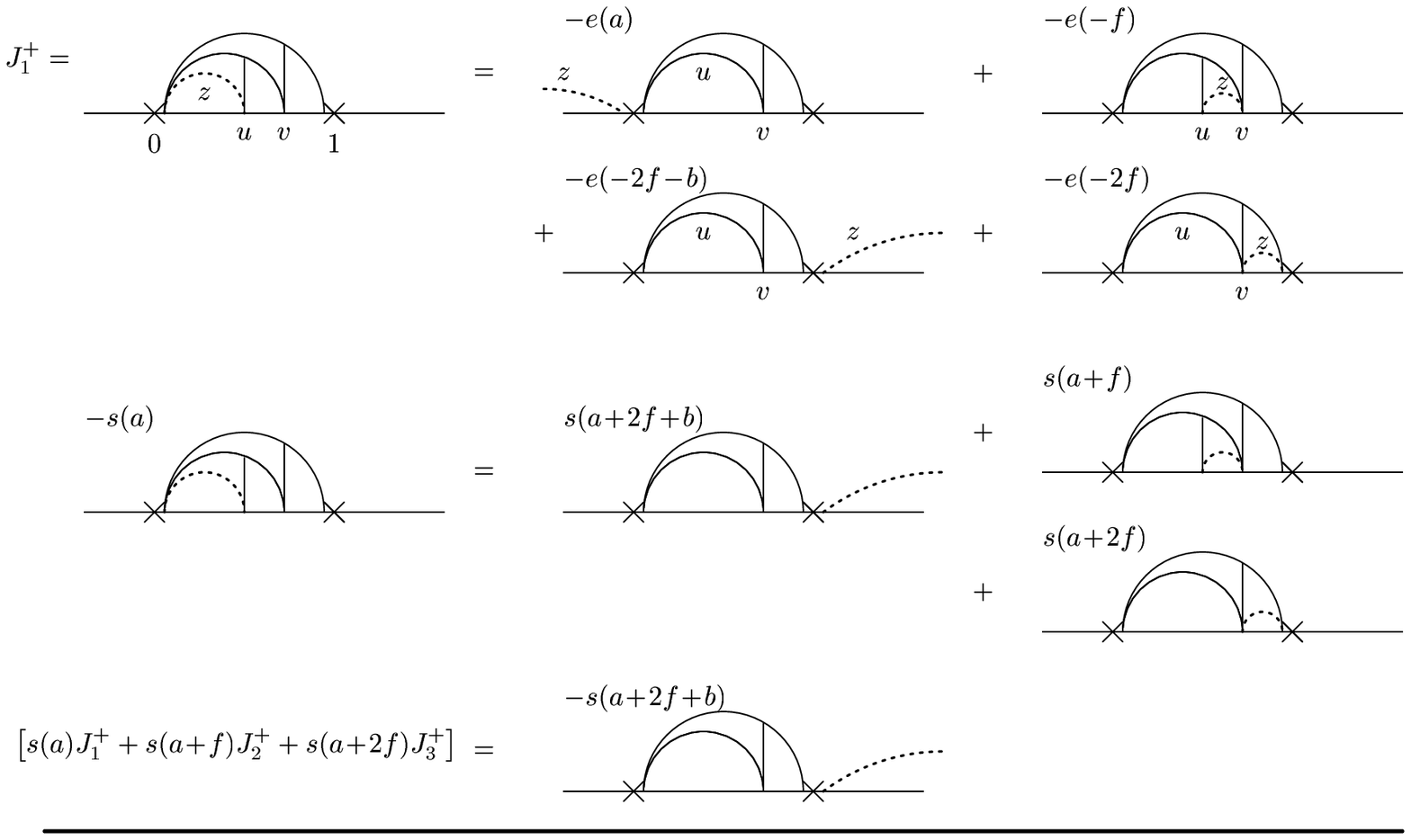}
\includegraphics[width=12cm]{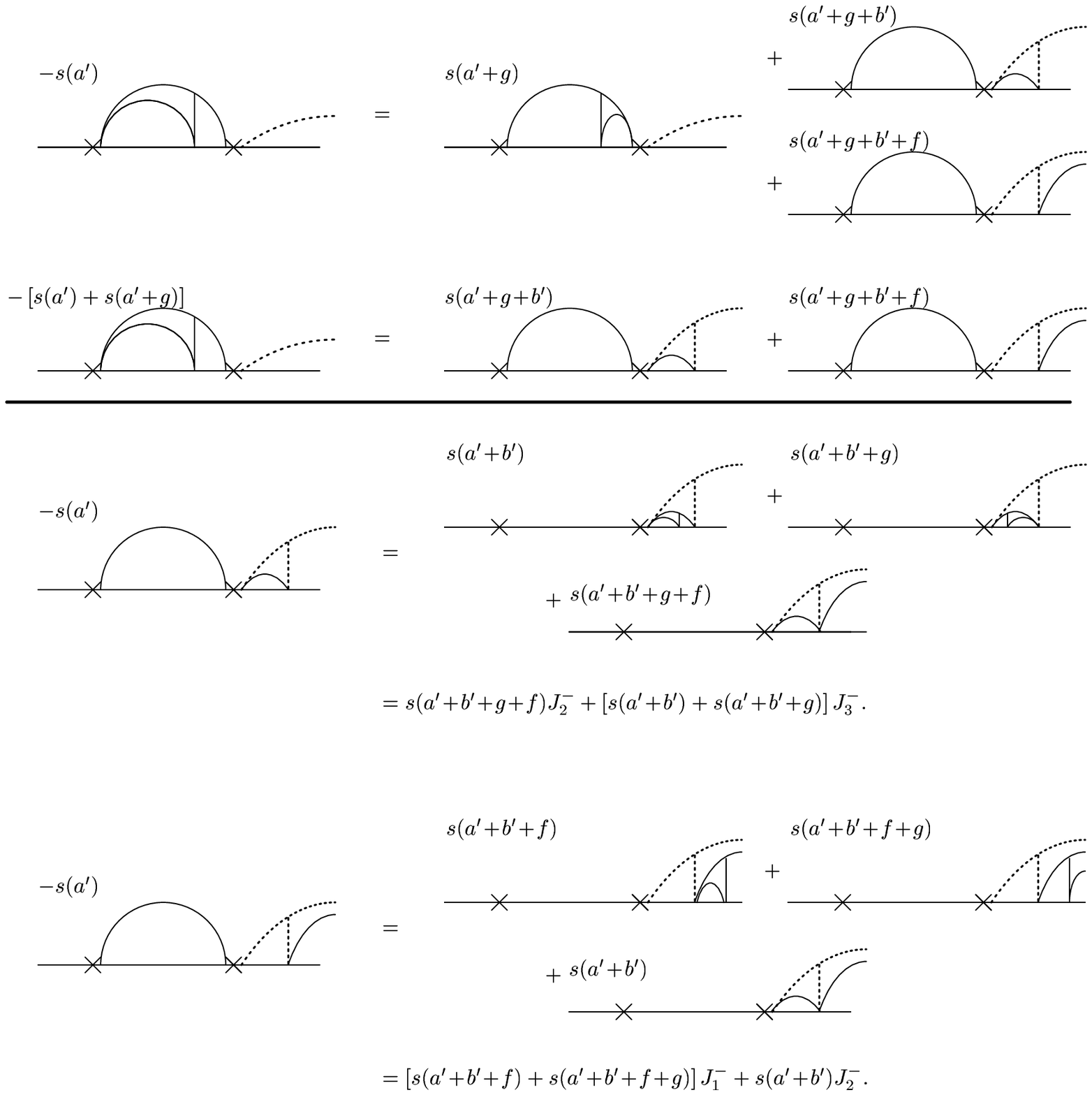}
\caption{The successive deformations of the paths on the three complex planes.
The notation $e(x)=\exp(\pi x)$ and $s(x)=\sin(\pi x)$ are used.
The semicircular shape represents the regularization of the interval (see Figure \ref{reg_circles}).
The dotted curves indicate the integrations over the variable $z$, 
while the bold curves indicate that of $u$ or $v$.
The vertical lines indicate the relative positions of the branch cuts induced for the other variables.}
\label{successive}
\end{center}
\end{figure}

Then the $z$-integral on a regularized interval $\mathrm{reg}\left[0,u\right]$ 
can be expressed by a certain linear combination of the integral on the intervals
 $\mathrm{reg}\left[u,v\right]$, $\mathrm{reg}\left[v,1\right]$ and $\mathrm{reg}\left[1,\infty\right]$
as shown in the first equality in Figure \ref{successive}.
Each coefficient comes from the pairing of the two segments 
which lie on opposite sides of the branch cuts.    
In this way, we can shift the integration path for each variables 
to the intervals in the positive real direction.

\newpage

\end{document}